%% file: LAE.tex
\documentclass[]{aa}
\pdfoutput=1
\usepackage{natbib}
\usepackage{epsfig}
\usepackage{epstopdf}
\usepackage{multirow}
\usepackage{caption}
\usepackage[varg]{txfonts}
\usepackage{pdflscape}

\usepackage{hyperref}
\usepackage{amssymb}
\usepackage{supertabular}

\def \aap  {A\&A}

\def \apjs  {ApJS}
\def \apjl  {ApJL}

\def \chisq  {\ifmmode  \chi^2   \else  $\chi^2$  \fi}  
\def \spose#1{\hbox  to 0pt{#1\hss}}  
\def \lta{\mathrel{\spose{\lower 3pt\hbox{$\sim$}}\raise  2.0pt\hbox{$<$}}}
\def \gta{\mathrel{\spose{\lower  3pt\hbox{$\sim$}}\raise 2.0pt\hbox{$>$}}}

\def \kms {\ifmmode  \,\rm km\,s^{-1} \else $\,\rm km\,s^{-1}  $ \fi }
\def \kpc {\ifmmode  {\rm~kpc}  \else ${\rm~kpc}$\fi}  
\def \pc {\ifmmode  {\rm~pc}  \else ${\rm~pc}$ \fi  }  
\def \Gyr {\ifmmode  {\rm~Gyr}  \else ${\rm~Gyr}$\fi}
\def \Msun {\ifmmode M_{\odot} \else $M_{\odot}$ \fi} 
\def \Lsun {\ifmmode L_{\odot} \else $L_{\odot}$ \fi} 
\def \Rsun {\ifmmode R_{\odot} \else $R_{\odot}$ \fi} 
\def \Msunpyr {\ifmmode M_{\odot}{\rm~yr}^{-1} \else $M_{\odot}{\rm~yr}^{-1}$ \fi} 
\def \hMsun {\ifmmode h^{-1}\,\rm M_{\odot} \else $h^{-1}\,\rm M_{\odot}$ \fi}

\def \LCDM {\ifmmode \Lambda{\rm CDM} \else $\Lambda{\rm CDM}$ \fi}
\def \sig8 {\ifmmode \sigma_8 \else $\sigma_8$ \fi} 
\def \OmegaM {\ifmmode \Omega_{\rm M} \else $\Omega_{\rm M}$ \fi} 
\def \OmegaL {\ifmmode \Omega_{\rm \Lambda} \else $\Omega_{\rm \Lambda}$\fi} 
\def \Deltavir {\ifmmode \Delta_{\rm vir} \else $\Delta_{\rm vir}$ \fi}
\def \rhocrit {\ifmmode \rho_{\rm crit} \else $\rho_{\rm crit}$ \fi}
\def \rhou {\ifmmode \rho_{\rm u} \else $\rho_{\rm u}$ \fi}
\def \zc {\ifmmode z_{\rm c} \else $z_{\rm c}$ \fi}

\def \rhos {\ifmmode \rho_{\rm s} \else $\rho_{\rm s}$ \fi} 
\def \rs {\ifmmode r_{\rm s} \else $r_{\rm s}$ \fi} 
\def \cvir {\ifmmode c_{\rm vir} \else $c_{\rm vir}$ \fi} 
\def \Rvir {\ifmmode r_{\rm vir} \else $R_{\rm vir}$ \fi}
\def \Vvir {\ifmmode V_{\rm  vir} \else  $V_{\rm vir}$  \fi} 
\def \Mvir {\ifmmode M_{\rm  vir} \else $M_{\rm  vir}$ \fi}  
\def \Nvir {\ifmmode N_{\rm  vir} \else $N_{\rm  vir}$ \fi}  
\def \Jvir {\ifmmode J_{\rm vir} \else $J_{\rm vir}$ \fi} 
\def \Evir {\ifmmode E_{\rm vir} \else $E_{\rm vir}$ \fi} 
\def \vvir {\ifmmode v_{\rm vir} \else $v_{\rm vir}$ \fi} 
\def \lam {\ifmmode \lambda  \else $\lambda$ \fi} 
\def \lamp {\ifmmode \lambda^{\prime} \else $\lambda^{\prime}$  \fi} 
\def \Vmax {\ifmmode V_{\rm  max} \else  $V_{\rm max}$  \fi} 
\def \Mdm {\ifmmode M_{\rm  dm} \else $M_{\rm  dm}$ \fi}

\def \Mgas {\ifmmode M_{\rm gas} \else $M_{\rm gas}$ \fi} 
\def \Mcg {\ifmmode M_{\rm cg} \else $M_{\rm cg}$\fi} 
\def \Mhg {\ifmmode M_{\rm hg} \else $M_{\rm hg}$ \fi} 
\def \Mdisc {\ifmmode M_{\rm disc} \else $M_{\rm disc}$ \fi} 
\def \Md {\ifmmode M_{\rm d} \else $M_{\rm d}$ \fi} 
\def \Mda {\ifmmode M_{\rm d,0\%} \else $M_{\rm d,0\%}$ \fi} 
\def \Mdb {\ifmmode M_{\rm d,20\%} \else $M_{\rm d,20\%}$ \fi} 
\def \Mdc {\ifmmode M_{\rm d,40\%} \else $M_{\rm d,40\%}$ \fi} 
\def \md {\ifmmode m_{\rm d} \else $m_{\rm d}$ \fi} 
\def \Mb {\ifmmode M_{\rm b} \else $M_{\rm b}$ \fi} 
\def \Mbh {\ifmmode M_{\rm b,pri} \else $M_{\rm b,pri}$ \fi} 
\def \Mbs {\ifmmode M_{\rm b,sat} \else $M_{\rm b,sat}$ \fi} 
\def \zo {\ifmmode z_{0} \else $z_{0}$ \fi} 
\def \rd {\ifmmode r_{\rm d} \else $r_{\rm d}$ \fi}
\def \rg {\ifmmode r_{\rm g} \else $r_{\rm g}$ \fi}
\def \rb {\ifmmode r_{\rm b} \else $r_{\rm b}$\fi}
\def \rs {\ifmmode r_{\rm s} \else $r_{\rm s}$\fi}
\def \rc {\ifmmode r_{\rm c} \else $r_{\rm c}$\fi}
\def \rvir {\ifmmode r_{\rm vir} \else $r_{\rm vir}$\fi}
\def \rbh {\ifmmode r_{\rm b,pri} \else $r_{\rm b,pri}$ \fi} 
\def \rbs {\ifmmode r_{\rm b,sat} \else $r_{\rm b,sat}$ \fi} 
\def \zp {\ifmmode z_{\rm phot} \else $z_{\rm phot}$ \fi}
\def \zs {\ifmmode z_{\rm spec} \else $z_{\rm spec}$ \fi}
\def \Lya{\ensuremath{\mathrm{Ly}\alpha\ }}
\def \LLya{\ensuremath{\mathrm{L}_{Ly\alpha}}}
\def \Lyap{\ensuremath{\mathrm{Ly}\alpha.\ }}

\def \Lyam{{\ensuremath{\mathrm{Ly}\alpha}}}

\begin{document}

\defcitealias{Karman2015}{Paper 1}
\defcitealias{Caminha2015}{C16}

\title{MUSE integral-field spectroscopy towards the Frontier Fields cluster Abell S1063: II. Properties of low luminosity Lyman $\alpha$ emitters at z$>$3}  

\author{W.  Karman\inst{\ref{inst1}}\thanks{karman@astro.rug.nl} \and K. I. Caputi\inst{\ref{inst1}} \and G.~B.~Caminha\inst{\ref{inst2}} \and M.~Gronke\inst{\ref{inst:oslo}} \and C.~Grillo \inst{\ref{inst3},\ref{inst:mil}} \and  I.~Balestra\inst{\ref{obs_munich},\ref{inst4}} \and P.~Rosati\inst{\ref{inst2}} \and  E.~Vanzella\inst{\ref{inst5}}   \and D.~Coe\inst{\ref{inst6}} \and M.~Dijkstra\inst{\ref{inst:oslo}}  \and A.~M.~Koekemoer\inst{\ref{inst6}} \and D.~McLeod \inst{\ref{roe}} \and A.~Mercurio\inst{\ref{inst9}} \and M.~Nonino\inst{\ref{inst4}}}
 
\institute{ Kapteyn Astronomical Institute, University of Groningen, Postbus 800, 9700 AV Groningen, the Netherlands\label{inst1} 
\and Dipartimento di Fisica e Scienze della Terra,
  Universit\`a degli Studi di Ferrara, Via Saragat 1, I-44122 Ferrara,
  Italy \label{inst2}
\and Institute of Theoretical Astrophysics, University of Oslo, Postboks 1029 Blindern, 0315 Oslo, Norway \label{inst:oslo}
\and Dark Cosmology Centre, Niels Bohr Institute,
  University of Copenhagen, Juliane Maries Vej 30, DK-2100 Copenhagen,
  Denmark\label{inst3}
\and Dipartimento di Fisica, Universit\`a degli Studi di Milano, via Celoria 16, I-20133 Milano, Italy \label{inst:mil}  
\and University Observatory Munich, Scheinerstrasse 1, 81679 Munich, Germany\label{obs_munich} 
\and INAF - Osservatorio Astronomico di Trieste, via
  G. B. Tiepolo 11, I-34143, Trieste, Italy \label{inst4}
\and INAF–Bologna Astronomical Observatory, via Ranzani 1, I-40127 Bologna, Italy \label{inst5}
\and Space Telescope Science Institute, 3700 San Martin
  Drive, Baltimore, MD 21208, USA \label{inst6}
\and SUPA, Institute for Astronomy, University of Edinburgh, Royal Observatory, Edinburgh EH9 3HJ, UK \label{roe}
\and INAF - Osservatorio Astronomico di Capodimonte, Via
  Moiariello 16, I-80131 Napoli, Italy \label{inst9}
}

\abstract{
In spite of their conjectured importance for the Epoch of Reionization, the properties of low-mass galaxies are currently still under large debate. In this 
article, we study the stellar and gaseous properties of faint, low-mass
galaxies at $z>3$. We observed the Frontier Fields cluster Abell S1063
with MUSE over a 2 arcmin$^2$ field, and
combined integral-field spectroscopy with gravitational lensing to 
perform a blind search for intrinsically faint \Lya emitters (LAEs).
We determined in total the redshift of 172 galaxies of which 14 are lensed LAEs at $z$=3-6.1. We increased the number of spectroscopically-confirmed
multiple-image families from 6 to 17 and updated our gravitational-lensing model
accordingly. 
The lensing-corrected \Lya luminosities are with $L_{\rm \Lya}\lesssim 10^{41.5}$ erg/s among the lowest for spectroscopically
confirmed LAEs at any redshift. We used expanding gaseous shell models to
fit the \Lya line profile, and find low column densities and expansion
velocities. This is to our knowledge the first time that gaseous properties
of such faint galaxies at $z\gtrsim3$ are reported. We performed SED modelling to broadband photometry from the
{\em U}-band through the infrared to determine the stellar properties
of these LAEs. The stellar masses are very low (10$^{6-8}$ \Msun),
and are accompanied by very young ages of 1-100 Myr. The very high specific star
formation rates ($\sim$100~Gyr$^{-1}$) are characteristic of starburst galaxies, and we find
that most galaxies will double their stellar mass in $\lesssim20$ Myr. The UV-continuum
slopes $\beta$ are low in our sample, with $\beta<-2$ for all
galaxies with $M_\star<10^8\Msun$. 
We conclude that  our
low-mass galaxies at $3<z<6$ are forming stars 
at higher rates when correcting for stellar mass effects than seen locally or in more massive galaxies. The young
stellar populations with high star-formation rates
and low \ion{H}{I} column densities lead to continuum slopes
and LyC-escape fractions expected for a scenario where low mass galaxies
reionise the Universe.
}
              
\date{Received ... /
Accepted ...}


\titlerunning{LAEs}
\authorrunning{W. Karman et al.}

\setcounter{footnote}{1}

\maketitle

\section{Introduction}
\label{sec:intro}

The evolution of the brightest galaxies in the Universe has now been studied in significant detail out to $z\sim$8 \citep[e.g][]{Bouwens2014,Salmon2015,Caputi2015}, and is in accordance with the now well-established $\Lambda$CDM model. The study of low-mass, faint galaxies at high-{\em z} is, instead, almost a completely unknown territory. Gaining a greater knowledge on these faint galaxies is important as they are the building blocks of the observed more massive galaxies at lower redshifts, and they are currently seen as the main candidates for reionizing the Universe at $z=6-10$ (\citealt{Wise2014,Kimm2014}, but see \citealt{Sharma2016}). 

Observationally, high-redshift, low-mass galaxies have been elusive to date. The Lyman break technique \citep[e.g.][]{Steidel1996,Steidel2003,Bouwens2011} and spectral-energy-distribution (SED) fitting codes \citep[e.g.][]{Caputi2011,Ilbert2013}, which are well proven for intermediate-mass and massive galaxies, are not applicable to these faint sources in all but the deepest multiwavelength studies \citep[e.g.][]{Ouchi2010,Schenker2013} or until JWST is operating \citep[e.g.][]{Gardner2009,Bisigello2016}. Therefore, other approaches are needed to understand the faint end of the galaxy population. One possible approach is looking for counterparts of absorbers in quasar lines-of-sights \citep[e.g.][]{Arrigoni2016}, but this is only feasible for bright quasars with long spectroscopic observations \citep[e.g][]{Rauch2008}. Fortunately, the \Lya line is redshifted in the optical domain for galaxies at z$\gtrsim$3. Although stars in massive galaxies are often surrounded by a dusty inter-stellar and circum-galactic medium which absorbs all \Lya photons \citep[e.g.][]{Laursen2009}, less-massive star-forming galaxies are often found with significant \Lya emission \citep[e.g.][]{Oyarzun2016}. Therefore, searching for galaxies with strong emission lines in the optical can be used to identify low-mass high-redshift galaxies.  

Another possibility are optical narrowband studies, which search for galaxies with strong emission lines \citep[e.g.][]{Nilsson2009,Nakajima2012,Matthee2016} by looking for sources with strong colours between the narrowbands and overlapping broadband observations. By applying additional colour cuts representative of high-redshift galaxies, reliable candidates for \Lya emitters (LAEs) can be found. However, it has been shown that low-redshift extreme line emitters can contaminate this sample \citep[e.g.][]{Atek2011,Penin2015}, and galaxies with intermediate \Lya line strengths will not survive the colour cuts. Another disadvantage of using narrow-band studies is that these selections are only useful for very narrow redshift ranges. 

Although \Lya has become the most important line to identify galaxies with redshifts between $2.5<z<7$ \citep[e.g.][]{Shimasaku2006,Dawson2007,Diaz2015,Trainor2015}, it is still unclear what governs whether a galaxy is a LAE or not. It has been found that LAEs are in general less dusty than LBGs \citep[e.g.][]{Atek2014}, but they have very similar stellar properties at fixed luminosity \citep{Shapley2001,Kornei2010,Yuma2010, Mallery2012,Jiang2016}. There is evidence however, that the prevalence of \Lya emission is much higher in less luminous systems \citep{Stark2010,ForeroRomero2012} and less massive systems \citep{Oyarzun2016}. A similar trend is also found for the equivalent width (EW) of \Lya, which anticorrelates with UV luminosity \citep[e.g.][]{Shapley2003,Gronwall2007,Kornei2010}. Further, the fraction of LBGs with \Lya emission increases with redshift out to $z\sim6$ \citep[e.g.][]{Ouchi2008,Cassata2011,Cassata2015,Pentericci2011,CurtisLake2012,Schenker2012,Henry2012}, but experiences a rapid decrease afterwards \citep[e.g.][]{Kashikawa2011,Caruana2012,Caruana2014,Ono2012,Schenker2012,Stark2010,Pentericci2014}. This drop has theoretically only been explained succesfully as arising from reionization \citep{Dijkstra2011,Jensen2013,Mesinger2015,Choudhury2015}, although additional processes might be involved \citep[e.g.][]{Dijkstra2014,Choudhury2015}.

While the broadband photometry can reveal much about the stellar and dust properties of galaxies, the \Lya line profile provides important information on the properties of the gas \citep[e.g.][]{Verhamme2006,Verhamme2008,Sawicky2008}. 
Since only \Lya photons shifted out of resonance can effectively escape the galaxy, moving gas clouds such as outflows allow \Lya photons to escape \citep[e.g.][]{Schaerer2011,Laursen2013,Dijkstra2014}. Dust absorbs the \Lya photons and emits them at longer wavelengths, while a patchy distribution of the surrounding medium allows the photons to escape. Therefore, by careful modelling of the \Lya line, one can learn about the properties of the gaseous medium in and surrounding galaxies. 
Recently, it has been demonstrated that galaxies with extreme optical and near UV emission lines are often exhibiting narrow \Lya emission \citep{Cowie2011,Henry2015,Izotov2016,deBarros2016,Vanzella2016b}. The fact that these galaxies are found to have \Lya emission both at low and high redshift, indicates that these so-called ``Green Peas'' might be good analogues of the high-redshift LAEs\citep[e.g.][]{Amorin2010,Amorin2015}. Another indication for a close resemblance between these galaxies is the finding that low stellar mass, high SFR, and low dust content correlate with \Lya emission both at low \citep[e.g.][]{Cowie2011,Henry2015} and high redshift \citep[e.g.][]{Jiang2016}. In addition, \citet{Vanzella2016a} and \citet{Izotov2016} found Lyman continuum leakage for two of these galaxies, making them important candidates for reionization.  

The Frontier Fields programme (hereafter FF, PI: J. Lotz; see \citealt{Lotz2016} and \citealt{Koekemoer2016}) provides an excellent opportunity to study intrinsically faint galaxies at high redshifts. Massive galaxy clusters provide a boost in depth thanks to the effect of gravitational lensing. The deep {\em HST} coverage over 7 different bands provides photometry for intrinsically faint sources which allows us to study their properties. Combining this deep gravitionally-lensed photometric survey with spectroscopy allows us to determine accurate stellar and gaseous properties down to an intrinsic faintness which is otherwise currently unachievable within a reasonable observing time. Abell S1063 (AS1063), the cluster studied here, is among the best studied FF clusters for which we have one of the best constrained and most precise strong lensing model available so far \citep[e.g.][]{Monna2014,Johnson2014,Richard2014a,Caminha2015,Diego2016}.

In \citet[][hereafter Paper I]{Karman2015}, we showed that using gravitational lensing in combination with the integral field spectrograph Multi Unit Spectroscopic Explorer (MUSE) we have been able to identify previously undetected, intrinsically faint LAE. In this work we expand on our previous results by adding observations of a second MUSE pointing covering the second half of the cluster, and using \Lya line profile modelling in combination with broadband photometry to study the properties of LAEs at $3<z<6$. In addition, we present an updated redshift catalogue using the full MUSE dataset.

The layout of this paper is as follows. In Section \ref{sec:data} we give a brief
overview of the MUSE performance and the obtained data, followed by the
data reduction process. In Section \ref{sec:results}, we describe our spectroscopic results, including
the determined redshifts and emission line properties. We used spectral energy distribution (SED) fitting to the broadband photometry
to study the stellar properties of these objects in Section \ref{sec:stars}. We summarise
and discuss our findings in Section \ref{sec:discussion}, and present our conclusions in Section \ref{sec:conclusions}. Throughout this paper, we adopt a cosmology with $H_0~=~70~$\kms~ Mpc$^{-1}$, $\Omega_M~=~0.3$, and $\Omega_\Lambda~=~0.7$. Unless we specify otherwise, all given star formation rates (SFRs) are derived from spectral energy distribution (SED) modelling. All magnitudes refer to the AB system, and we use a Chabrier initial mass function (IMF) over stellar masses in the range 0.1-100~\Msun.


\section{Observations}
\label{sec:data}

\subsection{Photometry}

The Hubble Frontier Fields programme\footnote{\url{https://archive.stsci.edu/prepds/frontier/}} (FF, PI: J. Lotz; see \citealt{Lotz2016} and \citealt{Koekemoer2016}) targets six galaxy clusters with large magnification factors, among which is AS1063.
The programme targets each cluster for a total of 140 orbits, divided over 7 bands in the optical and near infrared (NIR), reaching a 5 $\sigma$ depth
of $\sim 29$ mag in each of these bands. We used the available public {\em HST} data from this programme, retrieved from the Frontier Fields page at the STScI MAST Archive, to detect sources and measure locations and magnitudes of sources, adopting the current zeropoints, provided by the ACS and WFC3 teams at STScI, which are tabulated on the same MAST Archive page for these specific FF filters.
At the time of writing, the optical bands were fully observed for AS1063, while the NIR observations have had only a single orbit exposure. We used the v0.5 data products, which do 
not contain self calibration for this cluster. We used the images with a spatial resolution of 0.060\arcsec, in order to have a uniform pixel scale, without
oversampling the NIR images.

In addition to being a FF cluster, AS1063 is also part of the Cluster Lensing and Supernova Survey with Hubble \citep[CLASH][]{Postman2012} survey, which targets 25 gravitationally lensing clusters with {\em HST} in 16 bands.
We supplement our FF data with the CLASH data in 5 additional bands. These data are significantly less deep, but provide additional information for the brightest objects.
For all these filters, which are in addition to those used in the FF programme, we adopt the current zeropoints provided by the ACS and WFC3 teams at STScI\footnote{ACS zeropoints: \url{http://www.stsci.edu/hst/acs/analysis/zeropoints}}\footnote{WFC3 zeropoints: \url{http://www.stsci.edu/hst/wfc3/phot_zp_lbn}}.

As the LAEs discussed here all lie at $z>2.8$, the NIR images from {\em HST} do not cover the wavelength range above 4000 \AA\ restframe. Information at longer wavelengths
is therefore crucial to better constrain older stellar populations. We collected Hawk-I data in order to complement our
data at longer wavelengths.
The Hawk-I images \footnote{ESO Programme 095.A-0533, PI Brammer} were retrieved from the ESO Archive \footnote{\url{http://archive.eso.org}}. The whole dataset includes 997 images obtained in September 2015. After dark and flat correction, a first sky subtraction was performed without source masking. Sources extracted from these background subtracted images were used to solve the astrometry, where we used Scamp \citep{Bertin2006} in combination with a catalogue from an ESO-WFI-Rc stacked image as reference. Using {\sc Swarp} \citep{Bertin2002} we created a coadded image, which was used to create a segmentation map. We masked all source pixels in the original frame using the single-frame astrometric solution and the segmentation map, and estimated a new background from the masked image. Finally, we subtracted the estimated background and created a new final coadded image, with a 3$\sigma$ depth of 25.9 mag$_{\rm AB}$\footnote{After submission of this paper, \citet{Brammer2016} released a public version of the Hawk-I data. We performed a comparison of the data, and found a similar quality.}.


\begin{figure*}
\begin{center}
 \includegraphics[width=\textwidth]{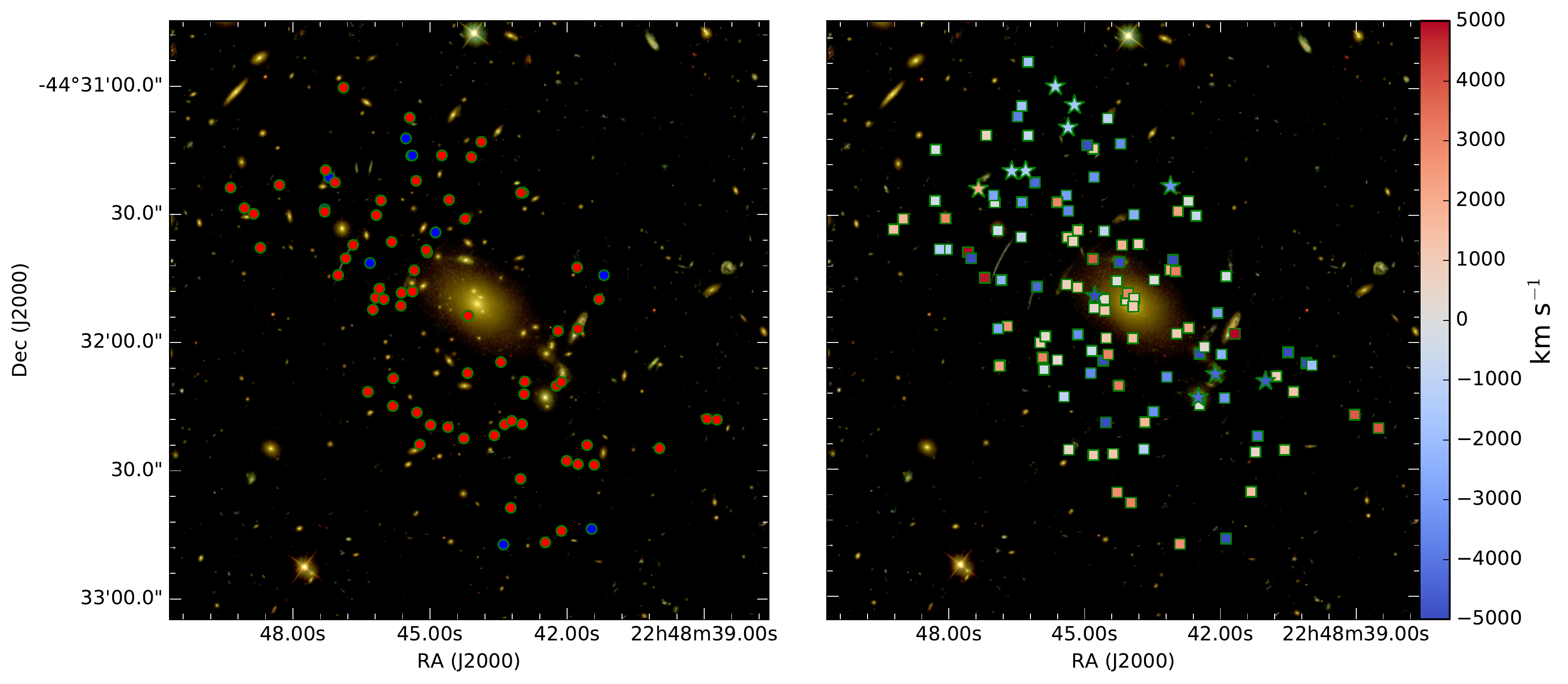}
\caption{Distribution map of the identified galaxies, in the fore and background (left) and in the cluster (right). The 
galaxies are overplotted on an {\em HST}-RGB image, consisting of the F435W (blue), F606W (green), and F814W (red) filters
from the FF programme.
In the left panel, the background galaxies are shown with red circles, while foreground objects are shown with blue circles.
In the right panel, squares correspond to passive cluster galaxies,
while stars indicate active cluster galaxies, where their classification is based on the presence
or absence of optical emission lines. The galaxies have been coloured according to their velocity
relative to the cluster ($z=0.3475$), with bluer colours meaning higher velocities towards us, and redder colours 
corresponding to higher velocities away from us, see also the colour bar on the right.\label{fig:fov_act}}
\end{center}
\end{figure*}

We extracted magnitudes from the optical and NIR images using {\sc SExtractor}. As most of these images have irregular morphologies due to lensing, see Fig. \ref{fig:stamps_814}, we adopt Kron-like apertures rather
than spherical apertures. We constructed a detection image for the FF photometry by combining the F435, F606, and F814W images, and required that each source is detected at more than 1 $\sigma$ in more than 8 connected pixels. For the CLASH images, we used the detection image provided by the CLASH collaboration as a detection image, due to a different spacing and resolution. We note that this might introduce an offset in the colours of the galaxies between CLASH and FF detections, but this effect will be small compared to the error bars obtained from the shallower CLASH observations. We tested the validity of using Kron-radii, different detection images resulting in possible colour differences due to our approach in Appendix \ref{sec:app_sed}. We used 32 deblending sub-threshold levels, with a relative
minimum contribution of 0.1\%. The background is calculated locally using the weight maps provided by the FF team.
We checked each individual detection if it was contaminated by other closeby galaxies, and removed detections when dubious, however we note that some galaxies might still suffer from contamination due to inaccurate background estimates. We noted that visually detected sources remained undetected by SExtractor in the F814W and Hawk-I Ks observations. We used more aggressive detection settings for these bands, and added the relevant detections to our catalog. 
 For the {\em HST} images, we compared
the errors provided by {\sc SExtractor} with those measured from the RMS images provided by the FF team. We found that multiplying the {\sc SExtractor} errors by a factor of 1.4 
reconciled the different methods. 

We also measured photometry in the available {\em Spitzer} Infrared Array Camera (IRAC) imaging in channel 1 ($\lambda$=3.6$\mu$m) and channel 2 ($\lambda$=4.5$\mu$m) \footnote{PI Soifer, programme ID 10170}, which we mosaiced.  This imaging covers a depth of typically $\sim$24.9 magnitudes at 5$\sigma$, although this is inhomogeneous across the imaging as a result of the increased crowding and intracluster light near the centre of the field of view.  These depths are also subject to being able to extract reliable photometry via deconfusion techniques.

The photometry in this imaging was measured using the deconfusion code {\sc TPHOT}\footnote{{\sc TPHOT} is publicly available for downloading from \url{www. astrodeep.eu/t-phot/}, see also \citet{Merlin2015}}.  Briefly, the user provides the code with spatial and surface brightness information for a catalogue of objects as detected in the high-resolution imaging (in this case, the {\em HST} F160W imaging).  The code convolves galaxy templates taken from the high-resolution image with a transfer kernel in order to create the corresponding template in the low-resolution image.  The fluxes of these low-resolution templates are all fitted together, in order to produce a best fitting model of the low-resolution image.  For further details, the reader is referred to \citet{Merlin2015}.

With this approach, we found two clear IRAC detections among our LAE sample, and an additional six with $\sim2-4\sigma$ detections. For the remaining objects, we used the locally estimated depth of the image to set an upper limit at 3 times the depth of the observation to better constrain the restframe optical properties. A caveat to our approach is the issue of excessive crowding in the cluster centre, where some of the candidates are situated. For these objects, extracting reliable photometry was particularly challenging, even when additionally attempting to fit the background to account for the cluster light, but we found that given their relatively large uncertainties, they had little effect on our results.

\input{stamps}

\subsection{Integral field spectroscopy}

The MUSE instrument mounted on the VLT \citep{Bacon2012} is a powerful tool to blindly look for LAEs behind clusters.
Its relatively large field of view (1 arcmin$^{2}$), spectral range (4750-9350 \AA), relatively high spatial 
(0.2\arcsec) and spectral ($\sim$3000) resolution, and stability allowed us to find LAEs down to an observed
flux of 10$^{-18}$ erg/s/cm$^2$ in a 1$\times$1\arcmin field with only 4 hours of exposure. 

AS1063 was targeted with MUSE to search for high-redshift galaxies \citepalias{Karman2015} and simultaneously aid in constraining the lens properties, see \citet[][hereafter C16]{Caminha2015} for a detailed
description of the used lensing models.
The data on the south-western half of AS1063 was
described in \citetalias{Karman2015} and consists of 8 exposures of 1400 seconds each\footnote{ESO Programme 060.A-9345, PI Caputi \& Grillo}, or a total integration time of 3.1 hours. In this paper we add the north-eastern half of AS1063 to the available data, which has 12 exposures of 1440 seconds, or a total integration time of 4.8 hours\footnote{ESO Programme 095.A-0653, PI Caputi}. We note that 4 of the later exposures were rated with a grade C, meaning the observational requirements were not met. However, we did include these exposures to our datacube, as they did not decrease the spatial resolution, and did improve the
depth of the final datacube.

Each pointing employed the same observing strategy, where we used observation blocks of 1440s 
which followed a dither pattern with offsets of a fraction of an arcsecond and rotations of 90 degrees
to better remove cosmic rays and to obtain a better noise map. 
We followed the data reduction as described in \citetalias{Karman2015} for both pointings, and 
refer to that paper for details. Here we provide only a brief description of the data reduction.
We used the standard pipeline of MUSE Data Reduction Software version 1.0 on all of the raw data.
This pipeline includes the standard reduction steps like bias subtraction, flatfielding, wavelength
calibration, illumination correction, and cosmic ray removal. We checked all wavelength calibrations for accuracy and verified
the wavelength solutions. The pipeline then combines the raw data into a datacube that includes
the variance of every pixel at every wavenlength. Consequently, we subtracted the remainder of the sky
at every wavelength by measuring the median offset in 11 blank areas at every wavelength, and subtracting
this from the entire field. We measured a spatial FWHM of 1.1\arcsec\ in the south-western datacube and a
FWHM of $<1.0\arcsec$ in the north-eastern datacube on a point like source selected from {\em HST} images 
in both pointings.

For each pointing, we used a spectrally collapsed image of the datacube to find sources. In addition to this, we visually inspected
the datacube to find sources with emission lines that were not visible in the stacked image. Further, we used the {\em HST}
images to look for bright galaxies not included in our list, or galaxies that were only visible in either or both of the
F606W and F814W bands, as this is often a good indication that the source is at high redshift. Finally, we used the
 predictions from our lensing models to search for additional images of lensed LAEs in MUSE observations. At each of
these positions, we then extracted a spectrum with an aperture of 1\arcsec\ radius to determine the redshift of the galaxy.

\section{Spectral analysis}
\label{sec:results}

\begin{figure*}
\begin{center}
 \includegraphics[width=\textwidth]{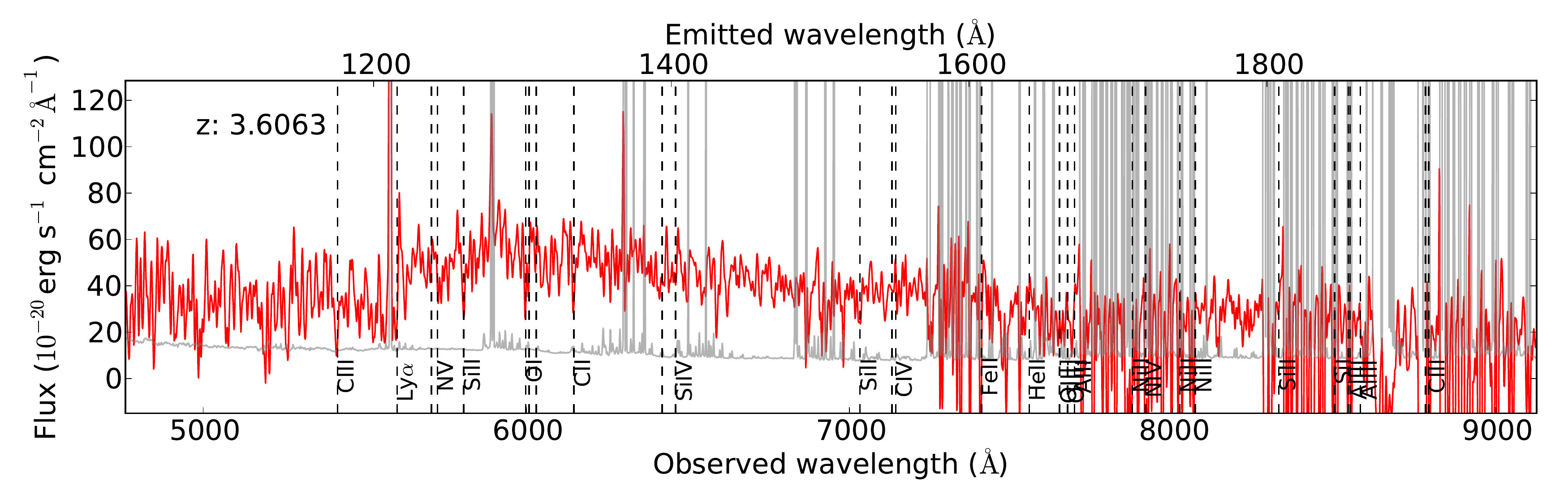}
\caption{Spectrum of object SW-70a, a newly identified LBG in the SW of AS1063. The spectrum has
been smoothed for illustrative purposes, the grey bands correspond to wavelengths with significant
sky interference.\label{fig:ceg1}}
\end{center}
\end{figure*}

We presented the redshifts obtained from the first south-western (SW) pointing in \citetalias{Karman2015}. Here we complement those measurements with the new redshifts determined for the north-eastern (NE) half of AS1063. In Appendix \ref{sec:appa} we provide a complete compilation of all the redshifts obtained from our two MUSE pointings. We determined redshifts for 3 additional high-redshift galaxies with multiple images, of which two were described in \citetalias{Caminha2015}. The third system is a $z=3.606$ LBG, with weak \Lya emission and two images within the south western MUSE pointing. This system, labelled as SW-70, has a clear continuum and several UV absorption features clearly visible in both images, see Fig. \ref{fig:ceg1}.

We selected all the LAEs that we found in the observations, see Table \ref{tab:LAE}, resulting in 6 
and 8 LAEs behind the south-western half and the north-eastern half of AS1063 respectively.
Two of these LAEs were discussed in more detail in \citet{Vanzella2016b} and
\citet{Caminha2016}. The first is an optically-thin, young, and low-mass galaxy that is a good candidate 
for a Lyman continuum emitter, which we studied using the expanded wavelength range and higher resolution spectroscopy
of X-SHOOTER. The second LAE is accompanied by an extended \Lya nebula, for which we found the most likely
origin is scattered \Lya photons emitted by embedded star formation.

We determined the redshifts for 116 objects in the NE of AS1063, belonging to 102 individual galaxies. We found 6 foreground objects, 74 galaxies that belong to the cluster, and 22 galaxies behind the cluster. We identified 10 galaxies that show multiple images, for a total of 25 images, including the two images of the quintiply lensed $z=6.11$ LAE which still lacked spectroscopic confirmation. Combining these redshifts with the SW MUSE observations, results in a total of 9 foreground objects, 121 cluster galaxies, and 42 background galaxies with MUSE redshifts in the central region of AS1063. We do not find any high-ionization UV emission lines for any of the new LAEs.

The total number of spectroscopically confirmed multiple image systems in AS1063 has been increased from 10 to 18, with 17 systems having at least 2 redshifts spectroscopically determined with MUSE, see Table \ref{tab:multim}. We find one additional faint line emitter which we associate with \Lya emission at $z=5.894$, which has no clear counterpart in the FF images. The lensing model predicts additional images outside of the observed MUSE field, but their magnifications are too low to be detected in the {\em HST} images. The addition of 2 and possibly 3 spectroscopically confirmed systems at $z>5$ should help to further constrain the cosmological parameters, see \citetalias{Caminha2015}, while the increased number of $z=3-4$ images will decrease the degeneracies and uncertainties in the models. We corrected all properties in the main body of this paper for gravitational lensing magnification, using the model described in Appendix \ref{sec:lensing}.

Due to lensing distortions, most of the galaxies discussed here have irregular morphologies in the image plane. To optimize 
the S/N of the \Lya line, we created a spatial mask for each source, and extracted the spectrum within this mask. Each mask was constructed
by collapsing the cube over the spectral width of the \Lya line, smoothing this stacked image by a 3 pixel wide boxcar function, and masking out every pixel with values $<5\sigma$ off the background. We verified that this effectively masked out nearby contaminating sources, while also selecting the entire region of \Lya emission.

\begin{figure*}
\begin{center}
 \includegraphics[width=\textwidth]{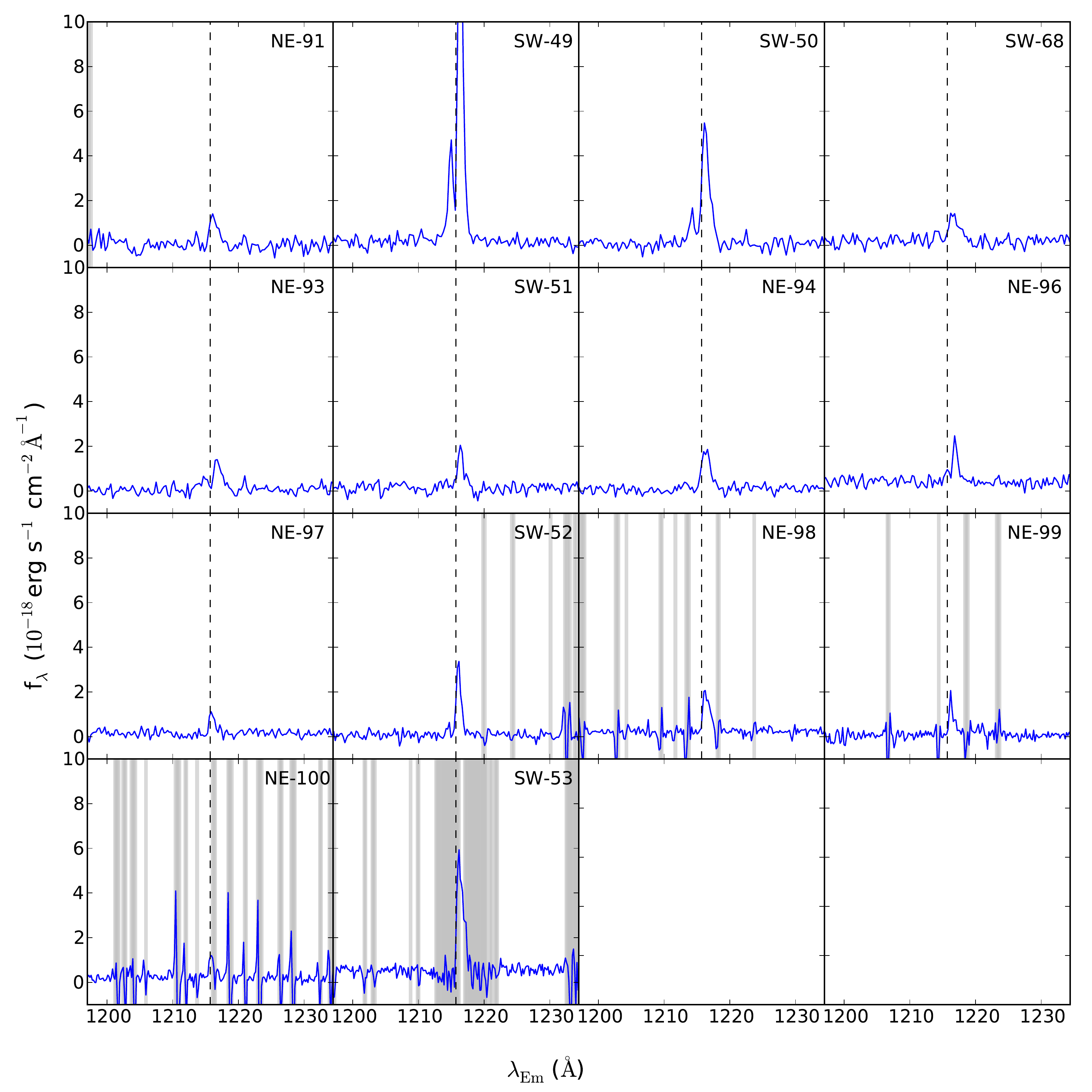}
\caption{\Lya lines for the LAEs extracted from the MUSE datacube, shown by the
blue lines. The spectra are shifted to restframe wavelengths, and the fluxes
are not corrected for the gravitational magnification. The grey bands show wavelengths with significant
sky interference, while the black dashed line shows the restframe wavelength of 
\Lyap The systemic redshifts of LAEs SW-49 and SW-68 have been determined from the narrow UV-emission lines,
while we adopted the redshifts based on the \Lya line for the other objects. \label{fig:lya_profiles}}
\end{center}
\end{figure*}

\begin{table}
\begin{center}
 \begin{tabular}{ccccc}
 {\bf ID}& {\bf RA} (J2000) & {\bf DEC} (J2000) & $z$ \\
  \hline 
NE-91 & 342.19238 & -44.52505 & 2.9760 \\
SW-49a & 342.17505 & -44.54102 & 3.1169$^{c,e}$ \\
SW-49b & 342.17315 & -44.53999 & 3.1169$^{a,b,c,e,f}$ \\
SW-50 & 342.16225 & -44.53829 & 3.1160 \\
SW-68a & 342.18745 & -44.53869 & 3.1166$^{a,h}$ \\
SW-68b & 342.17886 & -44.53587 & 3.1166$^{a,h}$ \\
NE-93a & 342.18283 & -44.52028 & 3.1690 \\ 
NE-93b & 342.19196 & -44.52409 & 3.1690 \\ 
SW-51 & 342.17402 & -44.54124 & 3.2275 \\
NE-94a & 342.18935 & -44.51871 & 3.2857 \\
NE-94b & 342.19615 & -44.52291 & 3.2857 \\ 
NE-96 & 342.19709 & -44.52483 & 3.4514 \\
NE-97 & 342.19100 & -44.52679 & 3.7131 \\ 
SW-52a & 342.18150 & -44.53936 & 4.1130$^{c}$ \\
SW-52b & 342.17918 & -44.53870 & 4.1130$^{c}$ \\
NE-98a & 342.19015 & -44.53093 & 5.0510 \\ 
NE-98b & 342.19085 & -44.53566 & 5.0510 \\ 
NE-99a & 342.18378 & -44.52122 & 5.2373 \\ 
NE-99b & 342.18874 & -44.52276 & 5.2373\\   
NE-100 & 342.19701 & -44.52212 & 5.8940\\ 
SW-53a & 342.18106 & -44.53462 & 6.1074$^{b,c,g}$ \\ 
SW-53b & 342.19088 & -44.53747 & 6.1074$^{b,c,g}$ \\ 
NE-118c$^\dagger$ & 342.18402 & -44.53159 & 6.1074 \\
NE-118d$^\dagger$ & 342.18904 & -44.53004 & 6.1074 \\ 
\hline
SW-70a & 342.18586 & -44.53883 & 3.6065 \\
SW-70b & 342.17892 & -44.53668 & 3.6065 \\
 \end{tabular}
 \caption{LAEs behind AS1063, see Table \ref{tab:redshifts} for quality flags and a cross correlation with
 multiple images. The last galaxy, SW-70, is no LAE, but a Lyman Break galaxy with minimal \Lya emission,
 and is therefore not considered in the remainder of this paper. Previous redshift
 determinations by: $^a$\citetalias{Caminha2015}, $^b$\citet[][]{Balestra2013}, $^c$\citetalias[][]{Karman2015}, $^d$\citet[][]{Richard2014a},$^e$\citet[][]{Vanzella2016b}, $^f$\citet[][]{Johnson2014}, $^g$\citet[][]{Boone2013}, and $^h$\citet[][]{Caminha2016}.\\
  $^\dagger$ NE-118 is the same image family as SW-53 but located in the NE rather than the SW. To avoid confusion with object NE-53, we listed the objects with the NE-118 identifier. \label{tab:LAE}}
\end{center}
\end{table}

In Fig. \ref{fig:lya_profiles}, we show the observed line profiles of all LAEs discovered in the
datacubes, uncorrected for magnification. It is clear from this figure that the observed fluxes vary
widely, from very bright  (SW-49) to very faint (NE-97) \Lya lines. We see that all profiles have an asymmetry typical for \Lya lines, and
most show a clear smaller blue peak or suggest the presence of a small blue peak.
The width of the lines varies amongst our sample, but we find that most LAEs have
narrow emission lines. Such narrow lines suggest the presence of low column density
gas, and are suggested to be candidate Lyman continuum emitters \citep[e.g.][]{Jones2013,Verhamme2015,Vanzella2016b,Dijkstra2016}.

\subsection{Physical properties deduced from Ly$\alpha$}
\label{sec:LAP}

\begin{figure}
\begin{center}
\includegraphics[width=\columnwidth]{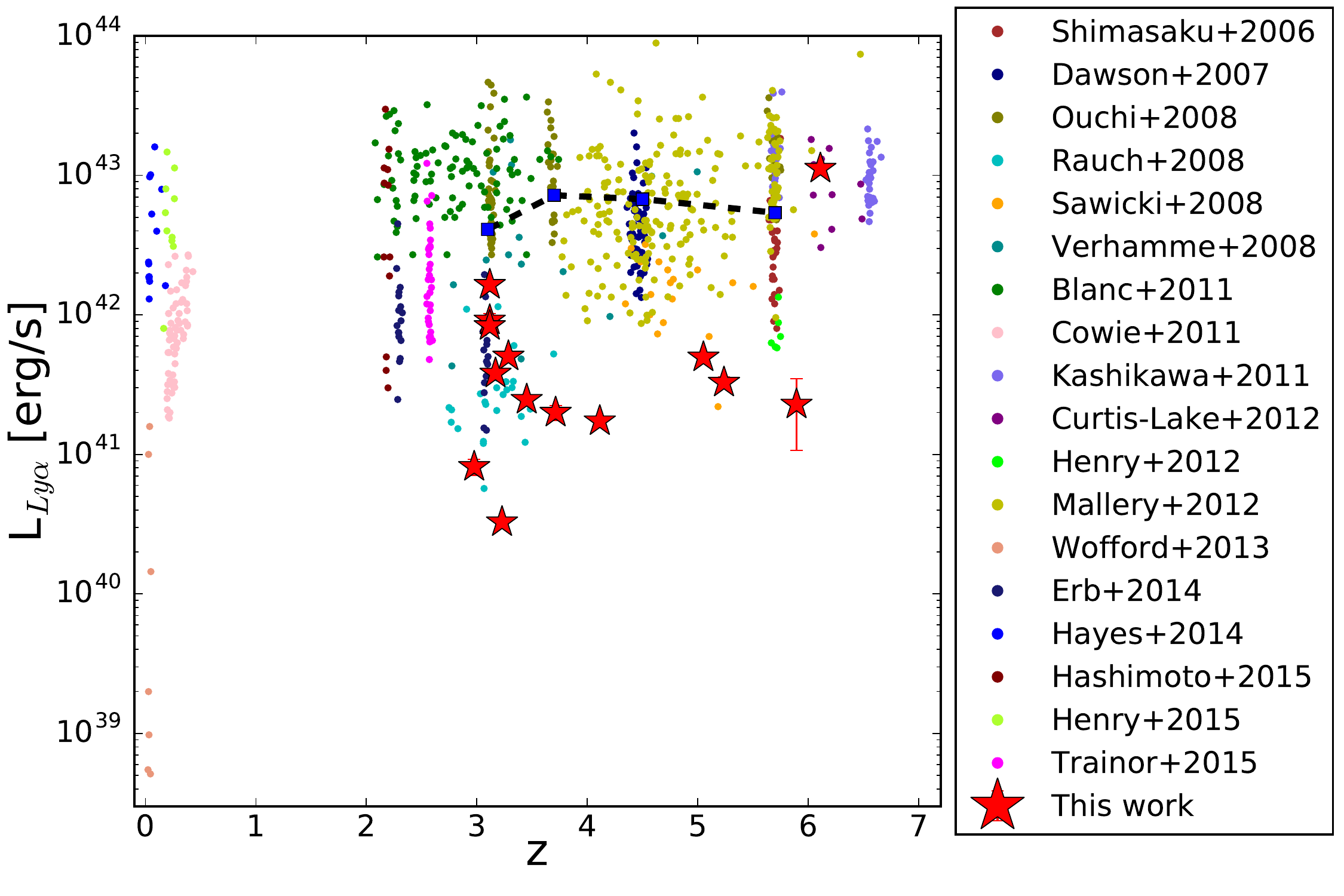}
\caption{The delensed luminosity of the \Lya line against the redshift of our
targets, marked by the red stars. We compare this to previously large
samples of spectroscopically-confirmed LAEs in the literature, which are 
shown by dots of different colours. We overplot the values of $L^\star$
at various redshifts from \citet{Ouchi2008} with a black dashed line.\label{fig:Lyalum}}
\end{center}
\end{figure}

The LAEs studied here belong to the intrinsically-faintest galaxies 
{\nobreak($f_\lambda = 36 - 2500 \times 10^{-20}$ erg s$^{-1}$ cm$^{-2}$)}
spectroscopically confirmed at these redshifts, see Fig. \ref{fig:Lyalum}.
We measured the \Lya luminosity by summing the flux over the spectroscopic
width of the \Lya line, and subtracting the average of uncontaminated 
spectral regions redwards and bluewards of \Lyap Subsequently, we used our lensing models
to correct the luminosities for the magnifications due to the galaxy cluster, see 
Table~\ref{tab:sed_ind} for the adopted magnification factors. The errors on the 
magnifications are typically of the order of 5-10\%, which are generally larger
than the photometric errors. For clarity,
we have not propagated the errors on our magnification into our error estimates of 
other properties in any table or figure. An additional error based on the magnification is therefore applicable
for flux derived properties, such as luminosities and stellar masses.

As we have multiple images for some galaxies, we can perform a test on our luminosities
and magnifications. We compared the obtained luminosities for these objects, and
used the mean luminosity when they agreed within 2$\sigma$. For those sources where
a larger difference was found, we reinvestigated the datacube, and found that 
the lower-luminosity objects were underestimated. For two of these,
the underestimation was due to the proximity of the edge, which resulted in only
a partial coverage. For 3 other objects, contamination by nearby cluster members
resulted in an oversubtraction of the continuum, while for 1 object a lower
S/N in combination with proximity to a skyline resulted in lower fluxes. 

\subsubsection{Line profile modelling}

In addition to \LLya, we used our spectra to obtain physical properties
of the gas surrounding these faint galaxies. We used the \Lya line fitting 
pipeline described in detail in \citet{Gronke2015} which consists of a pre-computed grid of \Lya radiative transfer models on an expanding shell and a Bayesian fitting framework.

The expanding shell model \citep[first used by][]{Ahn2003} consists of a central \Lya (and continuum) emitting source surrounded by an outflowing shell of hydrogen and dust. Such a model has six free parameters: two describing the photon emitting source (the intrinsic line width $\sigma_{\rm i}$ and equivalent width $EW_{\rm i}$), three for the shell content (the neutral hydrogen column density $N_{\ion{H}{I}}$, the dust optical depth $\tau_{\rm d}$ and the effective temperature $T$ which includes the approximate effect of turbulence) and the outflow velocity $v_{\rm exp}$. 

The pre-computed grid mentioned above consists of $10,800$ models\footnote{The spectra can be accessed online at
\url{http://bit.ly/man-alpha/}.} covering the three parameters $T$, $N_{\ion{H}{I}}$ and $v_{\rm exp}$ as they shape the spectrum in a complex, non-linear fashion \citep{Verhamme2015}. The grid was created using the radiative Monte Carlo code \texttt{tlac} \citep{Gronke2014} which traces individual photon packages in real- and frequency space \citep[for a comprehensive review on \Lya radiative transfer see, e.g.,][]{Dijkstra2014}.

The effect of the remaining three parameters is modelled in post-processing by assigning a weight to each individual photon packet, which means that the procedure affects the shape of the line and not only the normalization. This strategy does not only save computational time but allows to model these parameters continuously, and thus, leads to a more precise sampling of the likelihood when comparing the modelled data to observations.  

The actual fitting procedure is done by sampling the Gaussian likelihood using the affine invariant Monte-Carlo sampler \texttt{emcee} \citep{Foreman2013} using 400 walkers and 600 steps \footnote{For particularly difficult, multi-modal cases we used a parallel tempered ensemble MCMC sampler \citep[for a review see,][]{2005PCCP....7.3910E} with 20 temperatures, 50 walkers and 3000 steps.}. 
In addition to the minimal set of the six shell-model parameters, we also fit simultaneously for the redshift $z$ and the full-width at half maximum of the Gaussian smoothing kernel $FWHM$. Note that the former adds immense complexity to the fitting procedure as shifting $z$ by a small fraction can alter the quality of the fit tremendously. We used the redshift estimate from UV emission lines if available or otherwise the redshift of \Lya with an intrinsic uncertainty of $\sim200$ \kms (see \S \ref{sec:LAP}) as a prior. Alternatively, the latter, i.e. smoothing the spectrum, makes the likelihood function better behaved. However, the width of the smoothing kernel is a function of the actual size of the \Lya halo as well as the measurement aperture. Therefore, we used an allowed range for $FWHM$ corresponding to the wavelength-dependent spectral resolution of the MUSE instrument.  

\citet{Gronke2015} discussed the uncertainties of using \Lya line profile fitting for various effects, for example morphology and signal-to-noise ratio. They showed that the expansion velocity and column density can be recovered reasonably well in most cases, while degeneracies and uncertainties are more prominent among the other parameters. Therefore, we focus in this paper on these two
quantities, although we give the full fitting results in Appendix \ref{sec:app_lya}.
This is, to our knowledge, the first
time that the gaseous properties of such faint sources are studied through
\Lya modelling. Therefore, this presents one of the first studies to determine
the effect star formation has on gas in these faint galaxies.

\subsubsection{Shell properties}

We modelled the \Lya profiles of 12 LAEs using the approach described above
yielding excellent fits to the observed \Lya spectra (see Appendix~\ref{sec:app_lya}). We present 
the properties based on the \Lya line in table \ref{tab:Lya}, where
we already combined the multiple images into a single result. We modelled
the \Lya line profile of each image, and combined the results of each modelling
into a single result per LAE. We used the average of all images, after discarding
multiple images which were possibly affected by close galaxies or artefacts in the datacube. We did not include
2 LAEs in the modelling, as their \Lya lines were too faint and spectrally unresolved. 

\begin{figure}
\begin{center}
\includegraphics[width=\columnwidth]{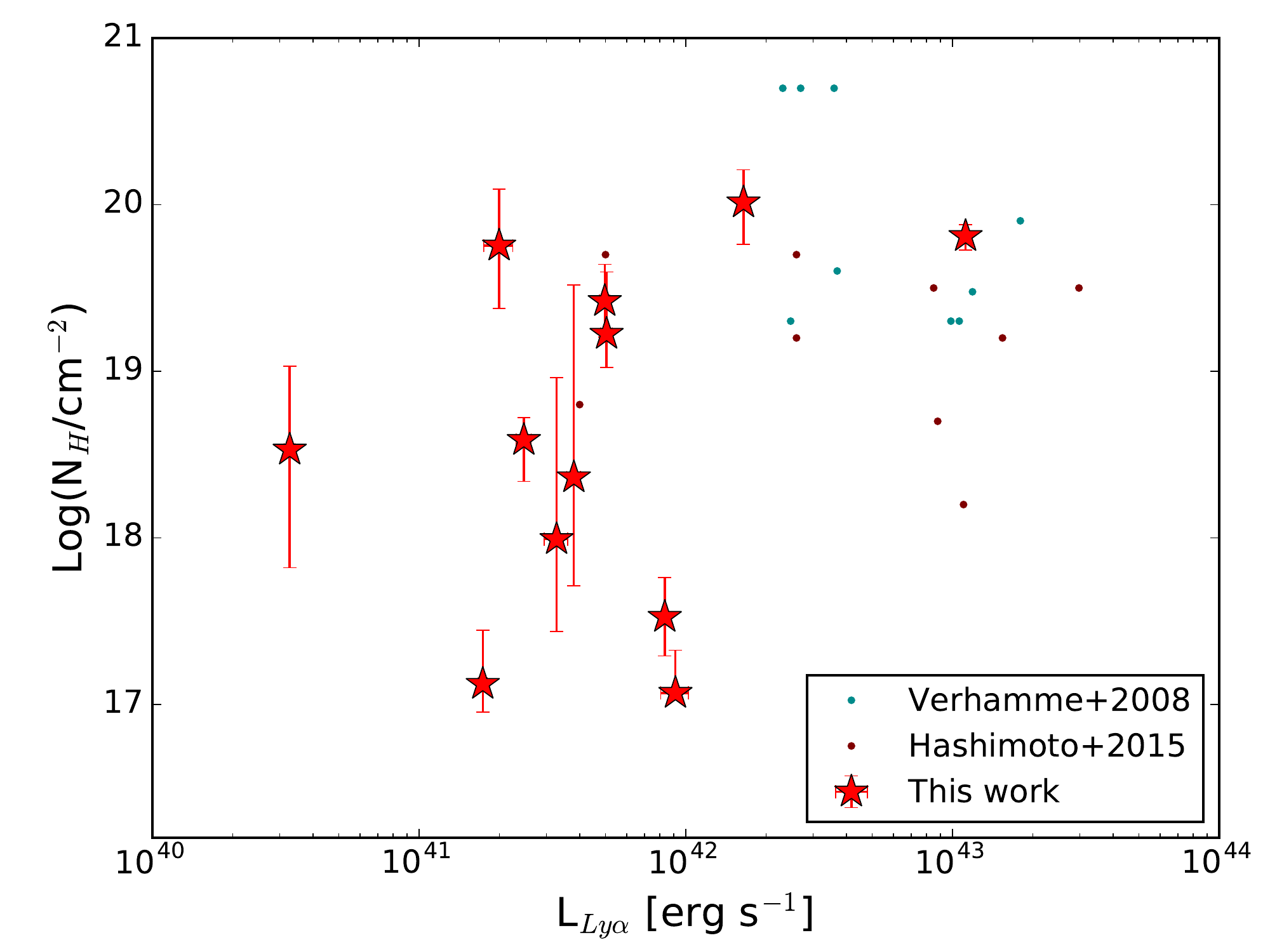}
\caption{The delensed luminosity in \Lya against the column density determined
from modelling the \Lya profile. Legend and symbols are as in Fig. \ref{fig:Lyalum}. \label{fig:LyaNH}}
\end{center}
\end{figure}

In Fig. \ref{fig:LyaNH}, we show that the column density of the expanding shell is 
low in all of the galaxies. In \citet{Vanzella2016b} we reported the results
of \Lya line modelling for SW-49 using higher resolution X-Shooter spectra and an 
updated shell-model fitting pipeline. We
find that the MUSE and X-Shooter results are broadly consistent within the error bars, 
although we used a larger database reaching lower column densities for the X-Shooter
spectrum. 

We compare the best fit column densities
in our sample to those found by \citet{Verhamme2008} and \citet{Hashimoto2015},
who also fit \Lya profiles with expanding shell models. With all except one LAE best 
fit with $Log(N_{\rm H}/cm^2)<20$ and only five LAEs with $Log(N_{\rm H}/cm^2)>19$, we find 
lower column densities than the other studies. However, the galaxies studied here
are also $\sim$1 dex fainter than the ones presented in \citet{Verhamme2008} and \citet{Hashimoto2015}.

These low column densities support the idea that for these faint
galaxies, the \Lya is not significantly broadened by scattering. 
This is consistent with the picture that \Lya escape
becomes easier for fainter galaxies with significant \Lya emission.

We find no trends in the velocity of the expanding shell with the \Lya luminosity, 
but we note that we find relatively low outflow velocities in all galaxies. 
The absence of a correlation between the \Lya luminosity and the expansion
velocity of the shell is unexpected, as both are considered to correlate with
the SFR \citep[e.g.][]{Weiner2009,Bradshaw2013,Chisholm2015}. A possible
explanation could be that the outflow speed does not follow the SFR
at low masses, see below, or that the \Lya luminosity depends more strongly on the
escape fraction of \Lya photons than on the SFR.

\input{tab_lya}

\section{Stellar properties}
\label{sec:stars}

\begin{table}
\begin{center}
 \begin{tabular}{l|ll}
  & Range & nr. steps \\ \hline
  Age & 0.01 Myr - 2.3 Gyr & 50 \\
  E(B-V) & 0 - 1.5 & 20 \\
  Z & 0.0004 - 0.02 (Z$_\odot$) & 4\\
  $\tau$ & 0.001 Gyr - 5 Gyr & 5\\
 \end{tabular}
 \end{center}
\caption{The parameter space used for constructing stellar templates used in our SED fitting. 
The stepsizes are logarithmic distributed for the ages, while we use an irregular spacing for
the stepsizes of E(B-V) where we finely sample the low values, and use larger steps for the
higher values. The metallicities correspond to the {\em m32, m42, m52} and {\em m62} models
of BC03.\label{tab:sed_props}}
\end{table}

We used the constructed photometric catalogue in combination with the spectroscopic redshifts
to perform a spectral energy distribution (SED) fitting on our selected sample. We used 
LePhare \citep{Arnouts1999,Ilbert2006} in combination with \citet[][hereafter BC03]{BC2003} templates to fit the photometry with stellar population models
(see Table \ref{tab:sed_props} for our model parameters).
The set of stellar populations consists of an exponentially declining star formation histories,
$SFR(t)=SFR_0\times e^{-t/\tau}$, with different values for $\tau$. In addition, we created templates
with three different metallicities, and ages up to the age of the Universe. We used a
  \citet{Calzetti2000} extinction curve to attenuate all stellar templates, with $0<E(B-V)<1.5$. We enabled adding
nebular emission lines to the templates based on their UV luminosity as described in \citet{Ilbert2009},
where the line fluxes of [\ion{O}{II}]~$\lambda3727$ are derived using the current SFR, and [\ion{O}{III}]~$\lambda\lambda4959,5007$, \ion{H}{$\alpha$},
and \ion{H}{$\beta$} are then scaled from locally derived line ratios. The addition of emission
lines to stellar-population templates is shown to be important in the high-redshift Universe \citep[e.g.][]{Schaerer2009,deBarros2014}.
 
We fitted the SED of each image and combined the results of the multiple images in a similar method as for the \Lya luminosities and
\Lya line fitting. We used the average results of multiple images when the quality of the photometric data of each image is similar,
but adopt the results of only the best constrained image if the difference is significant, for example for NE-94 we only used image a. 
For galaxies where we suspect contamination from nearby galaxies, e.g. SW-68b, we used only the images without
contamination. We performed tests on the reliability of our results in Appendix \ref{sec:app_sed}, and found that although
there are few constraints in the restframe optical, our results do not change significantly.

We present the results from the SED fitting in Tab. \ref{tab:sed}. 
We remind the reader that
most of the photometry is in the restframe UV, but that the Hawk-I and IRAC filters 
trace the restframe optical. We detected 6 LAEs in the K-band and 1 LAE in the 
IRAC filters, but the non-detections provide important upperlimits when determining
stellar masses and show that there is no hidden dominant old population of stars.

\input{tab_sed}

The masses we derived from our SED fitting are very low, varying from $\sim10^{6}\Msun$ to $\sim10^8\Msun$,
significantly lower than the stellar masses explored in previous studies of LAEs. This is not surprising, as these galaxies are
among the intrinsically faintest discovered so far with absolute UV-magnitudes ranging form -19 to -14, 
and illustrates once more the advantages of gravitational
lensing. We note that for 2 LAEs discovered here (NE-99 and NE-100), we recovered only a single detection through all deep FF filters, which is in the
filter containing \Lyap The completion of the NIR-imaging of AS1063 in the summer of 2016 should add 
more detections to these objects, and will better constrain the properties of these possibly even less massive galaxies. 

\begin{figure}
\begin{center}
\includegraphics[width=\columnwidth]{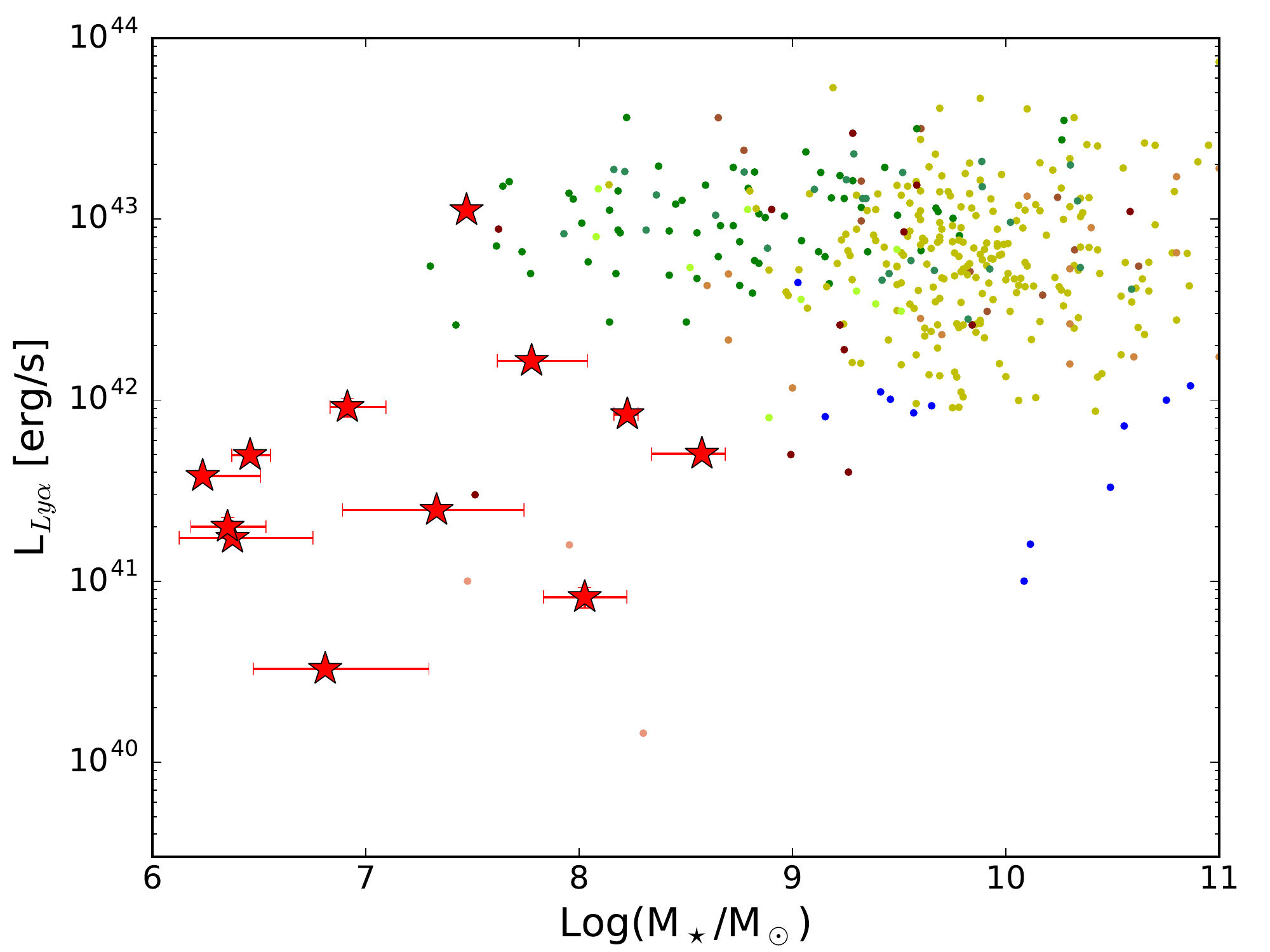}
\caption{The stellar mass versus the lens-corrected \Lya luminosity.
The LAEs described here are compared to a collection of previous LAE
studies, with the colours identical to Fig. \ref{fig:Lyalum}, 
where we have supplemented the results from \citet{Blanc2011} with those of
\citet{Hagen2014}.
\label{fig:mass_LLya}}
\end{center}
\end{figure}

In Figure \ref{fig:mass_LLya}, we compare the stellar masses of our galaxies to their \Lya luminosities. We see that the lower masses are paired with lower luminosities. The low luminosities and masses found here in comparison to previous literature results confirm again that these objects probe a new region of parameter space.

The presence of such narrow and strong \Lya emission is already a clear indication that there is little dust present. We find that the median marginalized E(B-V) values are all very small, with only 3 galaxies having E(B-V)$>$0.1, in agreement with previous studies \citep[e.g.][]{Atek2014}.

\begin{figure}
\begin{center}
\includegraphics[width=\columnwidth]{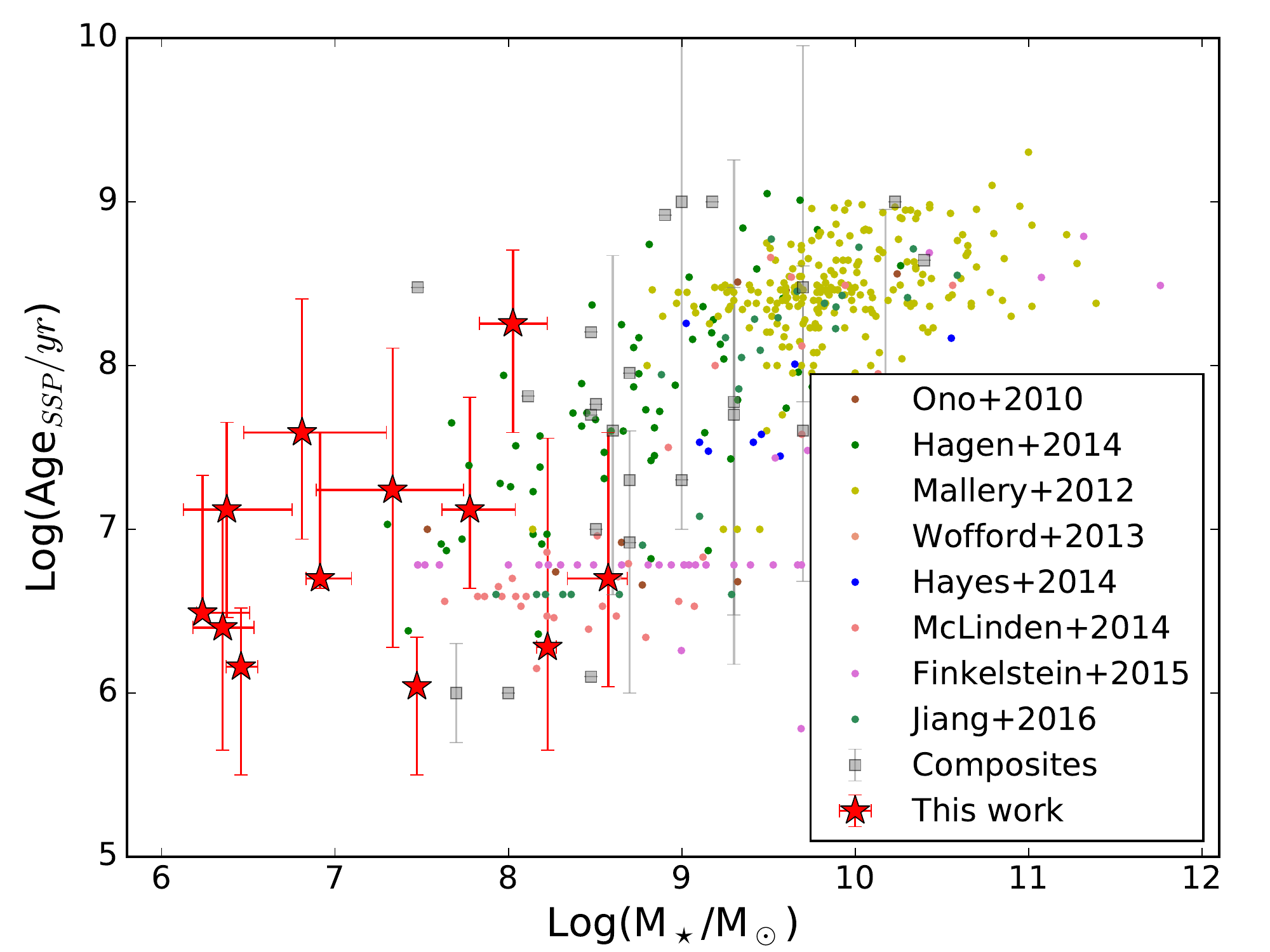}
\caption{The stellar mass versus the age of the stellar population,
as determined by SED-fitting of exponentionally declining star formation
histories. The coloured dots correspond to a variety of results from
previous spectroscopically-confirmed LAE studies, while the grey squares
correspond to the collection of composites assembled by \citet{McLinden2014}.
For comparison, we also plot a sample of non-LAE, represented by the pink
dots \citep{Finkelstein2015}.\label{fig:mass_age}}
\end{center}
\end{figure}

\begin{figure}
\begin{center}
\includegraphics[width=\columnwidth]{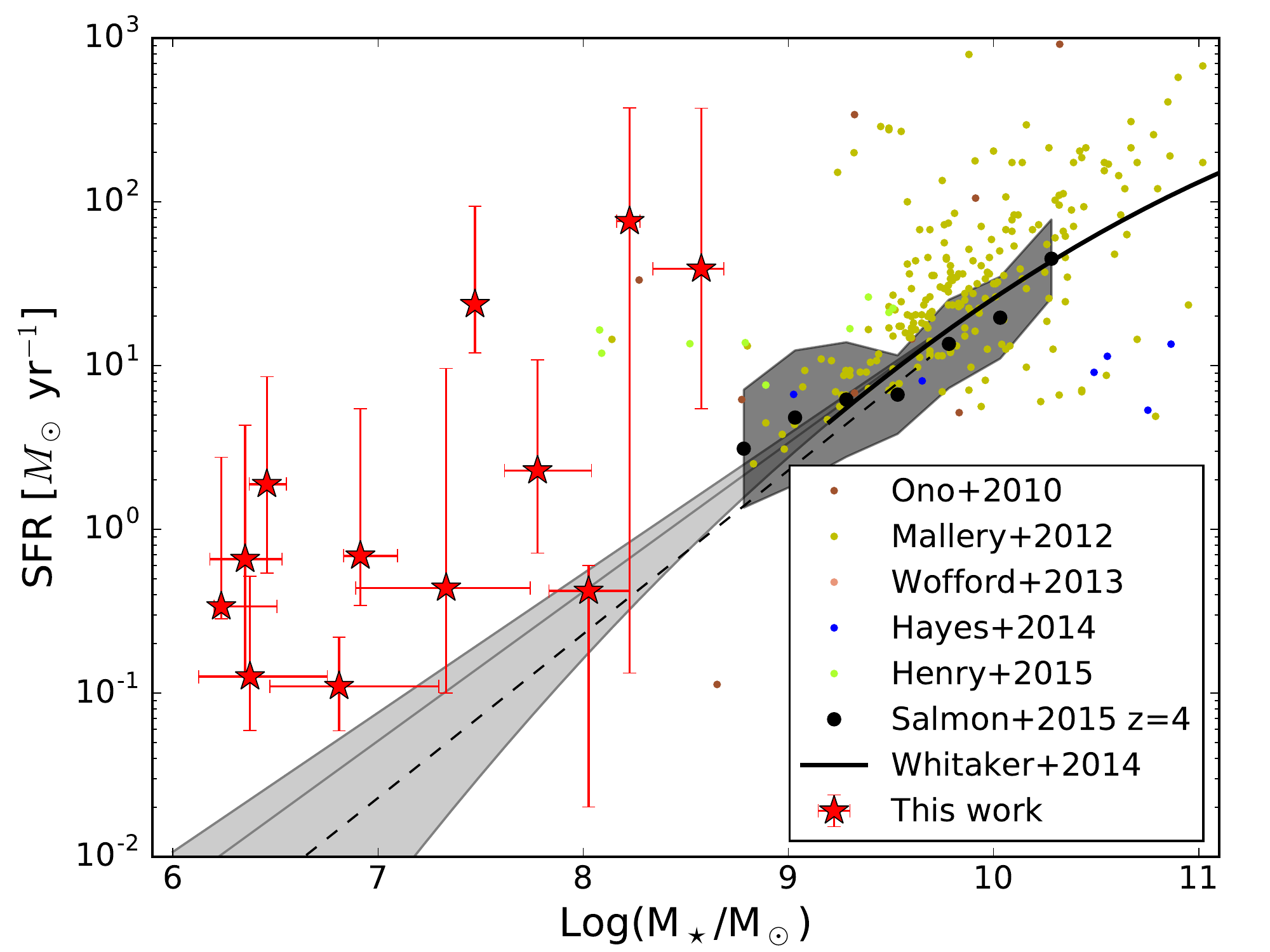}
\caption{Stellar mass versus the star formation rate. See Fig.
\ref{fig:z_ssfr} for a description of the different studies. The black dots
and dark grey region correspond to the SFR for galaxies in CANDELS at $z\sim4$ as determined by
\citet{Salmon2015}, while the black dashed line presents an extrapolation to 
lower masses based on these data. We show the SFR for a sample of $z=2.0-2.5$ galaxies from 
\citet{Whitaker2014} and three possible extrapolations to lower masses with the
 black line and three grey lines respectively. \label{fig:mass_sfr}}
\end{center}
\end{figure}

The ages of these very low mass objects are relatively low, i.e. 1-100 Myr, see Fig. \ref{fig:mass_age}. We find that only two galaxies have an age $>100$ Myr, and age seems to decrease with redshift. The young stellar ages indicate that these are systems that are rapidly building up their mass. This is confirmed by the SFR that we obtain given the low stellar masses, see Fig. \ref{fig:mass_sfr}, as with the current SFR, most galaxies will double their mass within $10^7$ years. As a consequence of these young ages, the models produced by different values of $\tau$ are very similar. Therefore, the current data are unable to distinguish between the different star formation histories. 

We compare the SED-derived SFR to the SFR extrapolated from the SFR-stellar mass relation determined for more massive galaxies, both at lower and similar redshift. We find that most of the LAEs fall above this relation if an extrapolation of the power-law relation described by \citet{Whitaker2014} is considered. Because very little is known about the SFR in low mass galaxies, this extrapolation is rather uncertain due to a degeneracy between the slope of the power-law and its zero-point. If we use the steep power-law relation described for $z\sim4$ galaxies by \citet{Salmon2015}, the number of LAEs above this relation will decrease by a factor 2. We note however, that many studies favor a slope of $\alpha=1$ \citep[e.g.][]{Gonzalez2010,Whitaker2014,Ilbert2015}, which would make most of these LAE starbursting galaxies.

\begin{figure}
\begin{center}
\includegraphics[width=\columnwidth]{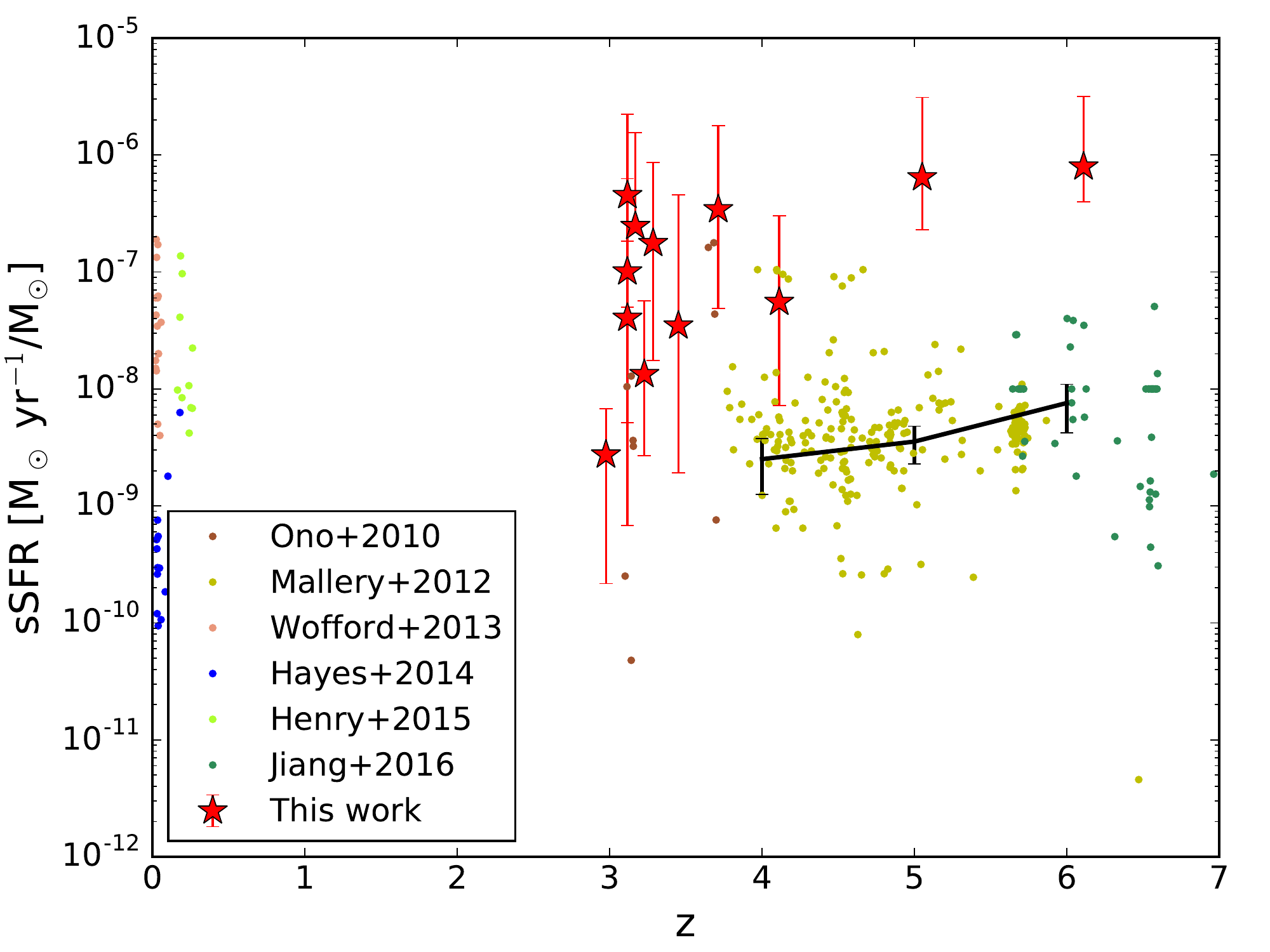}
\caption{The specific SFR as a function of redshift. The sSFR of our
sample, \citet{Ono2010a}, \citet{Mallery2012}, and \citet{Wofford2013} 
is calculated using only SED fitting, for \citet{Hayes2014}, \citet{Henry2015}
the sSFR is calculated using the SFR based on $H\alpha$, and the sSFR of
\citet{Jiang2016} is calculated using $SFR_{\rm UV}$. The black line shows the
sSFR of the CANDELS galaxies from \citet{Salmon2015}.\label{fig:z_ssfr}}
\end{center}
\end{figure}

\begin{figure}
\begin{center}
\includegraphics[width=\columnwidth]{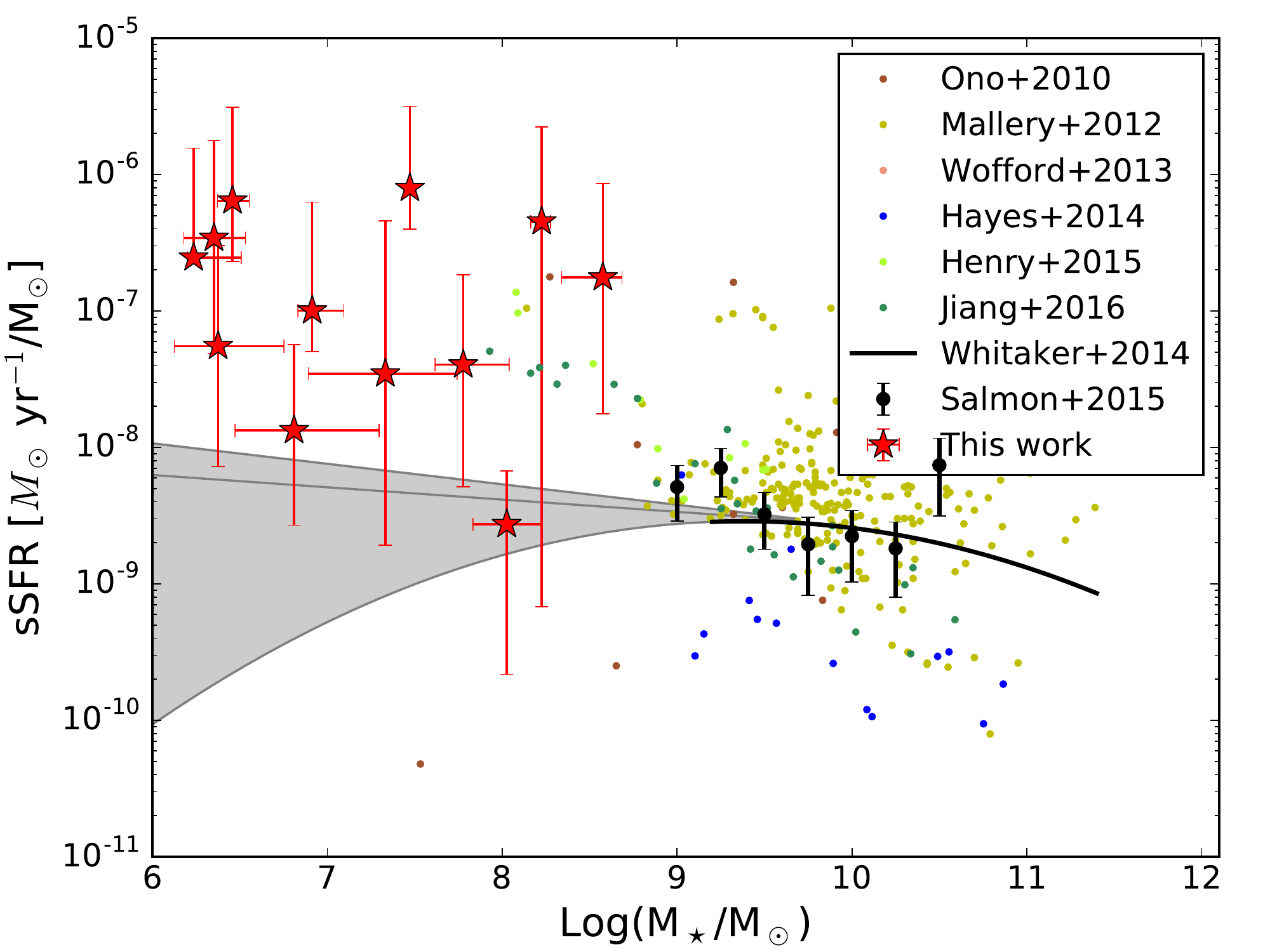}
\caption{Stellar mass versus the specific star formation rate. See Figs.
\ref{fig:z_ssfr} and \ref{fig:mass_sfr} for a description of the different studies and legend.
\label{fig:mass_ssfr}}
\end{center}
\end{figure}

To have a clearer understanding of whether these faint LAEs are normally star forming or starbursting, we looked at the specific SFR (sSFR=SFR/M$_\star$). We find that in our sample of LAEs the average sSFR is significantly higher than seen in other LAE samples, see Figs. \ref{fig:z_ssfr} and \ref{fig:mass_ssfr}. This shows that these galaxies are not only young, but are still actively forming stars. The sSFR classifies these galaxies as starburst galaxies, if one assumes the flat sSFR-$M_\star$ correlation found for stellar masses $M_\star<10^9$ \citep{Gonzalez2010,Whitaker2014,Ilbert2015}. However, whether this relation is flat at these redshifts is unknown, as this mass range has not been explored before even at low redshift.

Similarly high sSFRs are rarely found in any other sources. For more massive sources at high-redshift with young ages, a typical log(sSFR)=-8 is found \citep[e.g][]{Gonzalez2010,Stark2013,Oesch2015,Oesch2016,Tasca2015}, although \citet{Ono2012} and \citet{Finkelstein2013} report a similarly high sSFR at $z>7$. Extreme emission line galaxies (EELGs) at lower redshift show similar values on average, but the spread is large enough that the upper envelope contains some galaxies with log(sSFR)$>-7.5$ \citep[][]{Cardamone2009,Amorin2014a,Amorin2015,Maseda2014,Ly2014}. A comparable evolution of the sSFR with redshift as seen in normal galaxies \citep[see][for a recent compilation]{Speagle2014}, could bring the EELGs up to a similar sSFR found in our sample, strengthening the possible link between EELGs and high-redshift LAEs.

\begin{figure}
\begin{center}
\includegraphics[width=\columnwidth]{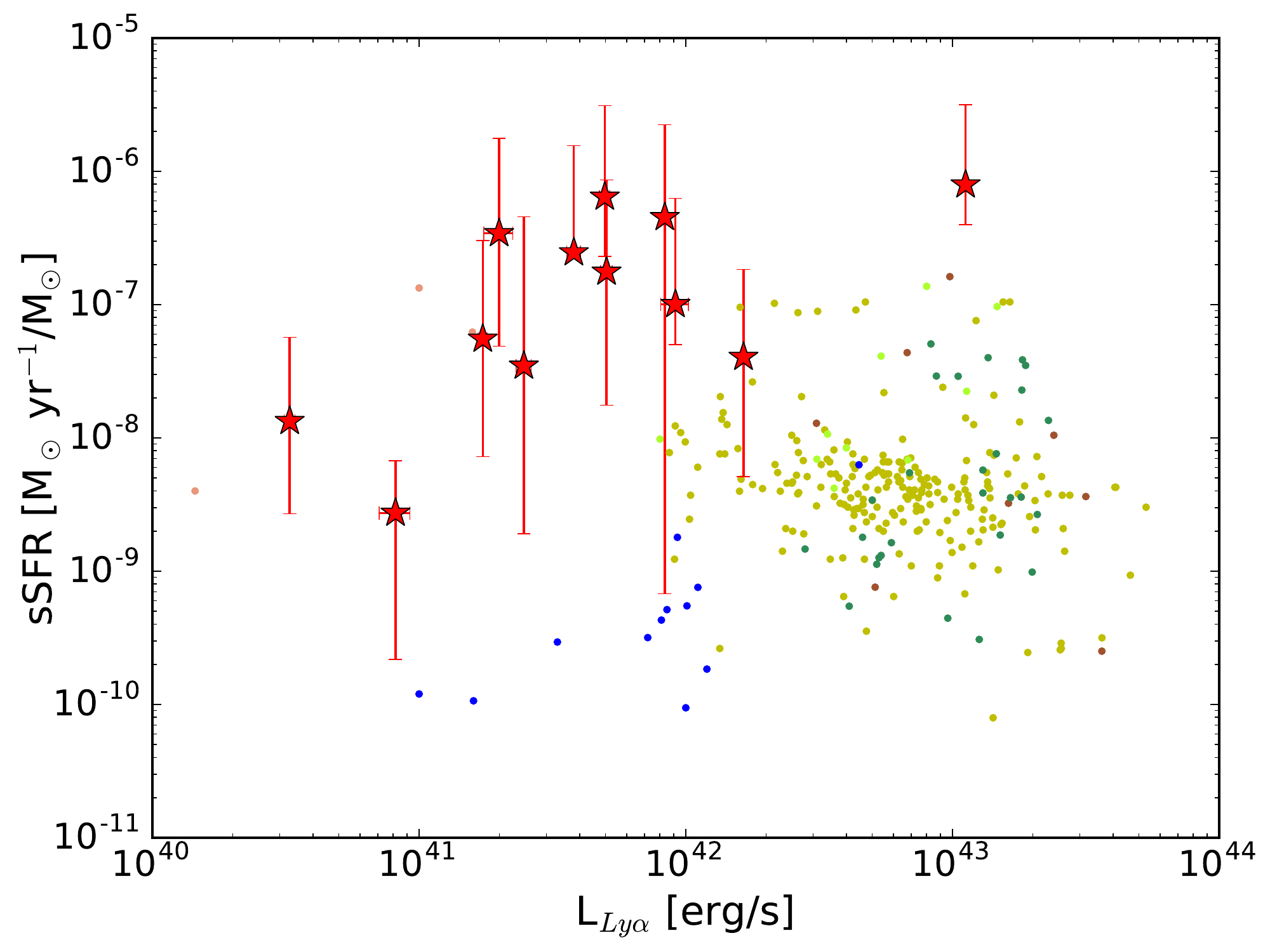}
\caption{\Lya luminosity versus the specific star formation rate. See Fig.
\ref{fig:z_ssfr} for a description of the different studies.\label{fig:llya_ssfr}}
\end{center}
\end{figure}

We compared the sSFR to all other parameters, but find no further correlations with neither any gaseous nor any stellar property. Even the apparent trend of an increasing sSFR with increasing \LLya within our sample disappears when other studies are added, see Fig. \ref{fig:llya_ssfr}. This suggests that the sSFR has no influence on the \Lya line profile or the physics that shape the line, nor is the sSFR influenced by any other stellar parameter than the stellar mass and age. It is noteworthy that in several of these plots we find higher sSFR within our sample when compared to other samples at a given property. This further suggests that the high sSFR is driven by the low stellar mass and young age.

Most galaxies, except two (SW-50 and SW-51), are best fit with a low metallicity, however a solar metallicity falls within our $1\sigma$ certainty range for most of the LAEs. A low metallicity for most of these galaxies would be in agreement with all previous results that these are recently formed, low-mass galaxies which are rapidly building up mass and have not been able to produce a large amount of metals.

\subsection{\Lya escape fraction}

To understand the evolution and formation of the \Lya luminosity function and the process of reionization it is important to determine how much of the \Lya flux escapes the galaxy. To estimate this, we follow previous studies, using the UV or SED derived SFR$_{UV/SED}$ and compare these to the SFR calculated from  \Lya (SFR$_{\Lyam}$) to define the escape fraction $f_{esc}$=SFR$_{\Lyam}$/SFR$_{SED}$. We determine SFR$_{\Lyam}$ following \citet{Kennicutt1998} and assuming case B recombination:

\begin{equation}
 SFR_{\Lyam} (\Msunpyr) = \LLya ({\rm erg\ s^{-1}}) / 8.7  \times 7.9 \times 10^{-42}. 
 \label{eq:sfrlya}
\end{equation}

\noindent We caution the reader that this relation has been calibrated on stable star forming galaxies and that this may not be applicable at very young ages. 

\begin{figure}
\begin{center}
\includegraphics[width=\columnwidth]{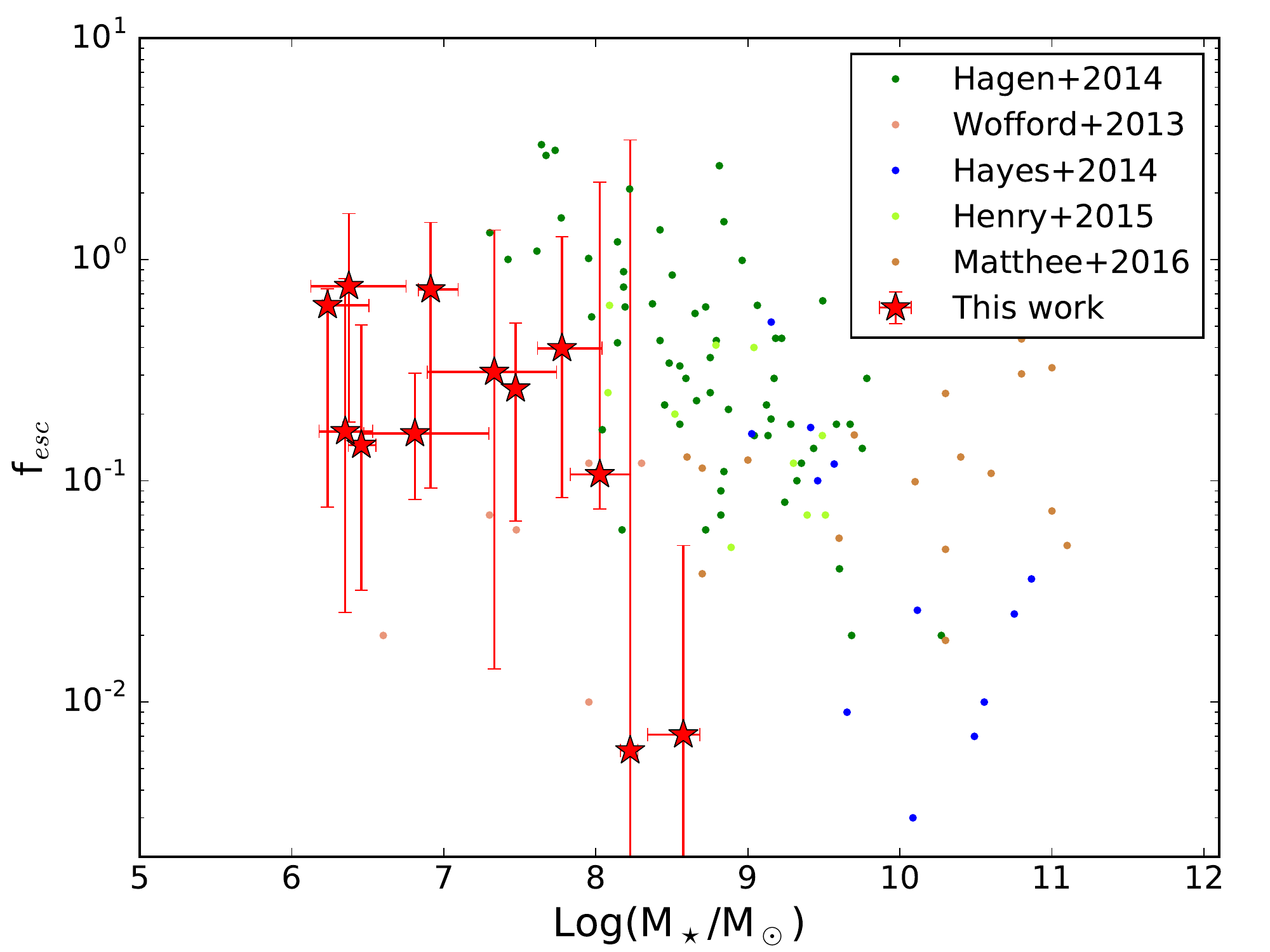}
\caption{The \Lya escape fraction versus the stellar mass. We 
followed most studies and used
the indirect $f_{esc}$=SFR$_{\Lyam}$/SFR$_{SED}$, while 
\citet{Hayes2014} and \citet{Henry2015}
used the flux ratio of \Lya to $H\alpha$ to directly derive $f_{\rm esc}$.\label{fig:mass_fesc}}
\end{center}
\end{figure}

In Fig. \ref{fig:mass_fesc} we show the \Lya escape fraction as a function of stellar mass. We find higher escape fraction for lower stellar masses, a trend which remains visible after other studies are included. We note, however, that this trend is also a natural effect of the selection bias when studying LAEs. Galaxies with low escape fractions will have very little or no \Lya emission, and will therefore remain undetected in emission line studies, whereas low mass galaxies will evade detection in the continuum. These effects would produce a trend similar to the one observed here, however, the dearth of massive galaxies with high-escape fractions is genuine and cannot be explained by selection effects.

\subsection{UV-continuum slope}

\begin{figure}
\begin{center}
\includegraphics[width=\columnwidth]{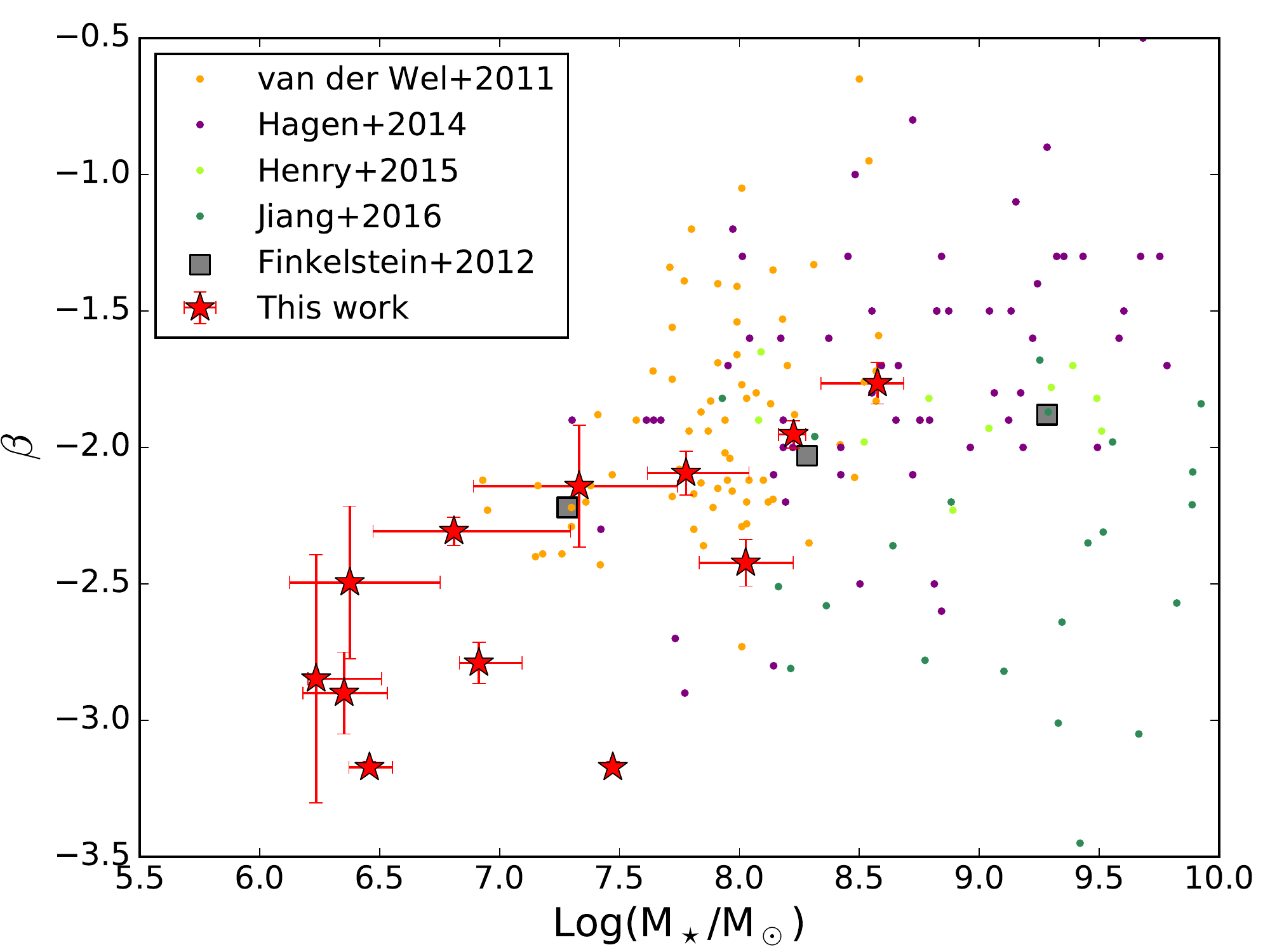}
\caption{The stellar mass versus the UV-continuum slope $\beta$.
We overplotted a number of LAEs from \citet{Hagen2014}, 
\citet{Henry2015}, and \citet{Jiang2016}, the median of CANDELS galaxies 
at $z=4$ from \citet{Finkelstein2012}, and a sample of EELGs from 
\citet{vdWel2011}. \label{fig:mass_beta}}
\end{center}
\end{figure}

We used the best-fitting SED model to measure $\beta$, the slope of the UV spectrum defined as $f_\lambda \propto \lambda^\beta$ \citep[e.g.][]{Meurer1999}, following \citet{Finkelstein2012}. We chose to use the best-fitting SED model to measure $\beta$ rather than observed colors as this was shown to be a more stable approach \citep[e.g.][]{Finkelstein2012,Bouwens2014}. We adopted the spectral windows defined by \citet{Calzetti1994}, and directly fitted a power law to the best-fitting template in these windows. We find a clear relation between the stellar mass and $\beta$, see Fig. \ref{fig:mass_beta}, in agreement with previous results \citep[e.g.][]{Bouwens2014}. For each galaxy, we derive the uncertainty of $\beta$ by repeating the full-SED modelling for 1000 mock galaxies, created by randomly disturbing its photometry taken from a normal random distribution based on the photometric errors. Because the minimum value of $\beta\approx-3.2$ for BC03 models with a Chabrier IMF, galaxies with photometric slopes bluer than this will almost always be best fitted with $\beta=-3.2$, even after scattering the photometry. This leads to unrealistically small errors for some of our UV-slopes, e.g. NE-98. 
We then fitted a Gaussian to the distribution of $\beta$ of all mock galaxies, and report the resulting $\sigma$ as the uncertainty in $\beta$. We calculated the slope of the $\beta$-mass relation to be 0.43 in our sample, comparable to the maximum slope of 0.46 at $z=7$ by \citet{Finkelstein2012}. We note however, that the majority of our samples lies at $3<z<4$, for which \citet{Finkelstein2012} found a slope of only 0.17. An important factor in this difference could be our selection bias, meaning that we only find galaxies with strong \Lya emission. 

\citet{Dunlop2012} argued that very blue UV slopes could be caused by low S/N photometry. In this study however, we do not suffer from uncertainties in redshift, which are the largest factor in the scatter of $\beta$ in photometric studies. Therefore, we do not expect to be affected by the same bias. The very blue slopes are also in agreement with the low stellar masses, low dust extinction, and young ages found in these LAEs, as also discussed for example by \citet{Dunlop2012}. 

\citet{Jiang2016} measured the UV-slope of their sample of LAEs, and they find a significant number of more massive LAEs with very blue slopes. This seems to be in disagreement with our and other previous results, as they would predict that more massive galaxies have redder continua due to a higher metallicity and larger amounts of dust. This is possibly due to low S/N photometric data available for the relevant sources in \citet{Jiang2016}, leading to a larger scatter found in the $\beta$ slope \citep{Dunlop2012}. Another possible explanation for this is that the number density of blue more-massive LAEs is significantly lower than those of slightly redder LAEs. Because we are only observing a rather small volume, this means that the very low number of blue massive LAEs in our sample is expected. 

We note that for a few of our LAEs, we find the steepest theoretically-allowed $\beta$ for pure stellar populations without a nebular contribution, i.e. $\beta=-3.17$. In these cases, the observations are representative of an even steeper slope, and, naturally, models with flatter continuum slopes are disfavoured by SED fitting. A steeper UV continuum than the theoretical maximum value for stellar templates $\beta<-3.2$ can be achieved by including the effects of a nebular continuum into the SED fitting \citep[e.g.][]{Schaerer2009, Zackrisson2013} or by using top-heavy IMFs or Pop III stars \citep[e.g.][]{Bouwens2010,Zackrisson2011}.

\begin{figure}
\begin{center}
\includegraphics[width=\columnwidth]{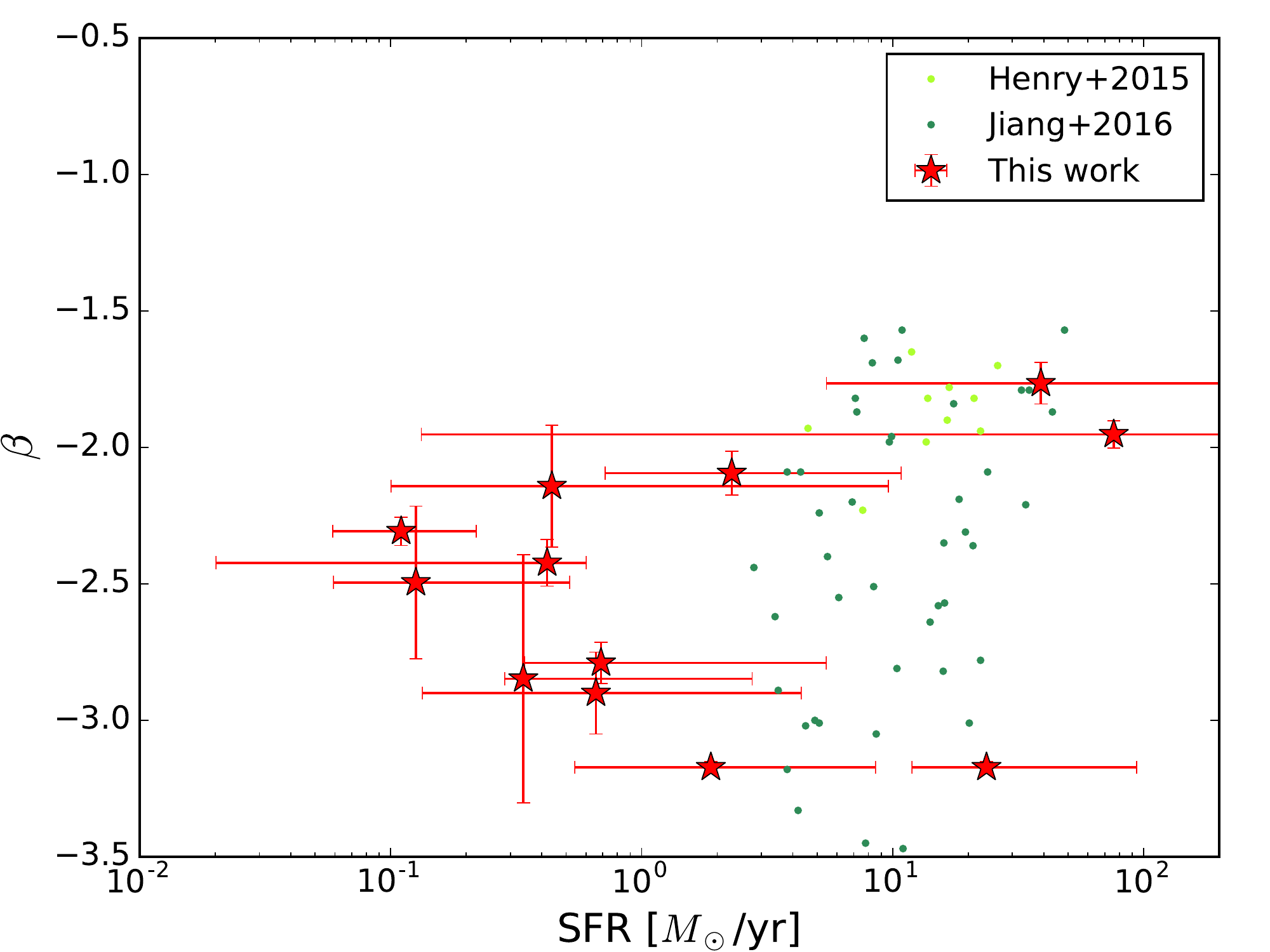}
\caption{The SFR versus the UV-continuum slope $\beta$.
We overplotted a number of LAEs from  
\citet{Henry2015} and \citet{Jiang2016}. \label{fig:sfr_beta}}
\end{center}
\end{figure}

A steep UV-continuum slope is associated with young and rapidly star forming galaxies without a significant amount of dust. Therefore, one would expect that $\beta$ might correlate with the SFR. We show the comparison between SFR and $\beta$ in Fig. \ref{fig:sfr_beta}, but we see that there is no obvious correlation between the two either in our sample, or in either of the other samples. The lack of a clear relation between SFR and $\beta$ in our sample could possibly be due to the large uncertainty in the SFR derived from SED modelling, however, it has also been found in other studies. A physical reason for the absence of a correlation could be that the metallicity, and therefore the dust opacity, is not as strongly correlated with SFR as with stellar mass. 

\subsection{Linking gas to stars}

\begin{figure}
\begin{center}
\includegraphics[width=\columnwidth]{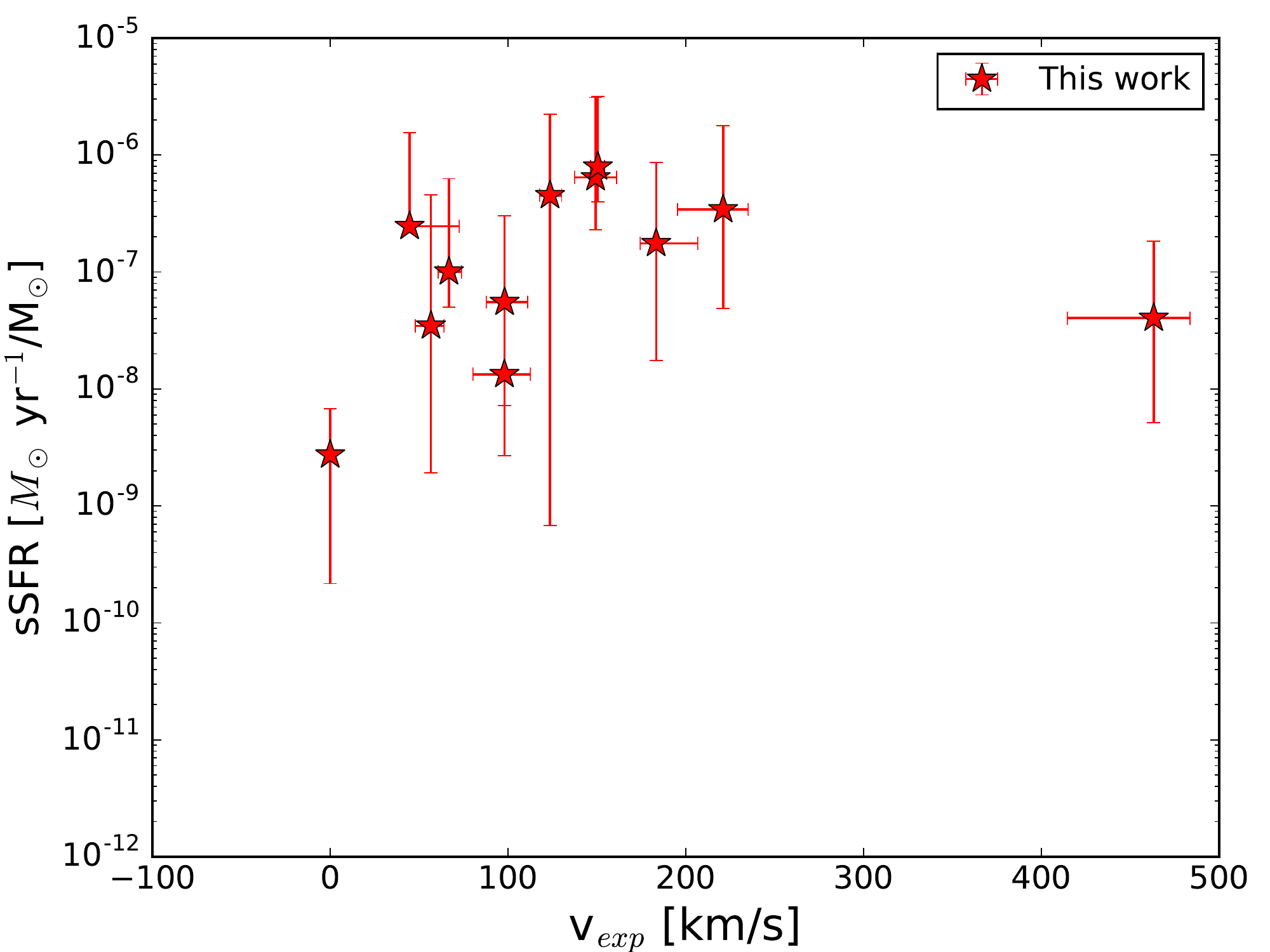}
\caption{The expansion velocity, derived from \Lya line modelling,
against the sSFR, derived from SED fitting. \label{fig:vexp_ssfr}}
\end{center}
\end{figure}

We now compare the stellar properties derived from the broadband photometry
to the gaseous properties derived from the \Lya line profile modelling. 
First, we compare the expansion velocity to the sSFR in Fig. \ref{fig:vexp_ssfr}.
We find no clear relation between these two quantities, see Sec. \ref{sec:discussion}
for a discussion on this.

\begin{figure}
\begin{center}
\includegraphics[width=\columnwidth]{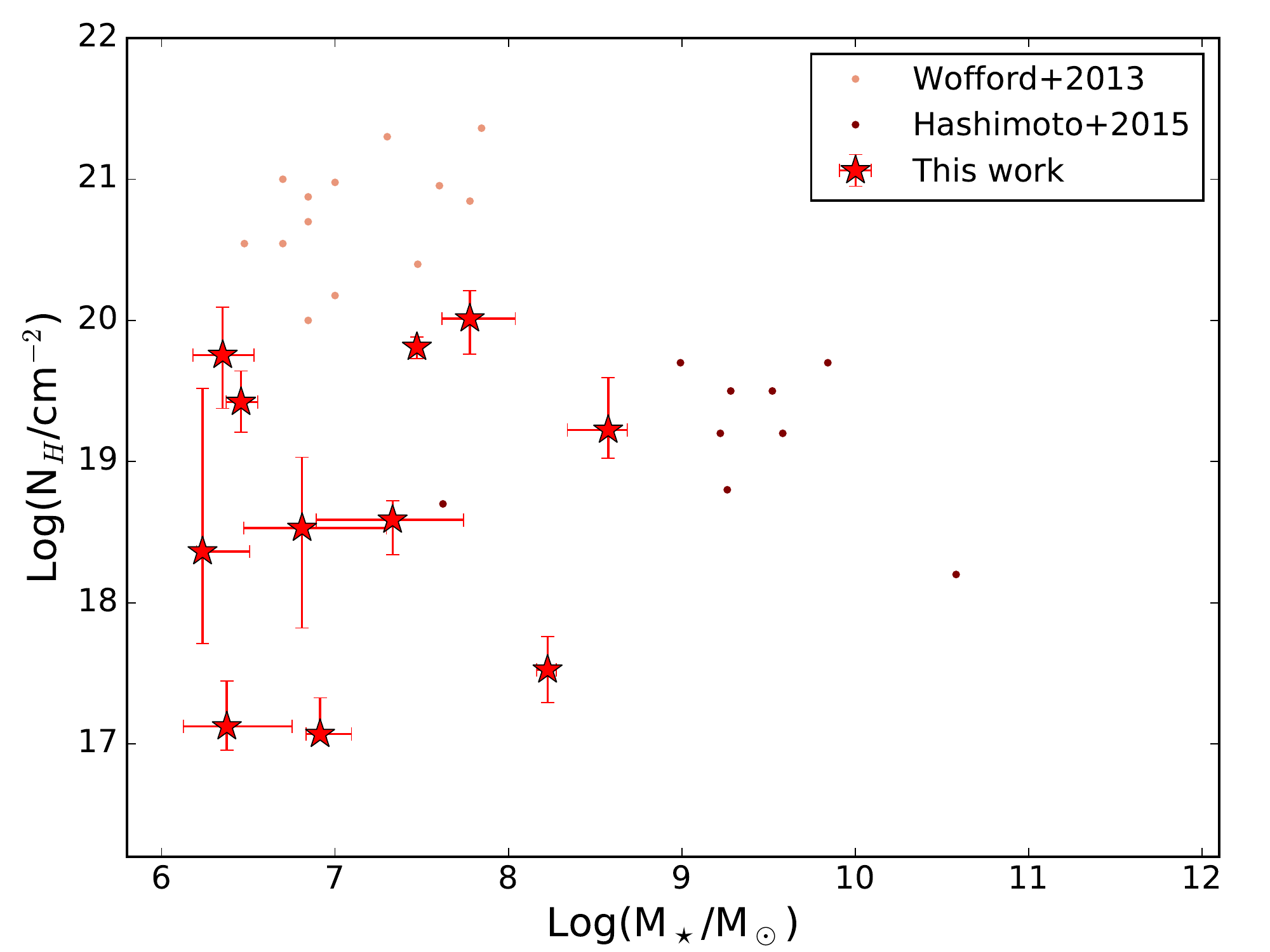}
\caption{Comparison of the stellar mass to the column density
based on \Lya line modelling. We also plot results from a sample
of LAEs analysed in a similar manner by modelling \Lya lines
using shell models \citep{Hashimoto2015}, and a sample of LBGs,
where the column density is derived from UV absorption line 
modelling \citep{Wofford2013}. \label{fig:mass_NH}}
\end{center}
\end{figure}

A more interesting result is found when we compare the
column density to the stellar mass, see Fig. \ref{fig:mass_NH}. On the one hand, we see that the
column density found by \citet{Hashimoto2015} using shell model fitting to $z\sim2.2$ LAEs is similar to those found here,
but at significantly higher stellar mass. On the other hand, the stellar masses
of the $z\sim0.03$ LAEs in \citet{Wofford2013}, who use UV absorption line modelling for LBGs, are similar to the ones discussed here, but the column
densities are two orders of magnitude larger. The difference could be a consequence of
the different redshifts, as low column density objects are at relatively
high-redshift when cosmic star formation was peaking, while the high column density
refers to galaxies at $z<0.06$. This suggests an evolution of the column density
of galaxies, although there are several caveats to be considered, see Sec. \ref{sec:discussion}.
We note, however, that none of the galaxies with column density measures from
\citet{Wofford2013} are LAEs, opposed to the sample of \citet{Hashimoto2015} and the one 
discussed here. The difference in column densities at the same mass could therefore simply mean that selecting by \Lya luminosity sets
a maximum value on the column density, which is in agreement with the finding of \citet{Hashimoto2015} that
the expansion velocity and column density in LAEs are significantly lower than those in LBGs.
It was shown by \citet{Schaerer2011} that a high expansion velocity $v_{\rm exp}>300 \kms$ is required
in galaxies with a high column density in order to allow the escape of \Lya photons. 
These high velocities are significantly higher than expected for $M_\star <10^8$~\Msun galaxies.

\section{Discussion}
\label{sec:discussion}

The absence of a relation between the outflow velocity and the SFR, stellar mass, and sSFR is apparently in contradiction with 
previous studies \citep[e.g.][]{Weiner2009,Bradshaw2013,Erb2014}. The difference in stellar masses could be the prime 
factor between these results, as a burst of star formation can have a much more destructive effect
on a low mass galaxy than on a massive galaxy. The low outflow velocities and young ages are in agreement
with a scenario where a short violent episode of star formation blows away a shell of gas at moderate velocity.
In more massive galaxies, the star formation is less episodic, and the velocity needed to expell gas from
the galaxy is higher. Therefore, one would expect to see a large scatter with on average very moderate velocities in 
low mass galaxies, and an increasing trend of outflow velocity with stellar mass at higher masses. We note
that all but one velocity reported here are below the velocity of the lowest-mass bin in \citet{Bradshaw2013}.

We used the expanding shell-model to fit the observed spectra with astonishing accuracy given the complexity of the spectral shape and the relatively few free parameters of the model. This finding is well aligned with previous studies \citep[e.g.][]{Verhamme2008,Schaerer2011,Hashimoto2015,Yang2016} who found the shell-model to represent also a good fit to the data. 

In spite of this success, the shell-model is clearly a simplification of the complex velocity and density fields existing in high-redshift galaxies and the physical meaning of the shell-model parameters is still unclear.  Recently, \citet[][]{Gronke2016} found that the shell-model parameters do not match the ones of a more complex, multi-phase medium. This hints towards a more subtle conversion between the shell-model and the actual physical parameters. For example, \citet{Gronke2016} 
showed that low column densities in shell models can reflect a medium consisting of optically thick clumps of gas with a low covering factor.

Independent of the question whether the parameters can be interpreted literally or are a more abstract quantity, we found several interesting correlations between them and other observed quantities. In particular, we found that the spectra can be reproduced best using lower neutral hydrogen column densities in shell models than in previous studies \citep{Verhamme2008,Hashimoto2015} which studied brighter galaxies than our sample. This fact combined with the relative large \Lya escape fraction for most of our objects \citep[the \Lya and LyC escape fractions are expected to correlate, see e.g.,][]{Yajima2014,Dijkstra2016}, and the low mass \citep[simulations suggest larger LyC escape fractions for lower mass galaxies, e.g.,][]{Paardekooper2015} makes our sample ideal candidates for LyC leaking galaxies.

When comparing properties as outflow velocity or column density to other studies,
one has to be aware of differences in methods. Modelling \Lya profiles is
very sensitive to neutral hydrogen, but as a large number of different properties
of neutral hydrogen are involved, degeneracies arise naturally. For example,
there are strong degeneracies between optical depth and EW, between $\sigma_i$ 
and the temperature of the gas. Perhaps
the most important degeneracy arises between $z$ and a combination of $v_{\rm exp}$ 
and $N_{\ion{H}{I}}$, i.e. a difference in the systematic redshift can 
be compensated by changing the outflowing velocity and the column density or
morphology or vice versa. The very consistent results of the \Lya modelling and the outflow velocity and column density determinations from other emission lines \citep[see,][]{Vanzella2016b} is encouraging. Results from relatively comparable
samples of galaxies with different techniques strengthen the finding of
low outflow velocities for faint LAEs \citep{Erb2014,Trainor2015}.
Running a large number of parametrizations of the properties in
combination with constraining the free parameters by other means helps to 
minimize the number of degeneracies. \citet{Gronke2015} showed
that the column density, expansion velocity, and $\sigma$ are well recovered by these models,
 indicating that for the properties discussed here the influence
of degeneracies is rather low. Other methods of determining the column 
density in outflows, for example through absorption line fitting
\citep[e.g.][]{Wofford2013,Karman2014}, are based on several assumptions 
such as temperature, metallicity, and electron density, which can lead to
systematic differences. Although there is currently no evidence for these
systematics, these concerns should be taken into consideration when comparing
different techniques. We therefore caution the reader when we compared
our column densities to those of \citet{Wofford2013}.

The simplified modelling thus introduces additional uncertainties.
As mentioned previously, if we are observing the galaxy through one of the holes of a patchy
distribution, the modelled column density will be underestimated \citep{Gronke2016}. Further,
as star formation occurs stochastic in these low mass galaxies 
\citep[e.g.][]{Cloet2014,Hopkins2014,Maseda2014,Shen2014,Dominguez2015,Guo2016}, it is expected that rather than a single
shell, multiple shells are present from the multiple episodes of star 
formation. It is unclear how much this affects the galaxies studied here,
as their ages indicate very young galaxies, preventing a large number
of episodes of star formation. 

The stochastic nature of low-mass galaxies poses another caveat, as we have
here modelled the galaxies with a single exponentially decaying star formation 
history. The very young ages could be a result of a strong recent episode of
star formation on top of a more evolved and older stellar population. However,
for most of the galaxies, the non-detections in the Hawk-I observations limit
a possible contribution of old stellar populations. At the moment,
the available data are unfortunately insufficiently deep to directly observe the
older population. This is because older populations dominate at longer wavelengths, for 
which data with similar depth and resolution are currently not achievable. 

Although we are including the strongest emission lines in our SED-fitting,
we are not including a nebular continuum. The best-fitting models match
the observations relatively well in the restframe FUV, but the nebular 
emission is normally more dominant in the optical and FUV. By ignoring
the nebular continuum, it is possible we slightly underestimate the ages,
find slightly bluer UV continua, and slightly lower masses, as we have
seen from a comparison of one galaxy within our sample 
\citep[see][]{Vanzella2016b}. For this galaxy (SW-49), we included nebular continuum
emission and a large number of absorption lines as described by \citet{Schaerer2009}.
Including this, we find an increase of 0.4 dex in the stellar mass, a similar age,
and an increase of 0.2 in $\beta$ compared to the values presented in Tab. 
\ref{tab:sed}. These differences are within our error bars, but 
agree with the general trends discussed above. Therefore, the found differences
are  not significant enough to change trends or conclusions in this paper.

\section{Summary and conclusions}
\label{sec:conclusions}

In this article, we used deep IFU and broadband observations in combination with 
gravitational lensing to study the properties of intrinsically faint LAEs at z$>3$. We targeted
two sides of the Frontier Fields cluster AS1063 for in total 10 hours of observation
time. We determined spectroscopic redshifts for 172 galaxies in a 2 arcmin$^{2}$ 
central field of the cluster. 
Among these redshifts are 17 multiple redshift families, of which 11 did not
have a redshift determination before. 
We extracted spectra for 14 LAEs, for which we observed in total 24 multiple
images. These LAEs were corrected for lensing magnification by an updated gravitational lensing
model, and are found to be among the intrinsically faintest observed at z$>3$. 
Without the help of lensing, the study of such a faint population of sources
with absolute UV-magnitudes ranging from -19 to -14 would have been possible only with the next generation of telescopes, 
 i.e. {\em James Webb Space Telescope} and European Extremely Large Telescope.

We modelled the \Lya line profiles using a spherically expanding gas shell
model, and found that the narrow lines of these faint LAEs are best-fit low column densities. The line
profiles are best fit with shells with low to moderate expansion velocities,
which is in line with their low luminosities. 

We obtained broadband photometry from the Hubble Frontiers Fields programme and the 
CLASH programme, and combined this with Hawk-I K-band and {\em Spitzer}
IRAC data. We used the derived photometric catalog to perform SED fitting on the LAEs,
and obtained stellar masses and ages, the UV continuum slope, and E(B-V).
We found low stellar masses for these objects, significantly lower than masses
reported for LAEs at any redshift before. In combination with the low stellar 
masses, these LAEs are characterised by young populations of tens of Myrs, 
and low dust obscuration. These properties are very similar to the low-mass
galaxies which are currently the main candidates for reionization, and provide
therefore excellent analogues.

That these galaxies are recently formed and rapidly
building up their stellar mass is clear from their sSFR, which is increased
by an order of magnitude in our sample compared to other LAE samples. We found
a comparable sSFR only in some very high-redshift extreme emission line 
galaxies. This shows that these galaxies are starburst galaxies, and
this rapidly forming nature might be characteristic for low mass galaxies
at z$>3$. We note however, that quiescent low-mass galaxies would not be 
detected in our observations. 

The UV continuum slope is very blue for these galaxies. We used the best-fitting
SED model to derive $\beta$, and found that $\beta<-2$ for all galaxies with 
M$_\star<10^8$ M$_\odot$. Such low values of $\beta$ are very rarely observed, and 
approach the theoretical limit of stellar population models. It has been argued,
however, that for example top-heavy IMFs or Pop III stars can decrease $\beta$ further \citep[e.g.][]{Bouwens2010,Zackrisson2011}, which
could explain the even steeper slopes found for some of the LAEs discussed here.

We did not find any correlation between the expansion velocity and other 
quantities. The absence of a relation between SFR and $v_{\rm exp}$ could
be due to the stochastic nature of SF in low mass galaxies. This would increase
the scatter such that any relation would be washed out in the mass range
studied here. In agreement with this, the maximum velocities derived here are 
less than those found in more massive galaxies.

The observed decrease of $\beta$ with stellar mass, in combination with the
previously found increase of occurence of \Lya emission with decreasing
stellar mass, suggests that low mass galaxies might have been effective 
at leaking ionizing radiation. If the escape fraction of Lyman continuum
photons follows these trends, this would confirm low mass
galaxies as the main drivers of reionization. The candidacy for Lyman continuum 
leakage is strengthened by the narrowness of the \Lya lines, which indicates
either a low column density or a low covering fraction. Both of these conditions
are predicted to enable Lyman continuum leakage. Because these faint LAEs are
probably good analogues to the galaxies responsible for reionization, it is
important to determine if these galaxies are indeed leaking Lyman continuum 
radiation.

\begin{acknowledgements}
 
The authors thank the anonymous referee for constructive comments
which helped improve the paper. The 
authors thank Ryan Trainor, Ryota Kawamata, and Kasper Schmidt for 
insightful comments on the manuscript.
Based on observations made with the European Southern Observatory 
Very Large Telescope (ESO/VLT) at Cerro Paranal, under programme IDs 60.A-9345(A),
095.A-0653, and 095.A-0533.
STScI is operated by the Association of 
Universities for Research in Astronomy, Inc. under NASA contract NAS 5-26555.
C.G. acknowledges support by VILLUM FONDEN Young Investigator Programme through grant no. 10123.
PR, AM and MN acknowledge support from PRIN-INAF 2014 1.05.01.94.02 (PI M. Nonino).
This research made use of APLpy, an open-source plotting package for Python hosted at \url{http://aplpy.github.com}.

\end{acknowledgements}

\bibliographystyle{aa}
\bibliography{muse.bib}

\appendix

\section{Redshifts in AS1063}
\label{sec:appa}

As described in Sect. \ref{sec:data}, we extracted spectra for all
sources for which we; 1) extracted a location in the 
spectrally stacked MUSE image using {\sc SExtractor} ; 2) found
emission lines in the MUSE datacube through visual inspection; 3)
multiple images of a source were expected from the lensing
model; 4) galaxies with a $M_{\mathrm F814W}<23.5$. We
repeated this analysis on the already published data of
the south-western half of the cluster \citepalias{Karman2015}, 
and added 13 cluster galaxies with spectroscopic redshift,
2 multiply lensed galaxies described by \citetalias{Caminha2015},
and 1 new multiply lensed galaxy described in the text to
the catalogue presented in \citetalias{Karman2015}.

We determined
redshifts in the same way as described by \citet{Grillo2016},
where redshifts were independently confirmed by 2 authors. Finetuning 
of the fourth decimal was done by a cross correlation with a template
using {\sc SpecPro} \citep{Masters2011}, where we added an 
additional higher-resolution GMASS template \citep{Kurk2013}.
Visual confirmation of the finetuning was performed for all spectra,
and, if necessary, redshifts were slightly adjusted in case of
residual offsets. We note that the quality flags used here follow those used in
\citet{Grillo2016}, and differ from those used in \citetalias{Karman2015}, 
decreasing in quality from 4 to 1. A QF of 4 is a highly secure
redshift ($\delta z<0.0004)$), QF=3 is a secure
redshift ($\delta z<0.001)$), QF=2 is an uncertain
redshift ($\delta z<0.01)$), while QF=1 indicates a tentative redshift.
If the redshift is based on emission lines, we use QF=92 in case
of a doublet or a clear line profile, while a single emission line
without a characteristic profile is assigned a QF=91. We labelled
all sources according to the region of detection (NE or SW) for clarity, 
keeping the IDs assigned previously by \citetalias{Karman2015} or
\citetalias{Caminha2015}.

\begin{figure}
\begin{center}
\includegraphics[width=\columnwidth]{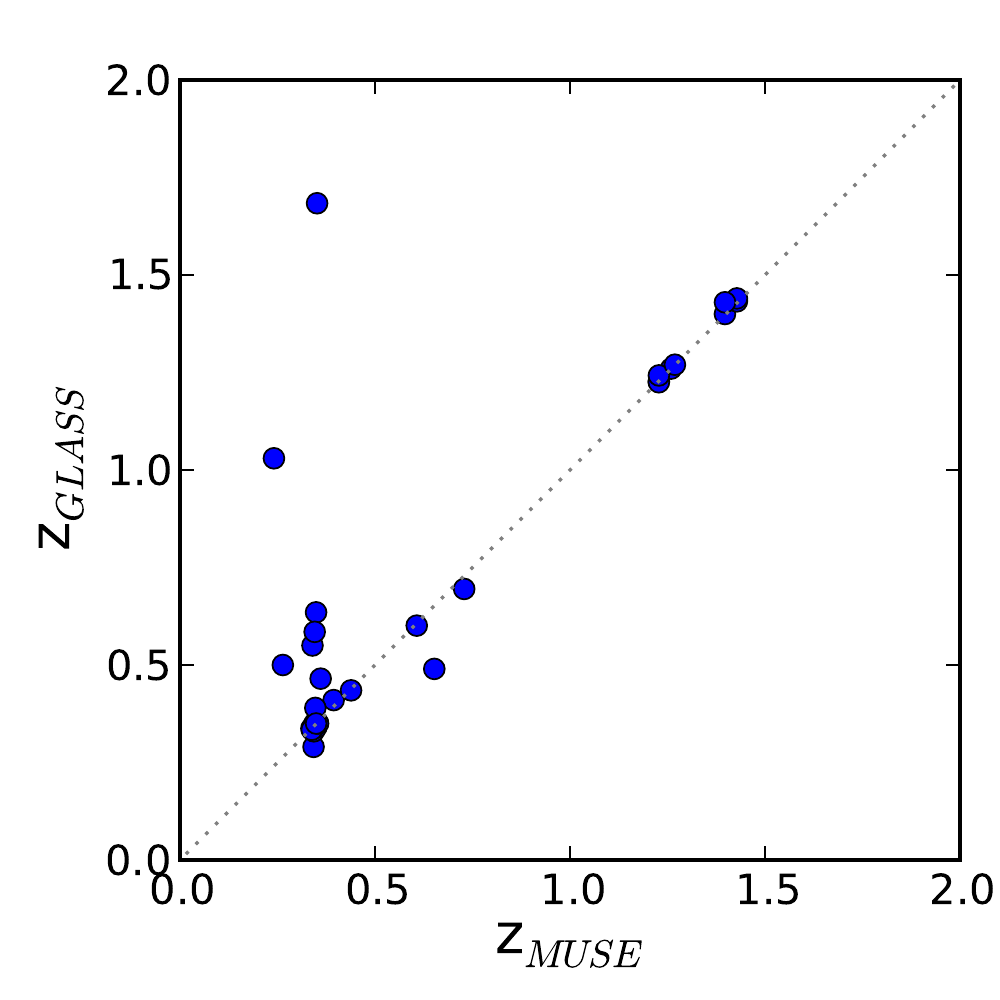}
\caption{Comparison of spectroscopic redshifts from this study and
GLASS \citep{Treu2015,Schmidt2016}. \label{fig:muse_glass}}
\end{center}
\end{figure}

We compared our redshifts to the release of the Grism Lens-Amplified Survey from Space 
survey \citep[GLASS;][]{Treu2015,Schmidt2016}. In general the redshifts agree
reasonably well, see Fig. \ref{fig:muse_glass}, and we find disagreeing redshifts
for only 12 galaxies which have low quality flags in GLASS. We note that the disagreeing
redshifts might be largely due to an incorrect cross matching, where they find lensed
background galaxies, while we indentified the lensing cluster members. In addition, we determined
redshifts for over 50 more galaxies in the inner part of the cluster, while six galaxies
with redshifts from GLASS are not identified in our catalog. We note that the area of GLASS
extends much further from the center of the cluster and the spectroscopy is performed
in the NIR, making the two catalogues complementary. Unfortunately, only one of
the LAE candidates at $z>6.5$ proposed by \citet{Schmidt2016} falls within the
observed area, and we do not detect any emission lines at this position. This
non-detection strenghtens the \Lya identification proposed by \citet{Schmidt2016}.

\input{AS1063_total_cat_v2}

\section{Lensing models}
\label{sec:lensing}

For an individual background galaxy which is multiply imaged, the positions of 
all images are determined by the relative distance of lens and object,
and the mass distribution in the lens. A larger number of multiple image
families with spectroscopic redshifts therefore allows one to more accurately
model the mass distribution of the cluster. We updated our
previous strong lensing model ID F1-5th, as presented in 
\citetalias{Caminha2015}, by using the new redshift information. 
In the updated model we included the extra constraints from 6 additional background
sources at different redshifts (sources NE-87, NE-93, NE-94, NE-98, NE-99, 
and NE-91, see Tabel \ref{tab:multim}), totalling 15 multiple image families over
the redshift range from $z=0.72$ to $z=6.11$.
Here, we describe the model shortly, but we refer the interested reader to
\citetalias{Caminha2015} for more details.

We modelled the galaxy cluster using three mass components. The first
component is a smooth and extended exponent corresponding to the global
dark matter halo, the hot gas present in the cluster, and the intra-cluster
light. We model this component by a pseudo-isothermal elliptical mass 
distribution \citep[hereafter PIEMD][]{Kassiola1993}. The PIEMD 
parametrization has six free parameters: the centre position (2), the
ellipticity, the orientation, a fiducial velocity dispersion, and the 
core radius. 

The second and third mass components correspond to the BCG and the other
galaxy cluster members. The cluster galaxies are modelled using a dual
pseudo-isothermal mass distribution \citep{Eliasdottir2007,Suyu2010} with
zero ellipticity and core radius and a finite truncation radius. To
reduce the number of free parameters, all cluster members except for the
BCG have a scaled velocity dispersion and radius, based on their luminosity
in the {\em F160W} band. This leaves only 2 free parameters for the full
set of cluster members. Because the BCG has a large influence on the
central volume of the cluster, we do not scale its velocity dispersion and
radius based on its luminosity, but leave these as additional free parameters
together with its ellipticity and orientation.

Using these mass components, we model the strong lens using the
publicly available software {\em lenstool} \citep{Kneib1996,Jullo2007}. This
software minimizes the distance between the observed and predicted positions
of multiple images based on a Bayesian Markov chain Monte Carlo technique, 
by varying the free parameters given by the mass distributions.

\begin{table}[!h]
\begin{center}
 \begin{tabular}{ccccc}
 {\bf ID}& {\bf RA} (J2000) & {\bf DEC} (J2000) & $z$ \\
  \hline 
2a & 342.19588 & -44.52895 & 1.2278$^{a,b}$ \\ 
2b & 342.19450 & -44.52698 & 1.2279$^{a,b}$ \\ 
2c & 342.18642 & -44.52116 & 1.2277$^{a,b}$ \\ 
2d & 342.19520 & -44.52786 & 1.2279 \\ 
3a & 342.19271 & -44.53119 & 1.2592$^{a,b}$ \\ 
3b & 342.19213 & -44.52983 & 1.2593$^{a,b}$ \\ 
3c & 342.17983 & -44.52156 & -- \\ 
4a & 342.19317 & -44.53653 & 1.3972 \\ 
4b & 342.18783 & -44.52731 & 1.3972$^{a,b}$ \\ 
4c & 342.17921 & -44.52359 & 1.3972$^{a,b}$ \\ 
6a & 342.18847 & -44.53998 & 1.428$^{a,b,c}$ \\ 
6b & 342.17585 & -44.53254 & 1.428$^{c}$
\\
6c & 342.17420 & -44.52831 & 1.428$^{c,d}$
\\
7a & 342.18006 & -44.53842 & 1.035$^{c}$ \\ 
7b & 342.17554 & -44.53590 & 1.035$^{c}$ \\ 
7c & 342.17191 & -44.53023 & 1.035$^{c}$ \\ 
8a & 342.18186 & -44.54050 & 1.837$^{a}$ \\
8b & 342.17424 & -44.53711  & 1.837$^{a}$ \\
8c & 342.16938 & -44.52726 & -- \\
11a & 342.17505 & -44.54102 & 3.116$^{c,e}$ \\
11b & 342.17315 & -44.53999 & 3.116$^{a,b,c,e,f}$ \\
11c & 342.16557 & -44.52953 & -- \\
13a & 342.19369 & -44.53014 & 1.2583 \\ 
13b & 342.19331 & -44.52942 & 1.2583 \\
14a & 342.19088 & -44.53747 & 6.107$^{b,c,g}$ \\ 
14b & 342.18106 & -44.53462 & 6.107$^{b,c,g}$ \\ 
14c & 342.18904 & -44.53004 & 6.107 \\ 
14d & 342.17129 & -44.51982 & 6.107$^{b,g}$\\
14e & 342.18402 & -44.53159 & 6.107 \\  
18a & 342.18150 & -44.53936 & 4.113$^{c}$ \\
18b & 342.17918 & -44.53870 & 4.113$^{c}$ \\
20a & 342.18745 & -44.53869 & 3.118$^{a,h}$ \\
20b & 342.17886 & -44.53587 & 3.118$^{a,h}$ \\
20c & 342.17065 & -44.52209 & --\\
21a & 342.18586 & -44.53883 & 3.606 \\
21b & 342.17892 & -44.53668 & 3.606 \\
87a & 342.18429 & -44.52529 & 0.7287 \\ 
87b & 342.18894 & -44.52864 & 0.7287 \\ 
87c & 342.19010 & -44.53010 & 0.7287 \\ 
91a & 342.19238 & -44.52505 & 2.9760 \\
91b & 342.18151 & -44.52025 & -- \\
91c & 342.19838 & -44.53575 & -- \\
93a & 342.18283 & -44.52028 & 3.169 \\ 
93b & 342.19196 & -44.52409 & 3.169 \\ 
94a & 342.18935 & -44.51871 & 3.2857 \\ 
94b & 342.19615 & -44.52291 & 3.2857 \\  
98a & 342.19015 & -44.53093 & 5.0510 \\
98b & 342.19085 & -44.53566 & 5.0510 \\ 
99a & 342.18378 & -44.52122 & 5.2373 \\ 
99b & 342.18874 & -44.52276 & 5.2373\\ 
100a & 342.19701 & -44.522121 & 5.8940\\ 
100b & 342.19010 & -44.517886 & -- \\ 
 \end{tabular}
 \caption{Spectroscopically confirmed multiple images in AS1063, see Table \ref{tab:redshifts} for quality flags. Previous redshift
 determinations by: $^a$\citetalias{Caminha2015}, $^b$\citet[][]{Balestra2013}, $^c$\citetalias[][]{Karman2015}, $^d$\citet[][]{Richard2014a},$^e$\citet[][]{Vanzella2016b}, $^f$\citet[][]{Johnson2014}, $^g$\citet[][]{Boone2013}, and $^h$\citet[][]{Caminha2016}.\label{tab:multim}}
\end{center}
\end{table}

Our best-fitting model recovers the position of the multiple images with a precision
of $\lesssim 0.3\arcsec$, demonstrating that our model provides a good description
of the observations.
We calculated the magnification factors, $\mu$, using the median magnification
from a MCMC with 10000 points. We found values varying from $\mu=3$ to 
magnifications of $\mu\approx50$, see Appendix \ref{sec:app_sed} for an overview
of all magnification factors.

\section{Individual SED fitting results}
\label{sec:app_sed}

\input{SED_figures}

In Sect. \ref{sec:stars}, we showed and discussed the properties of the
LAEs after combining the individual results. Here, we present the individual
SED-modelling results of all LAEs and the LBG, as shown in Table \ref{tab:sed_ind} and in Fig. 
\ref{fig:ind_seds}. In addition, we present the magnification factors $\mu$ of 
all of the LAEs  in Table \ref{tab:sed_ind}.

\begin{figure*}
\begin{center}
\includegraphics[width=\textwidth]{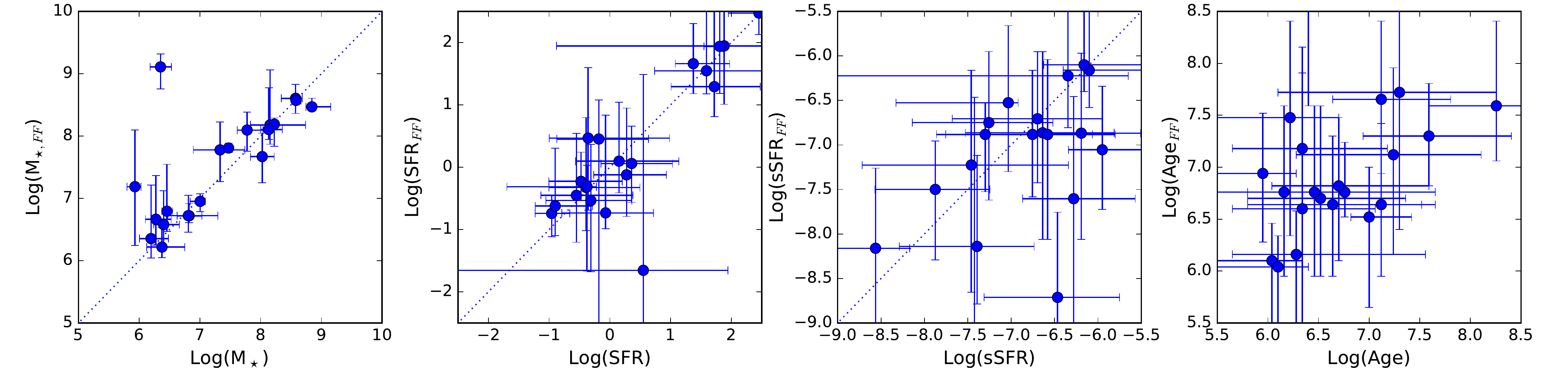}
\includegraphics[width=\textwidth]{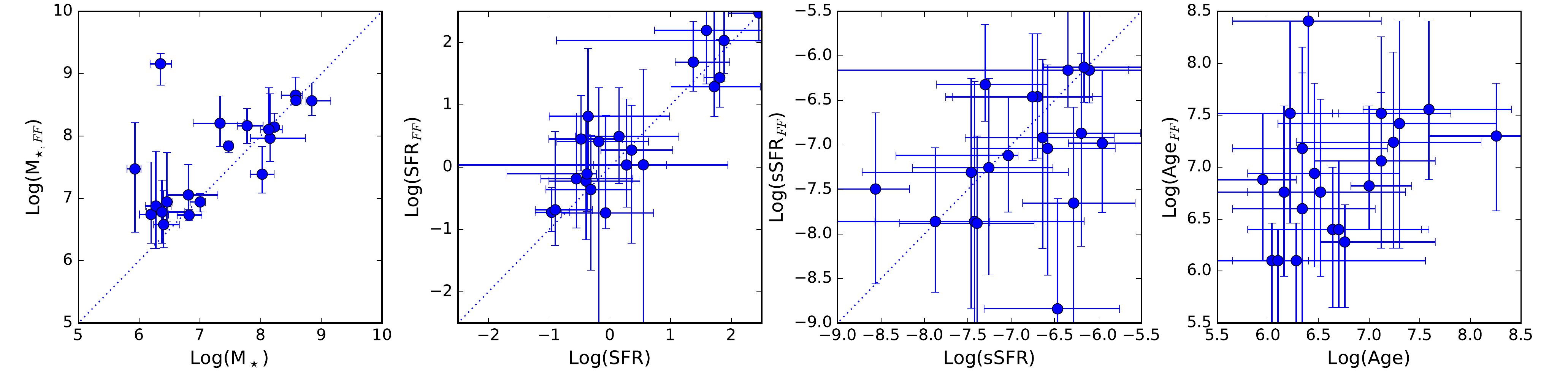}
\caption{Comparison of the SED-fitting results obtained using the Kron-radii photometry using Kron-radii on the x-axis
versus results obtained using a 2\arcsec aperture (top row) or 3\arcsec (bottom row)  on the y-axis. The panels
show, from left to right, the stellar mass, the SFR, the sSFR, 
and stellar age. \label{fig:comp_apertures}}
\end{center}
\end{figure*}

Using Kron-radii on different detection images can lead to colour differences when the PSF
is not corrected for. As the PSF of {\em HST} is relatively small, circular apertures of 2\arcsec to 
3\arcsec should not suffer from this effect
We have checked this by comparing the results of the SED-fitting obatined using different choices of apertures, and the results are shown in Fig. \ref{fig:comp_apertures}. Despite some scatter, most of the results are consistent within 1$\sigma$. We note that the estimates of ages show the largest scatter. However, the uncertainties are large on this parameter and the derived ages are still consistently young ($<100$ Myr).
The one outlier is NE-97, which suffers from a significant increase in errors in the {\em HST} photometry compared to the IRAC photometry when using circular 2\arcsec\ to 3\arcsec\ apertures. We note however, that this IRAC detection is less reliable than the results suggest, as we failed to visually confirm the presence of this source in the IRAC maps. We therefore deem the results of the fitting using these circular apertures less reliable than the original values obtained using Kron-radii, which was dominated by the {\em HST}-photometry. Because the estimates for the stellar properties remain within 1$\sigma$ of the values used in the main body of the paper, we conclude that
the results obtained with the Kron-radii are reliable and that are conclusions hold.

\begin{figure*}
\begin{center}
\includegraphics[width=\textwidth]{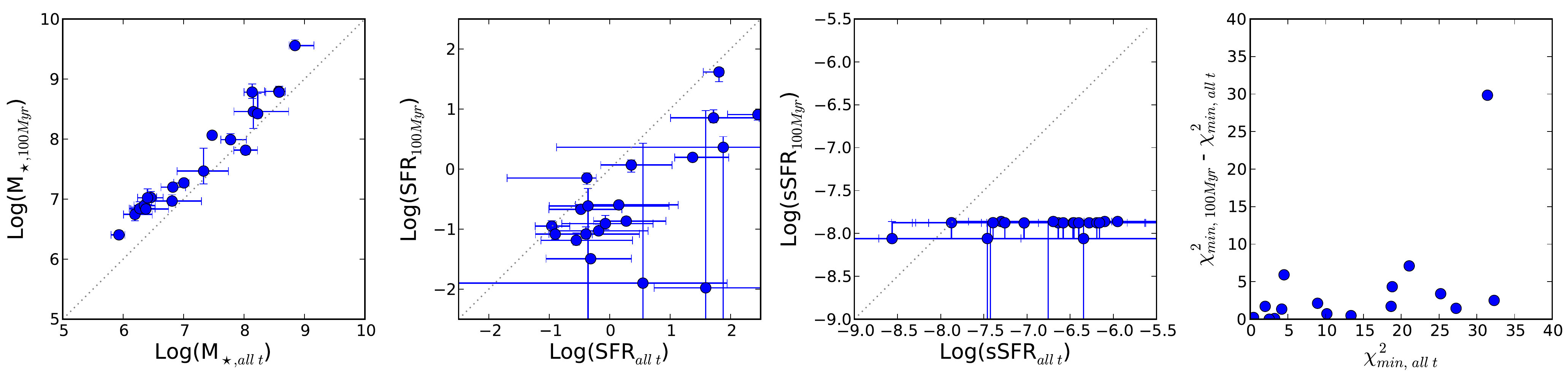}
\caption{Comparison of the SED-fitting results obtained when leaving the age free (x-axis)
versus those obtained for a fixed stellar age of 100 Myr (y-axis). The panels
show, from left to right, the stellar mass, the SFR, the sSFR, 
and the difference in $\chi^2$. \label{fig:comp_100myr}}
\end{center}
\end{figure*}

\begin{figure*}
\begin{center}
\includegraphics[width=\textwidth]{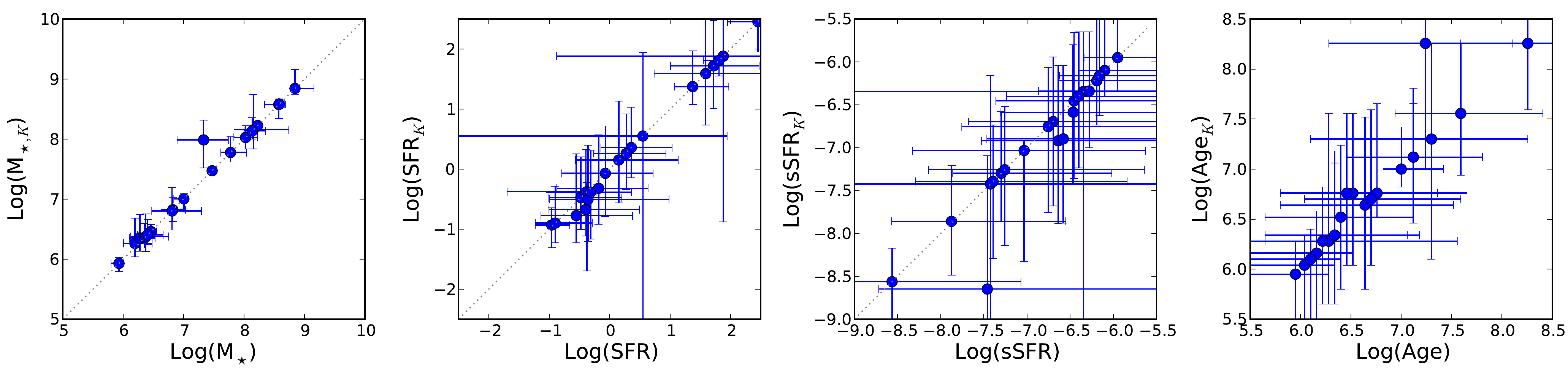}
\caption{Comparison of the SED-fitting results obtained using the original photometry (x-axis)
versus those obtained using a catalogue where mock 3$\sigma$ K-band detections were inserted for
non-detections (y-axis). The panels
show, from left to right, the stellar mass, the SFR, the sSFR, 
and stellar age. \label{fig:comp_Kband}}
\end{center}
\end{figure*}

We see that the magnification corrected stellar masses of different images are 
all within 2$\sigma$ of each other, except for SW-52a. The increased difference 
between these two images is due to the addition of a Hawk-I detection for SW-52b.
This extra datapoint constrains the mass of this LAE to a significantly higher
value. We find a similar result for the stellar age, SFR, and sSFR, with only 
SW-52a showing a discrepant result. We therefore conclude that by taking the
average of the different lens-corrected results provides reliable estimates
for the stellar parameters, except for SW-52a where we adopt the values obtained
from SW-52b.

In the last column of Table \ref{tab:sed_ind}, we show the determined escape
fraction of \Lyap We find large differences between the escape fractions
calculated for different images, although these are often within the error 
estimates. It is peculiar to see that SW-52b has $f_{\rm esc}>1$, indicating
that either the \Lya flux is boosted \citep{Laursen2013,Gronke2014}, or that
the SFR based on the SED modelling is underestimated. We find the latter more 
likely given that the SFR$_{\rm SED}$ of SW-52a is significantly higher. 

Most of the photometric data used in this paper is in the UV-restframe,
and as a result it is dominated by young stars of which the mass-to-light
ratio can vary significantly. To test the reliability of our SED fitting we
performed two tests. 

First, we constrained the stellar age of our SEDs to 
100 Myr. We found that the stellar masses increase when setting the age rather
than leaving it as a free parameter, see Fig. \ref{fig:comp_100myr}. However,
we find that the decrease is only $\sim0.5$ dex for the lowest stellar masses
in our sample and decreases with increasing stellar mass. As expected, the SFRs are lower
when setting the age to 100 Myr but for most galaxies they agree within 2$\sigma$.
The older stellar age have lower sSFRs by construction, and it is therefore
not surprising that we find significantly lower values. It is important to note
that the $\chi^{2}$ values of galaxies with ages well below 100 Myr and small 
errors increase significantly. 

Second, we used the K-band data to probe the maximum effect adding restframe
optical data would have. For those galaxies where we have no K-band detection,
we insterted a K-band magnitude equal to the $3\sigma$ depth in our catalogue. 
This corresponds  to a conservative upperlimit and therefore illustrates the 
maximum effect an older or redder population might have. We show in Fig. 
\ref{fig:comp_Kband} that there is very little difference between the results
when we include mock $3\sigma$ K-band detections and the original photometry.

These tests show that although we are mostly using 
UV restframe data, the properties derived are reliable and our results
remain intact even using different assumptions. We note that a red and
old population might still be present as three datapoints are not able
to securely determine this, but it does constrain their maximal contribution.

\input{tab_sed_ind}

\section{Individual Lyman $\alpha$ modelling results}
\label{sec:app_lya}

We modelled the profile of the \Lya for each LAE with a clear asymmetric profile
or S/N$>8$ of the \Lya flux individually\footnote{For completeness, we also modelled NE-98b although this object does not satisfy our S/N criterion. However, we could not derive any meaningful results from the \Lya modelling and therefore, the results of the \Lya modelling for this object were not used anywhere else in the paper.},
and we show the results in Figs. \ref{fig:lya_112} to \ref{fig:lya_804}. We
used our spectra to constrain the intrinsic redshift based on UV emission lines
whenever possible, and we measured the FWHM due to the instrument and weather
conditions at every wavelength. We note that there is no clear peak in the
FWHM distribution after fitting, and we can therefore not better constrain the values of the FWHM over
the narrow, but limited allowed range. However, because there does not seem to be correlation with any
of the other parameters, this does not have any implications on the results in this paper. Although the majority of \Lya models are well 
fit by the models, it is clear that in some spectra, for example NE-98b, NE-99a, 
and NE-99b, the sky interference forces the models to an incorrect expansion 
velocity. We have masked out the LAEs where this was a problem in the main
body of the paper.

\input{tab_lya_ind}

\input{Lya_triangles}

\end{document}

%% file: stamps.tex
\begin{figure*}
 \includegraphics[height=0.20\textwidth]{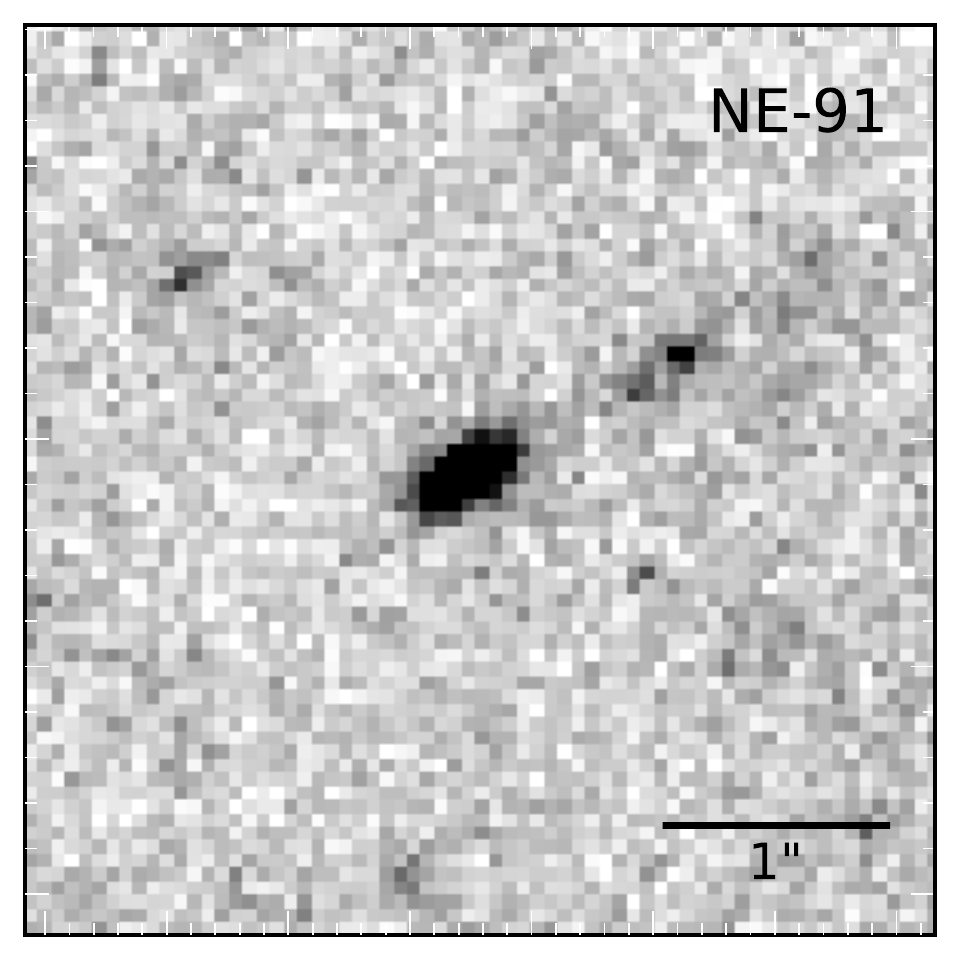}
 \includegraphics[height=0.20\textwidth]{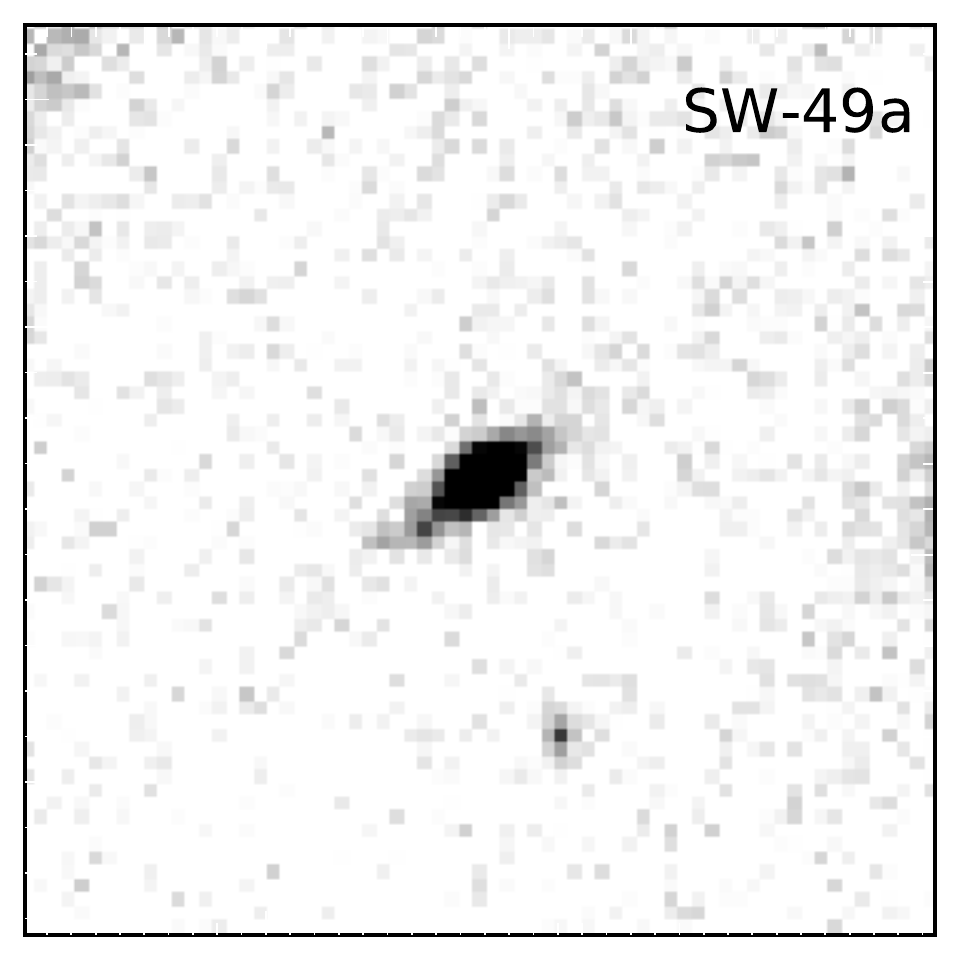} 
 \includegraphics[height=0.20\textwidth]{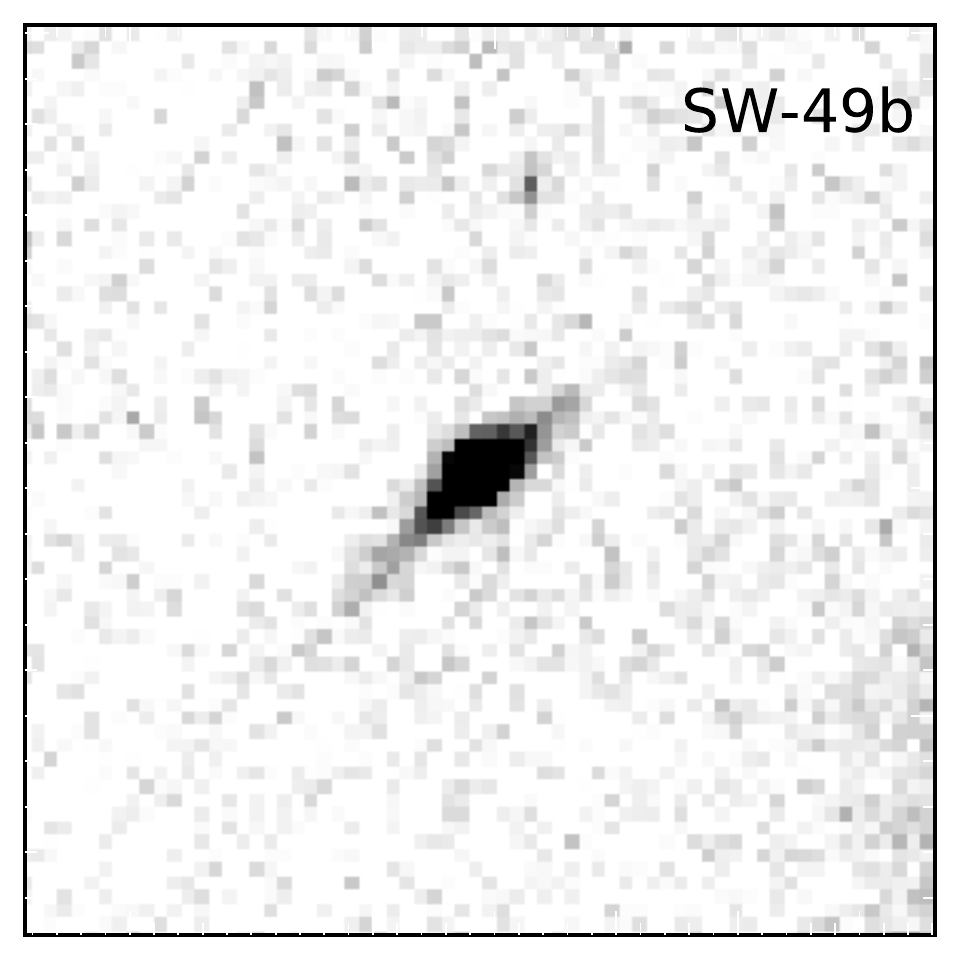}
 \includegraphics[height=0.20\textwidth]{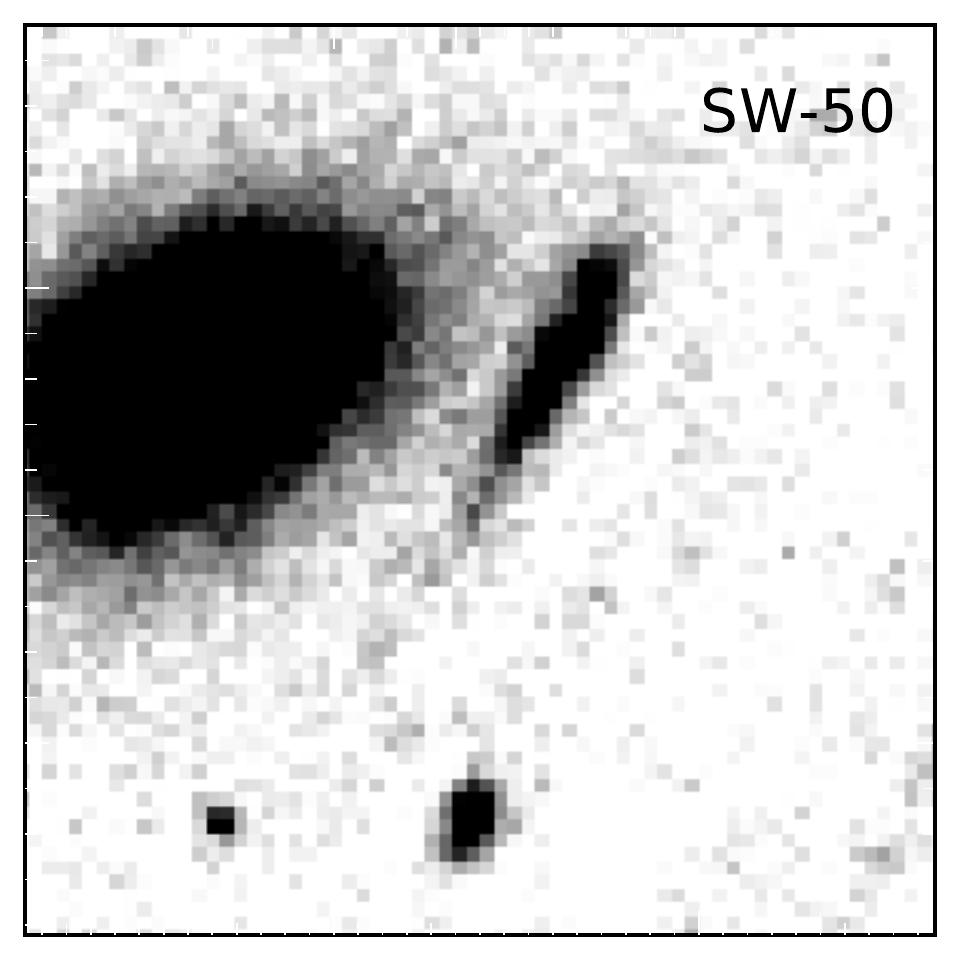}
 \includegraphics[height=0.20\textwidth]{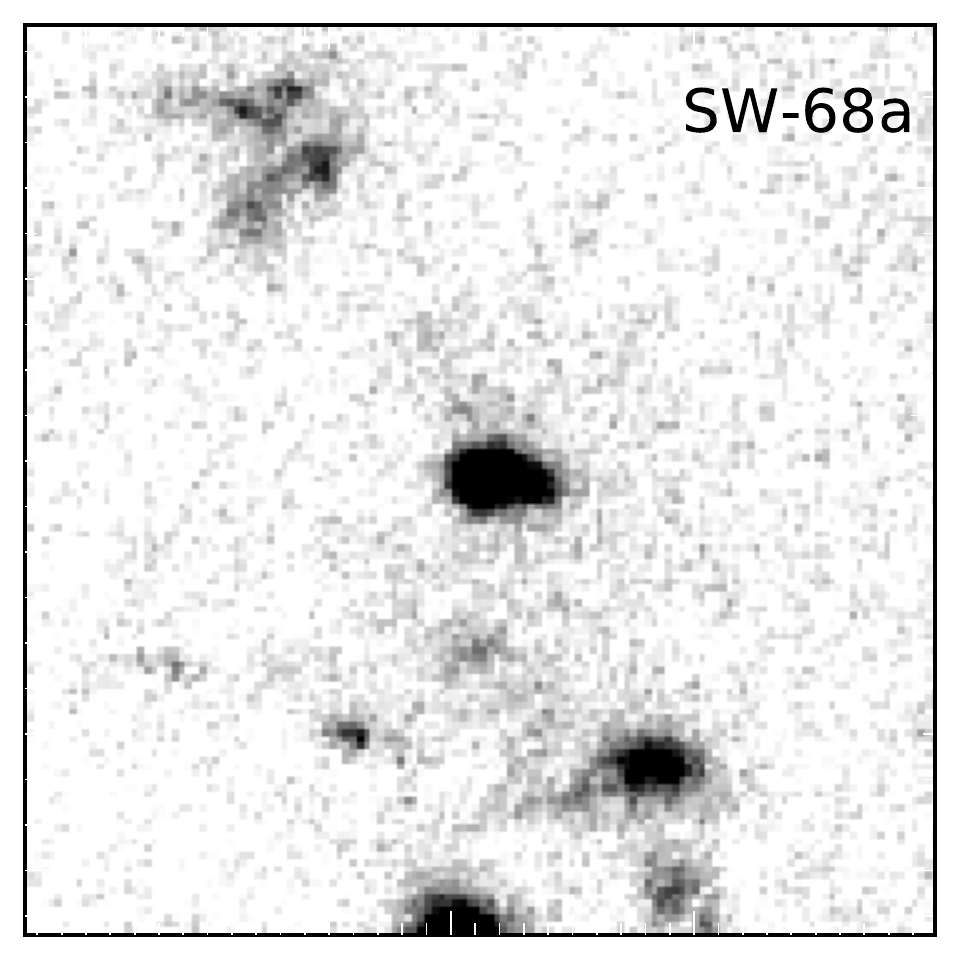} 
 \includegraphics[height=0.20\textwidth]{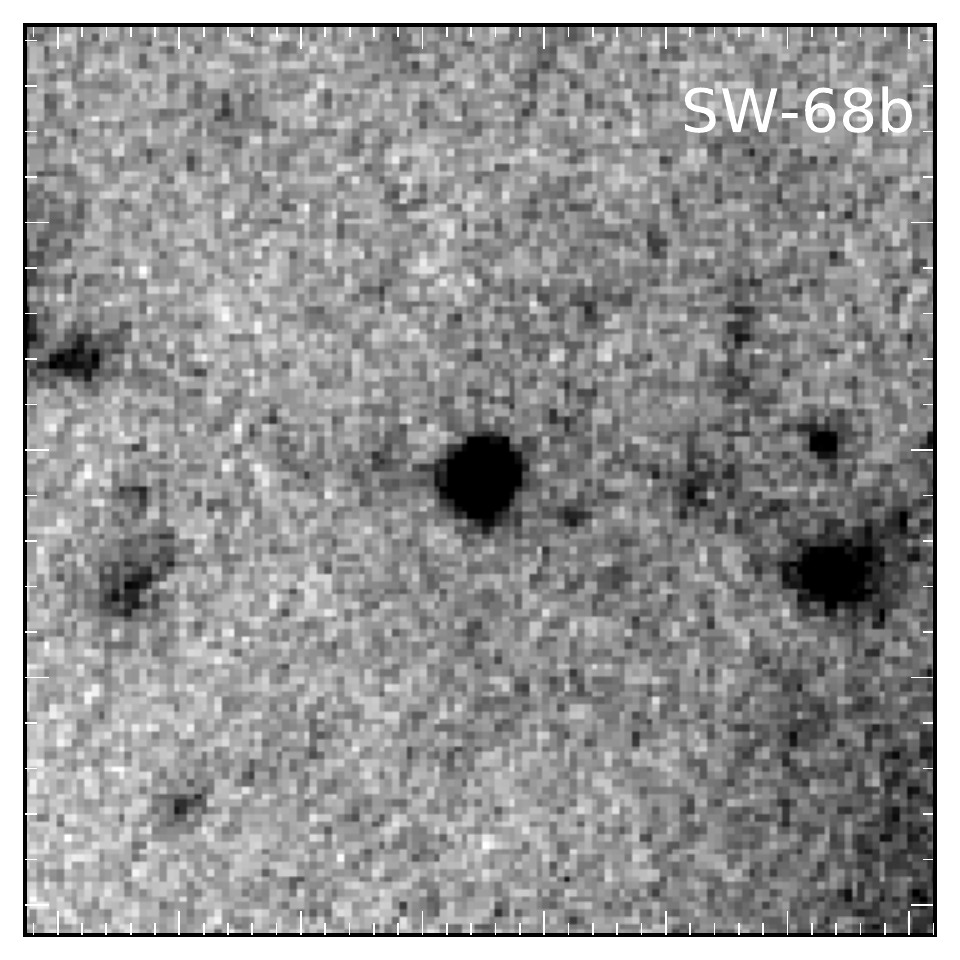}
 \includegraphics[height=0.20\textwidth]{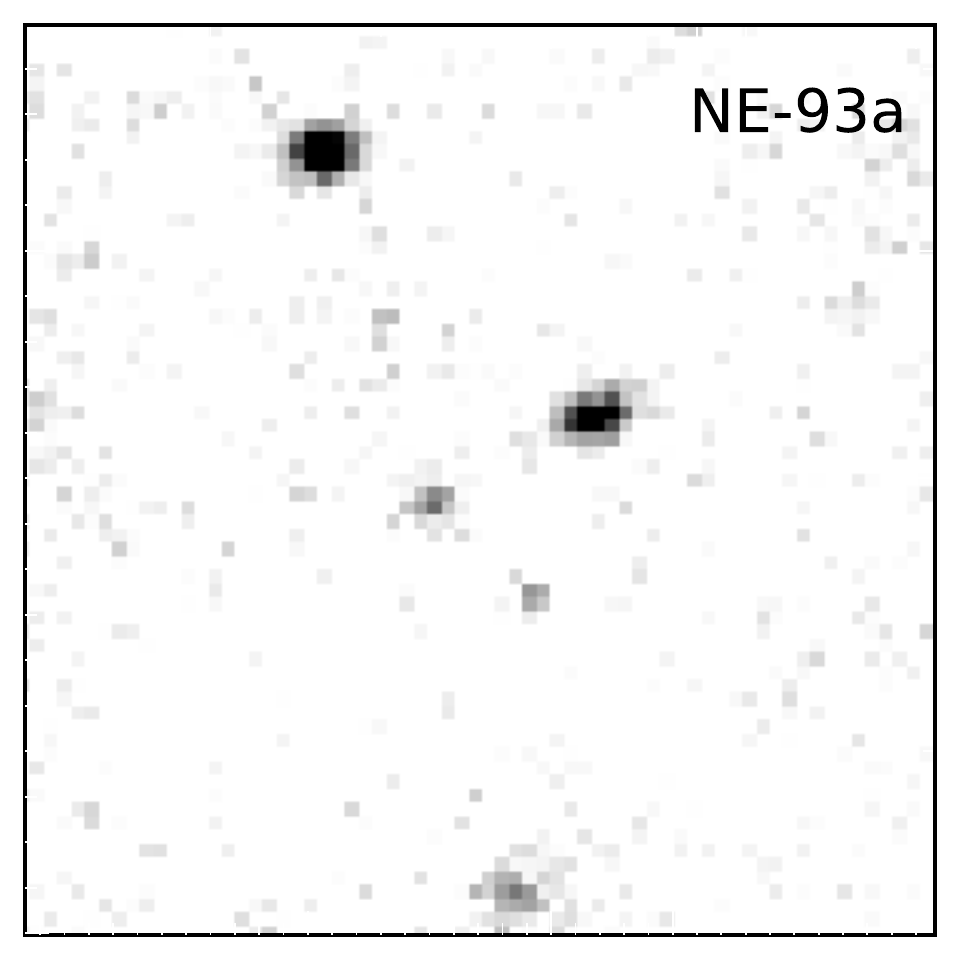}
 \includegraphics[height=0.20\textwidth]{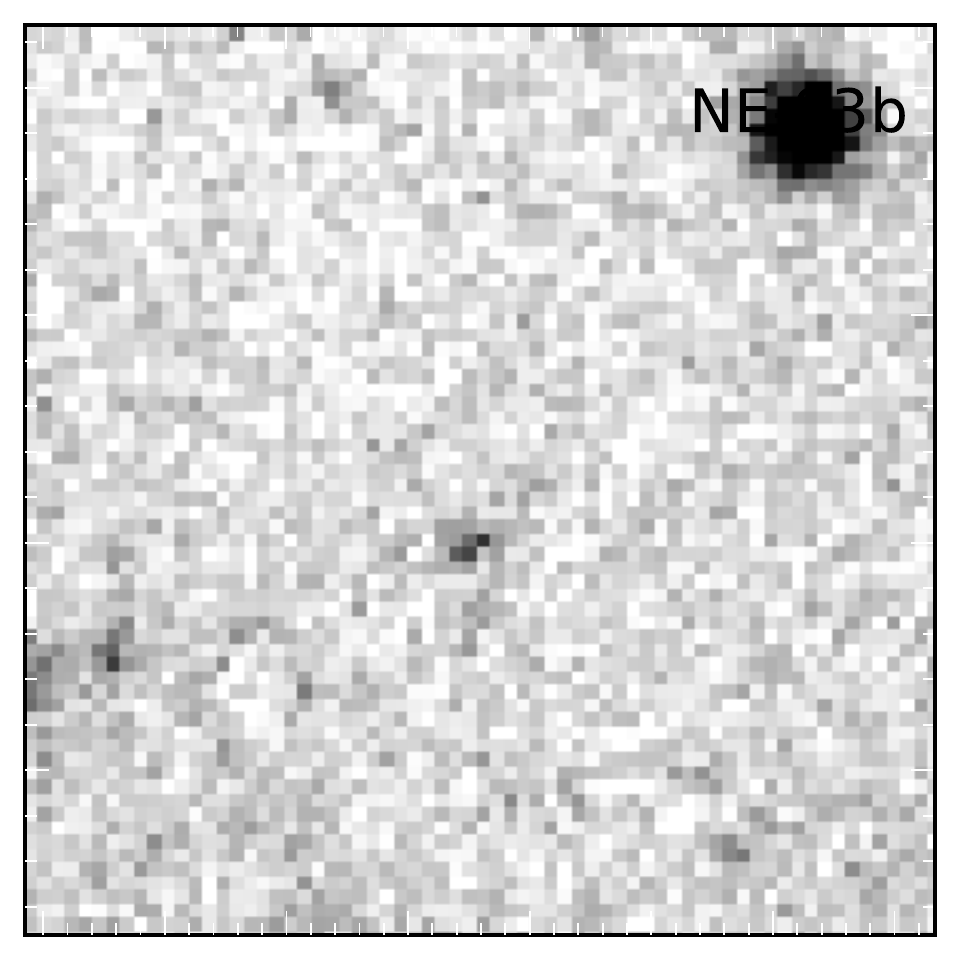}
 \includegraphics[height=0.20\textwidth]{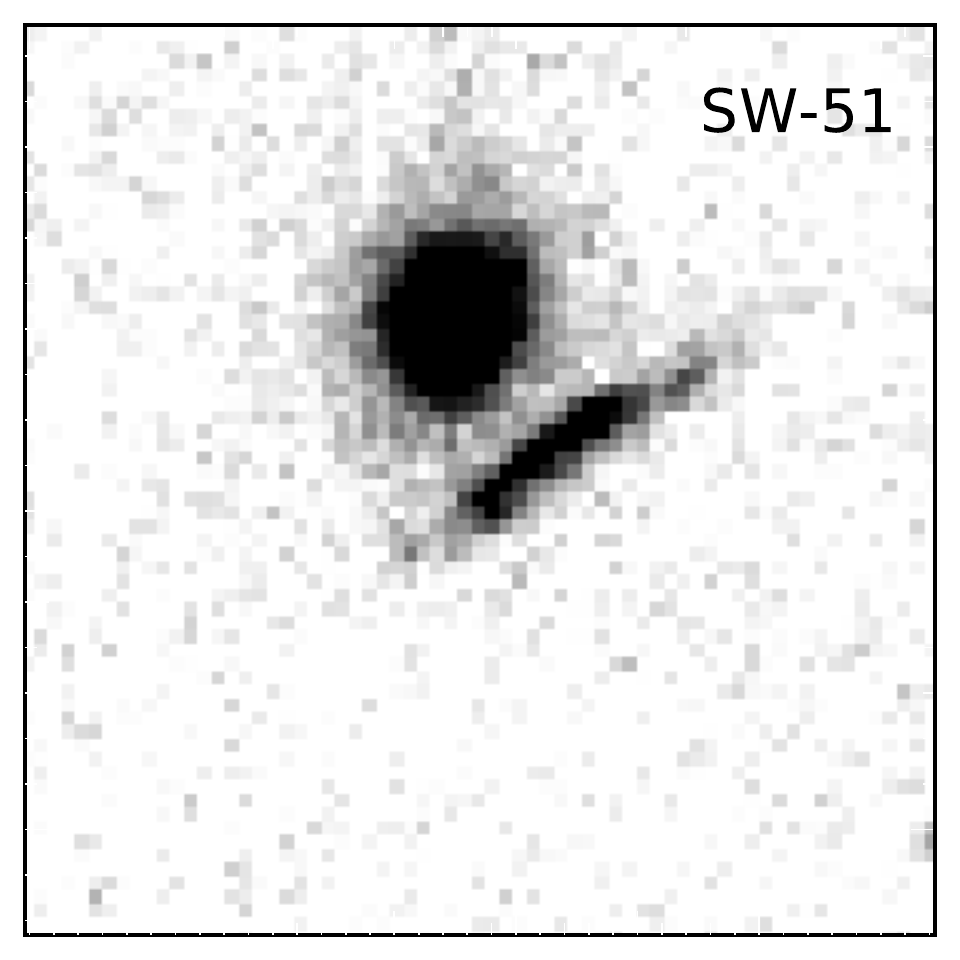}
 \includegraphics[height=0.20\textwidth]{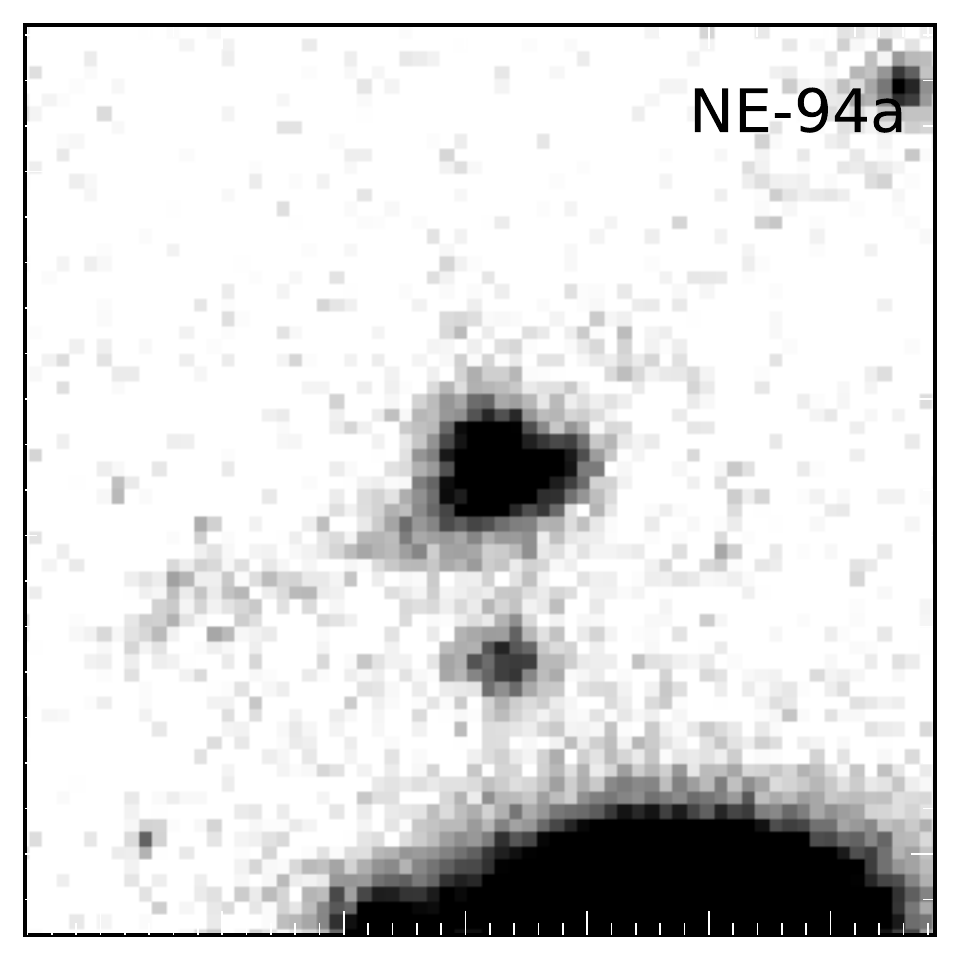}
 \includegraphics[height=0.20\textwidth]{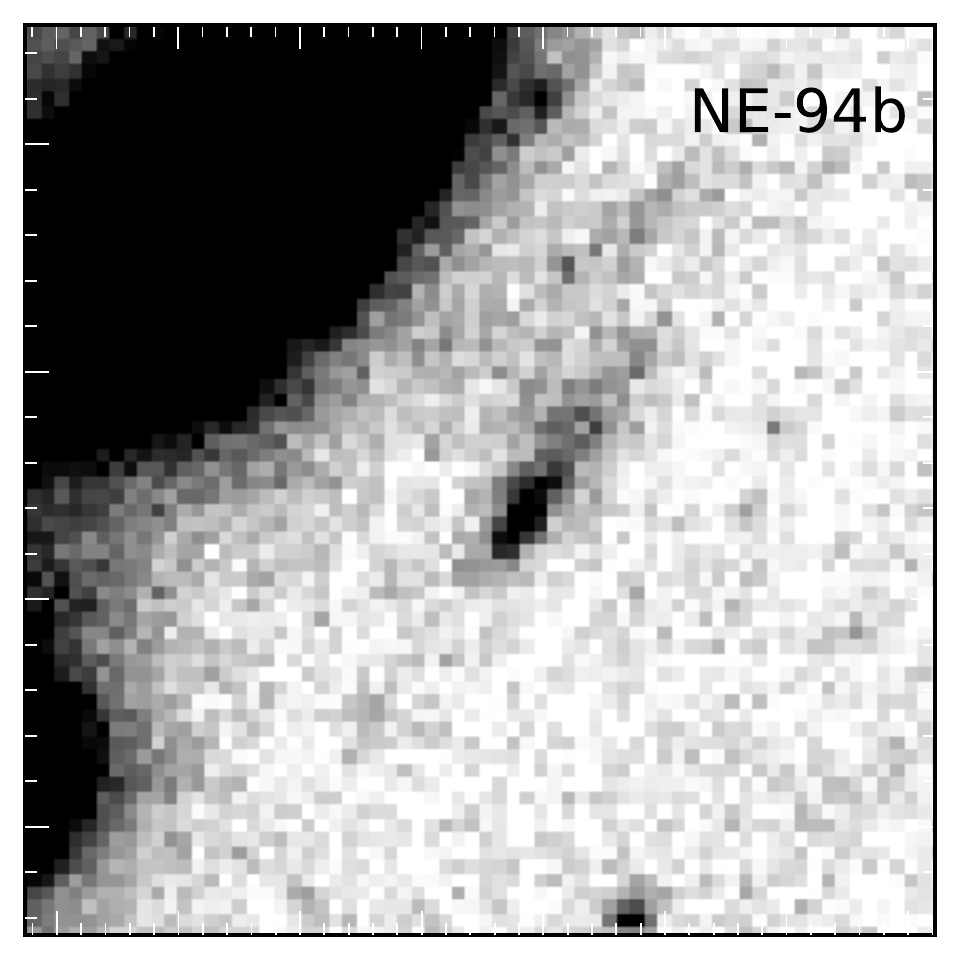}
  \includegraphics[height=0.20\textwidth]{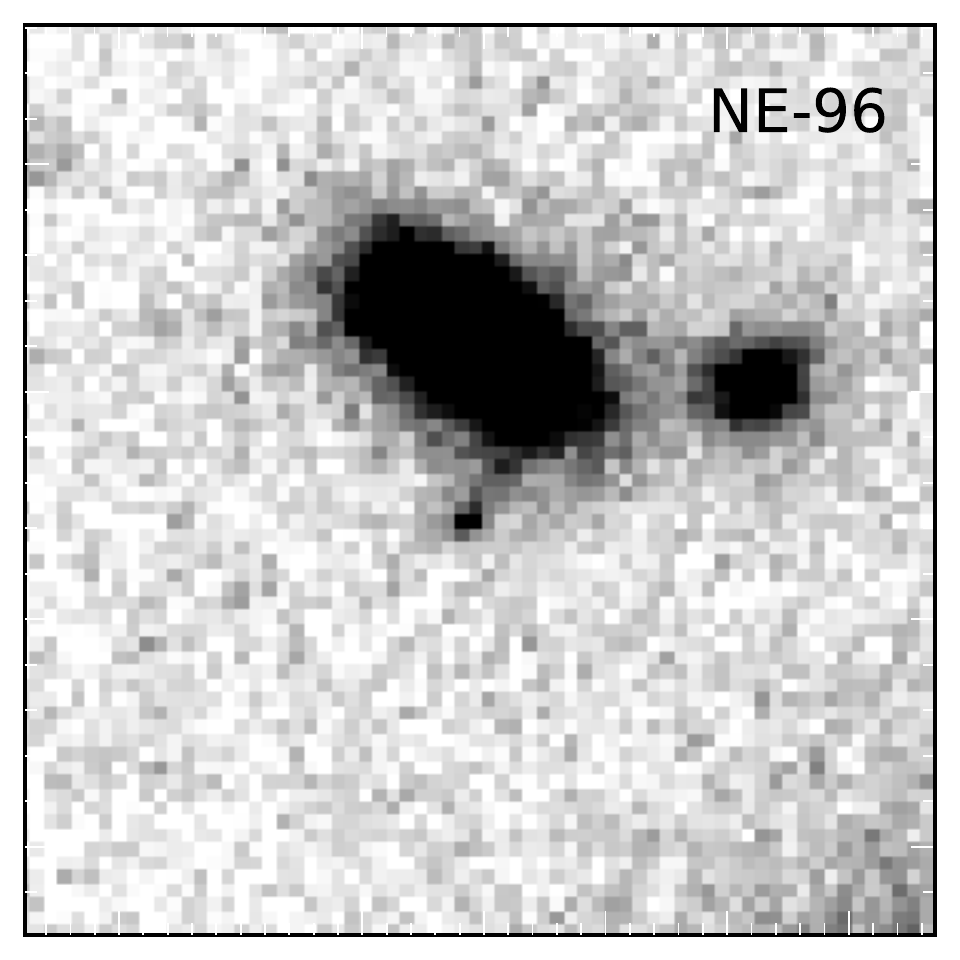}
 \includegraphics[height=0.20\textwidth]{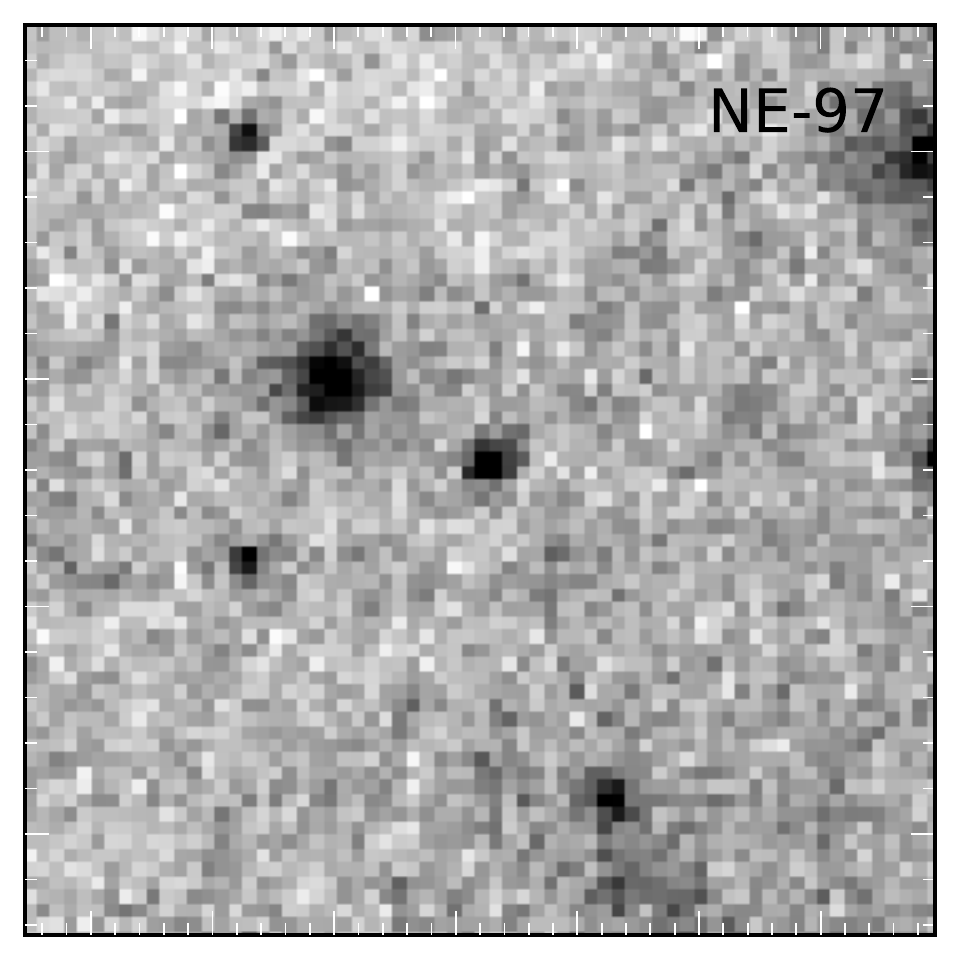}
 \includegraphics[height=0.20\textwidth]{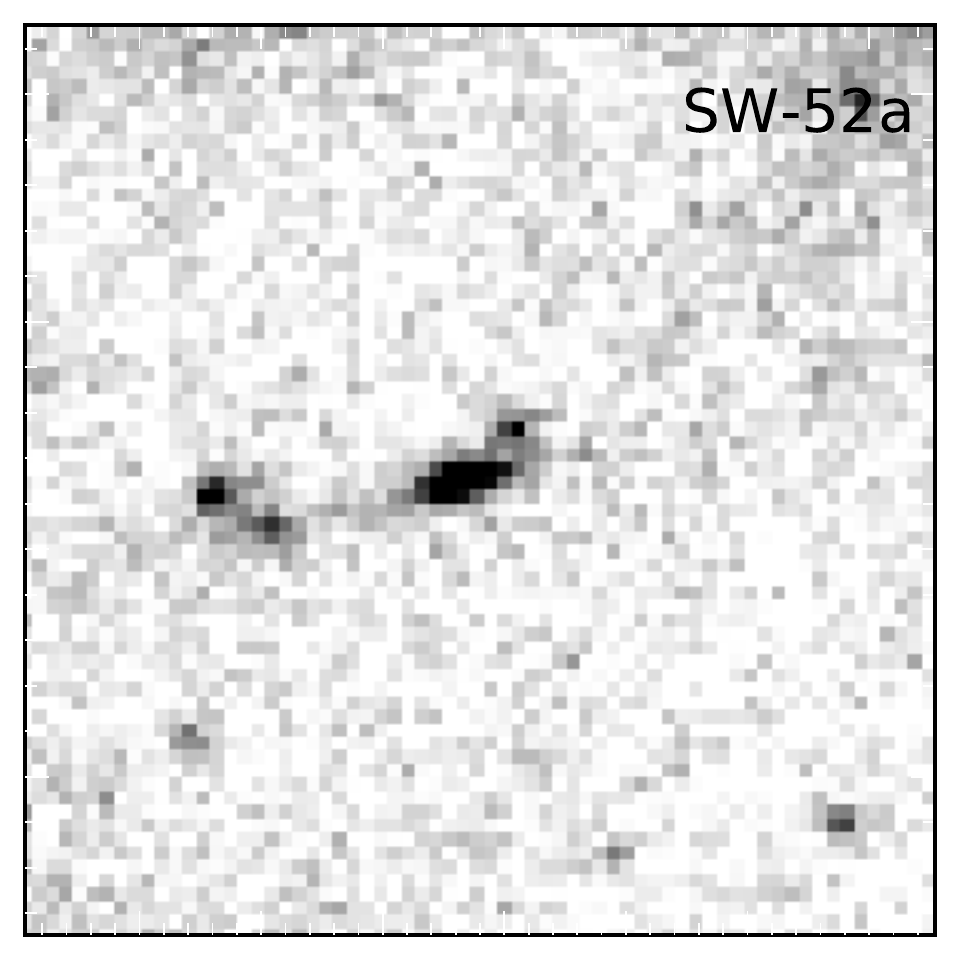} 
 \includegraphics[height=0.20\textwidth]{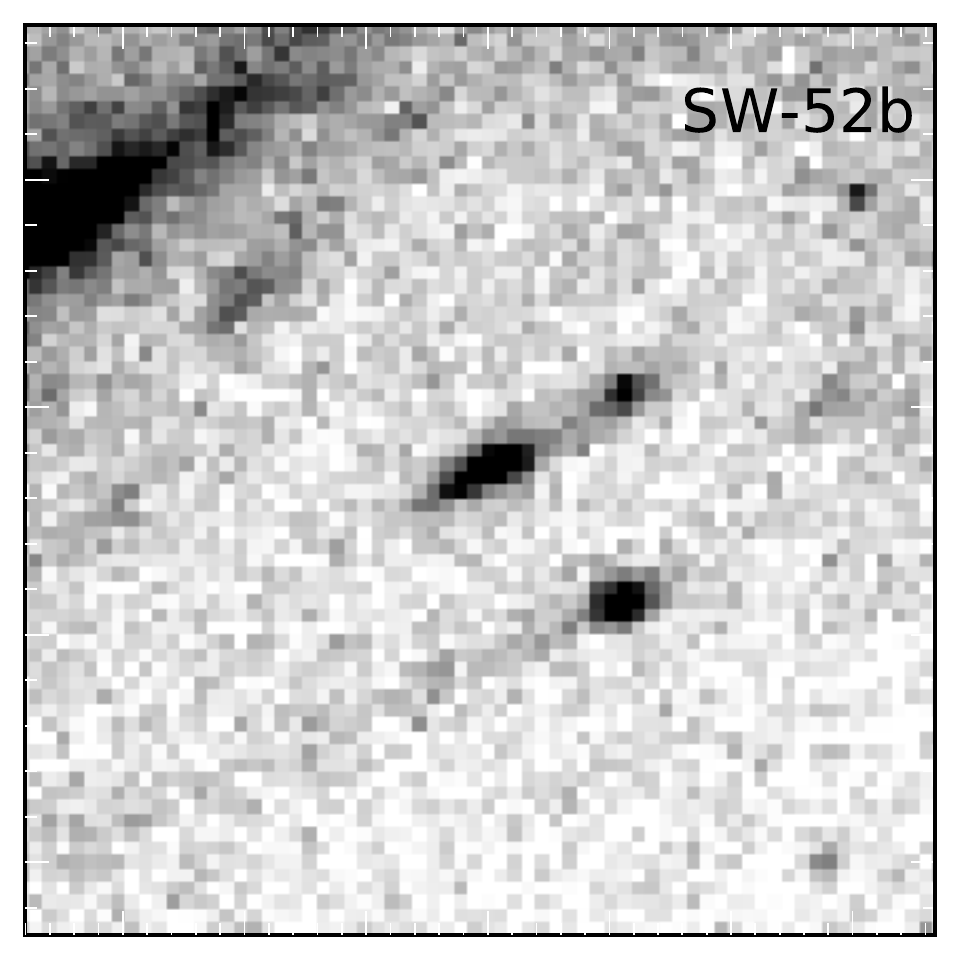}
  \includegraphics[height=0.20\textwidth]{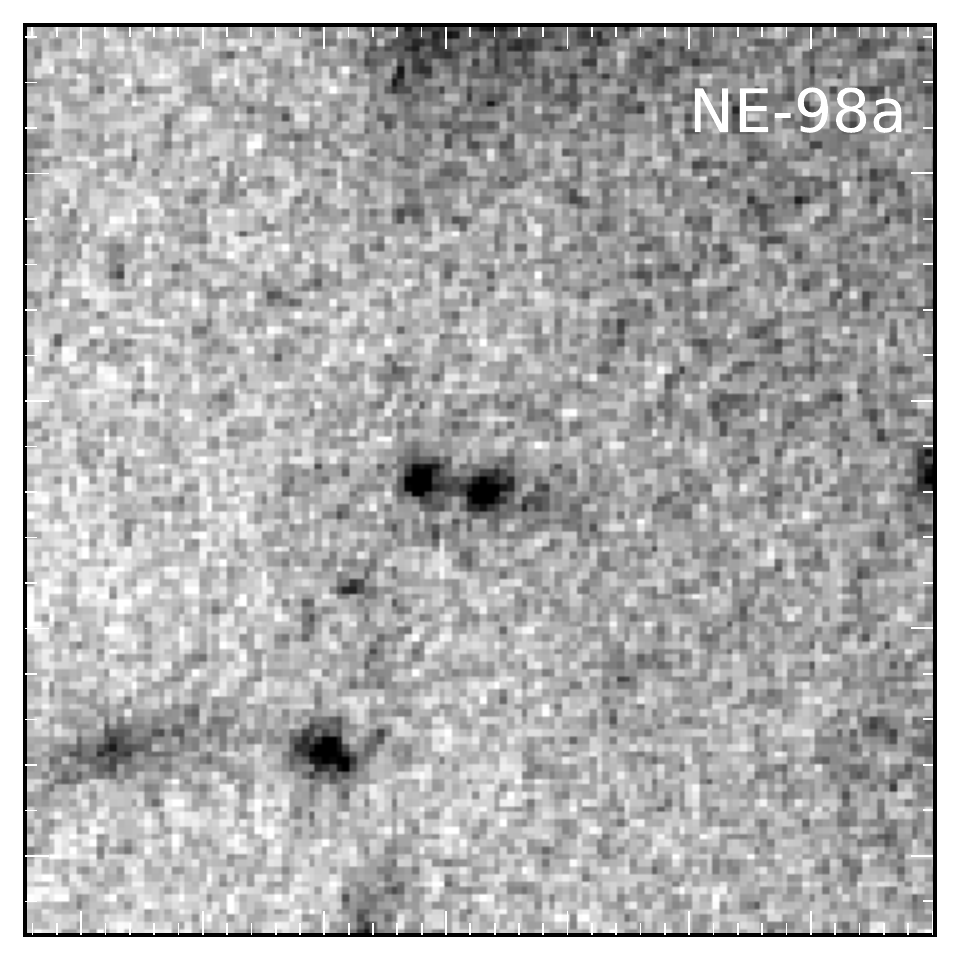}
 \includegraphics[height=0.20\textwidth]{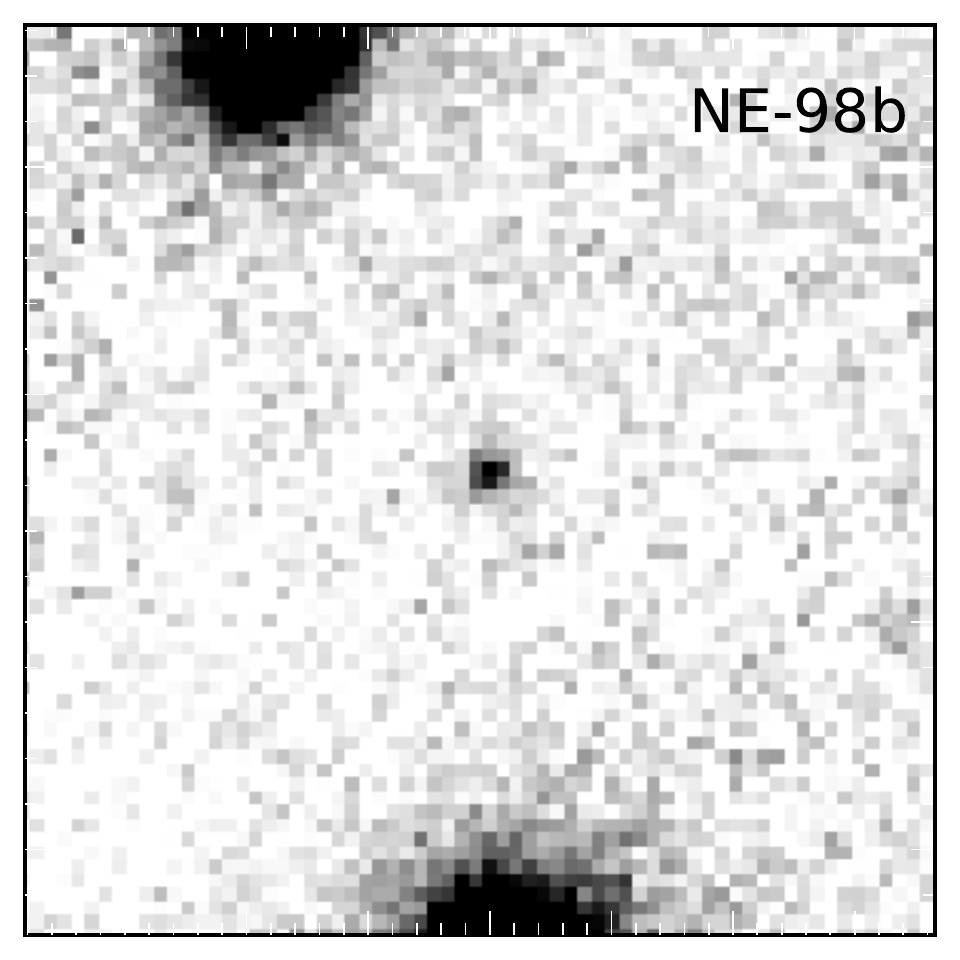}
  \includegraphics[height=0.20\textwidth]{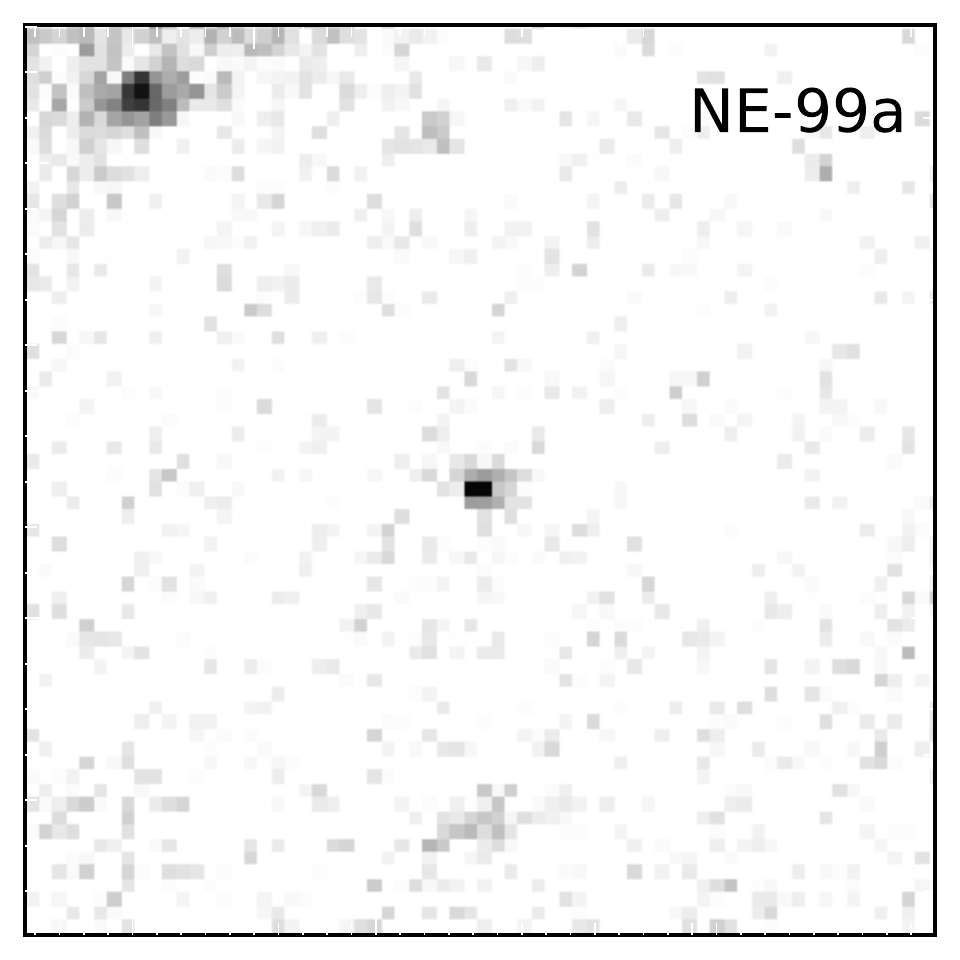}
 \includegraphics[height=0.20\textwidth]{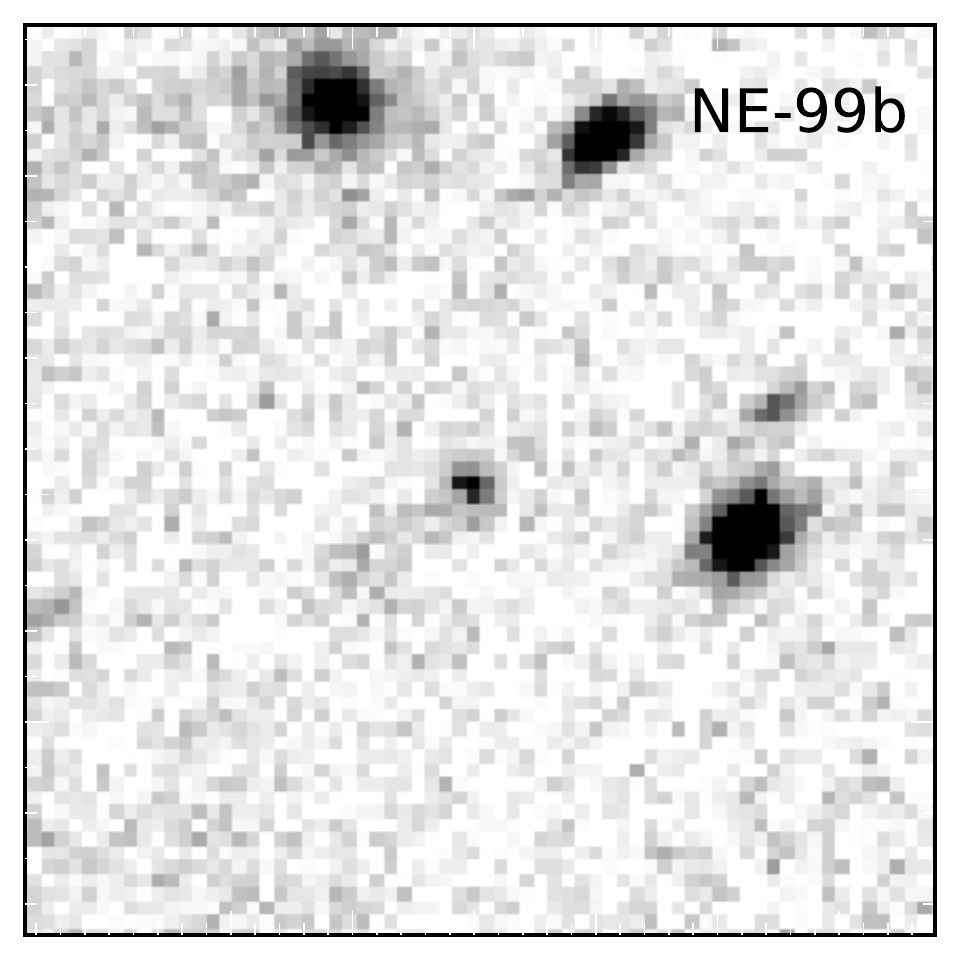}
 \includegraphics[height=0.20\textwidth]{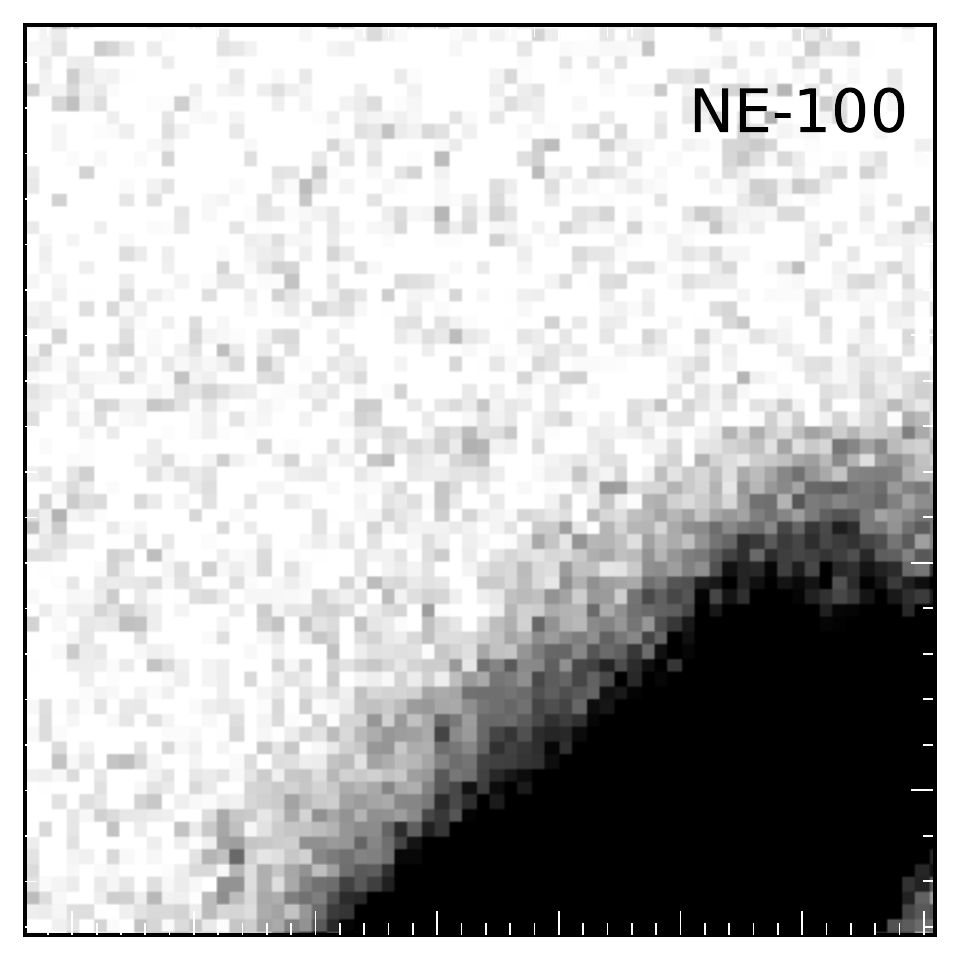}
  \includegraphics[height=0.20\textwidth]{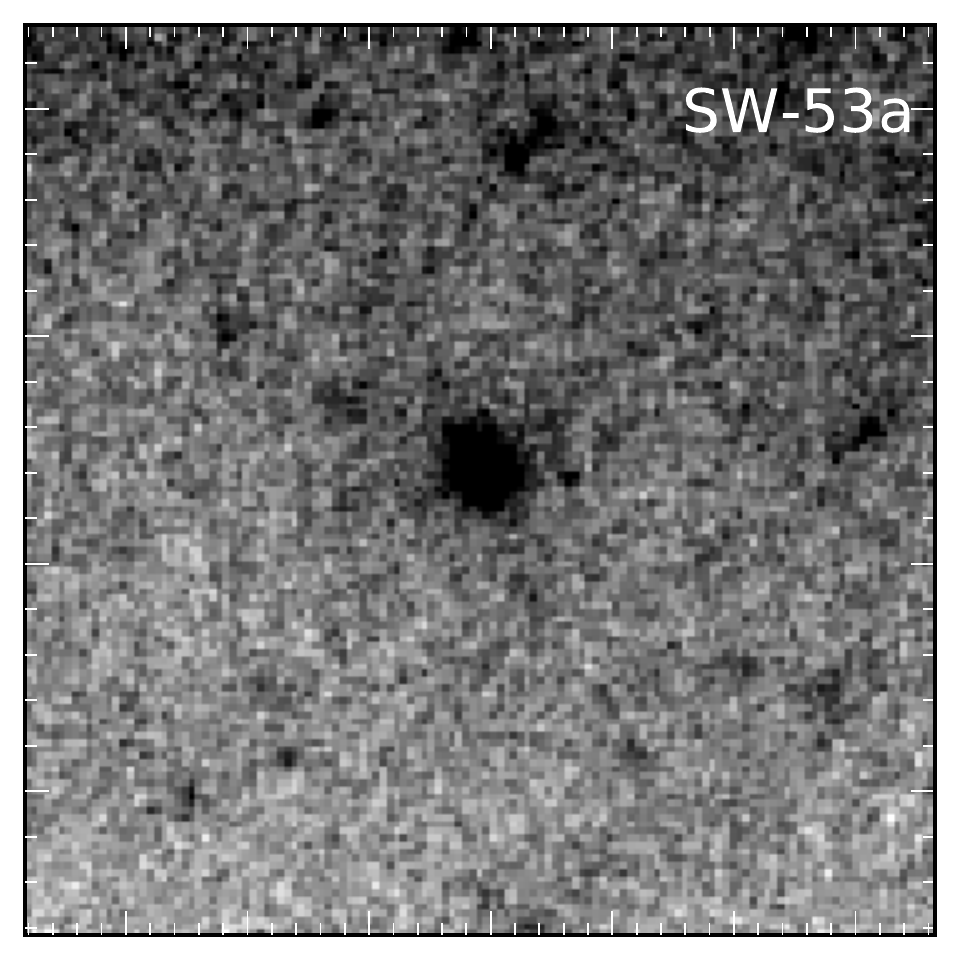} 
  \includegraphics[height=0.20\textwidth]{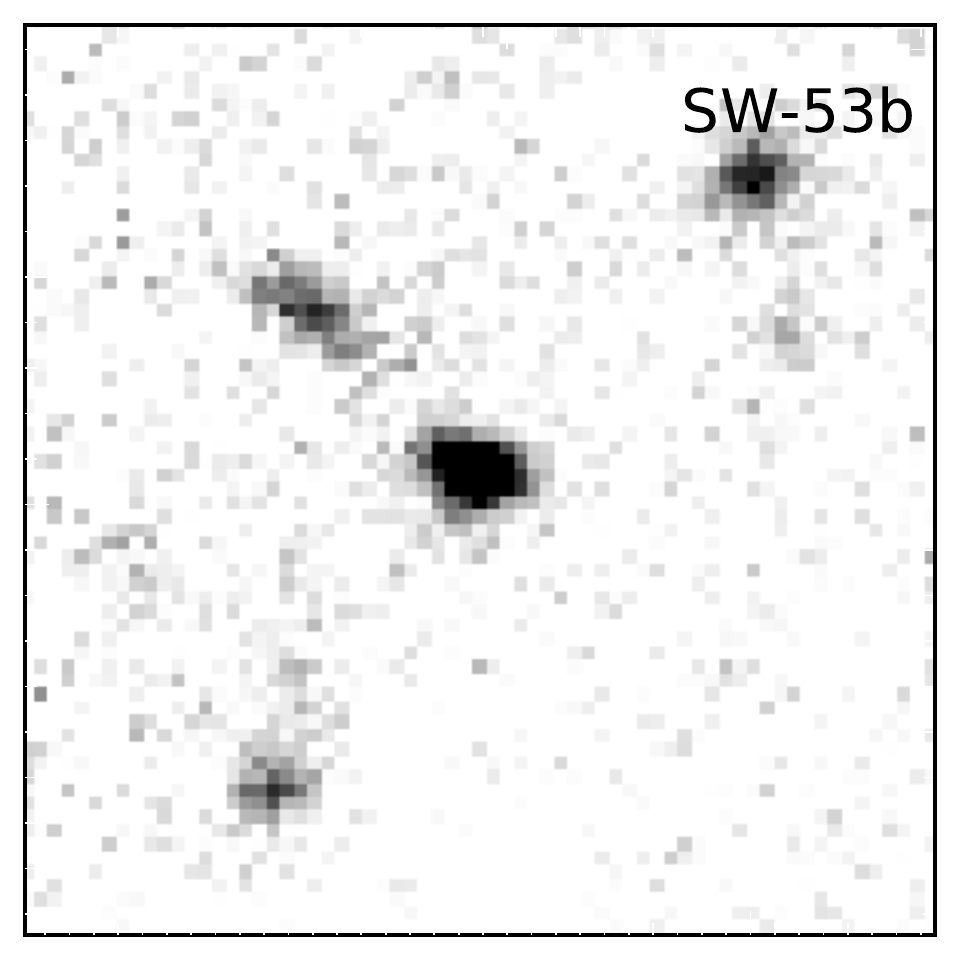}  
  \includegraphics[height=0.20\textwidth]{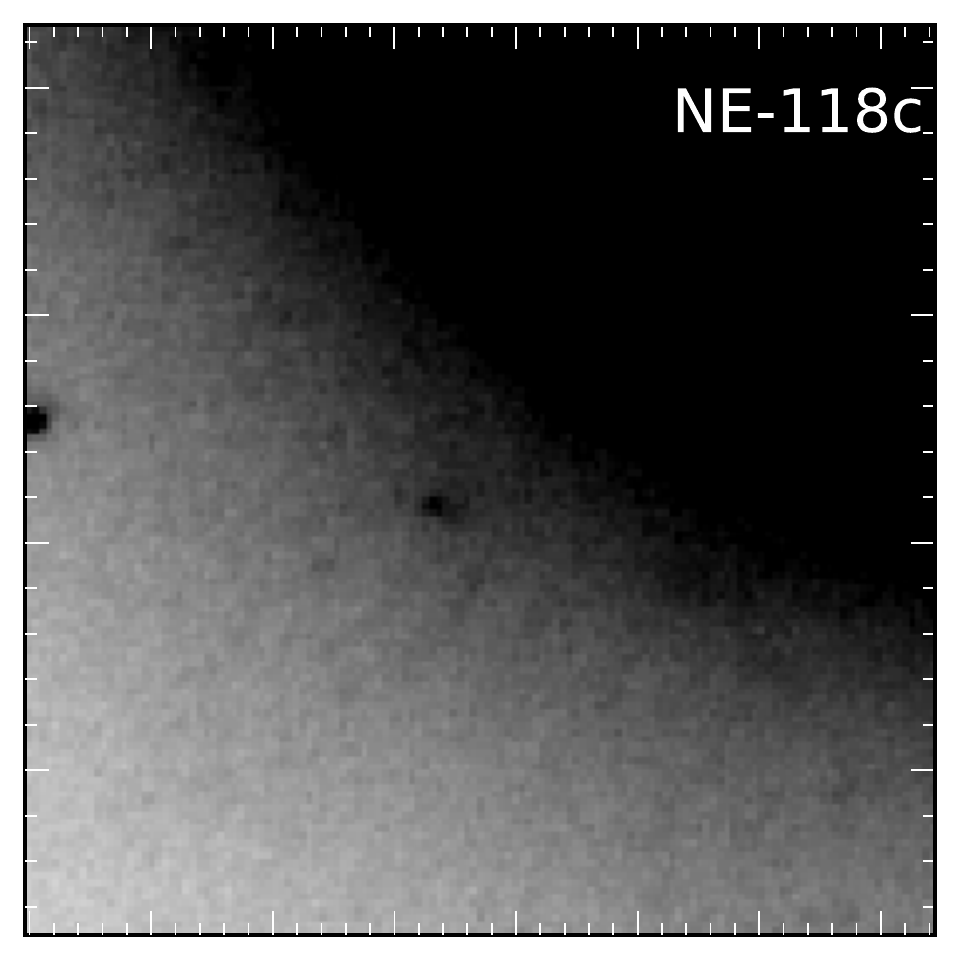} 
 \includegraphics[height=0.20\textwidth]{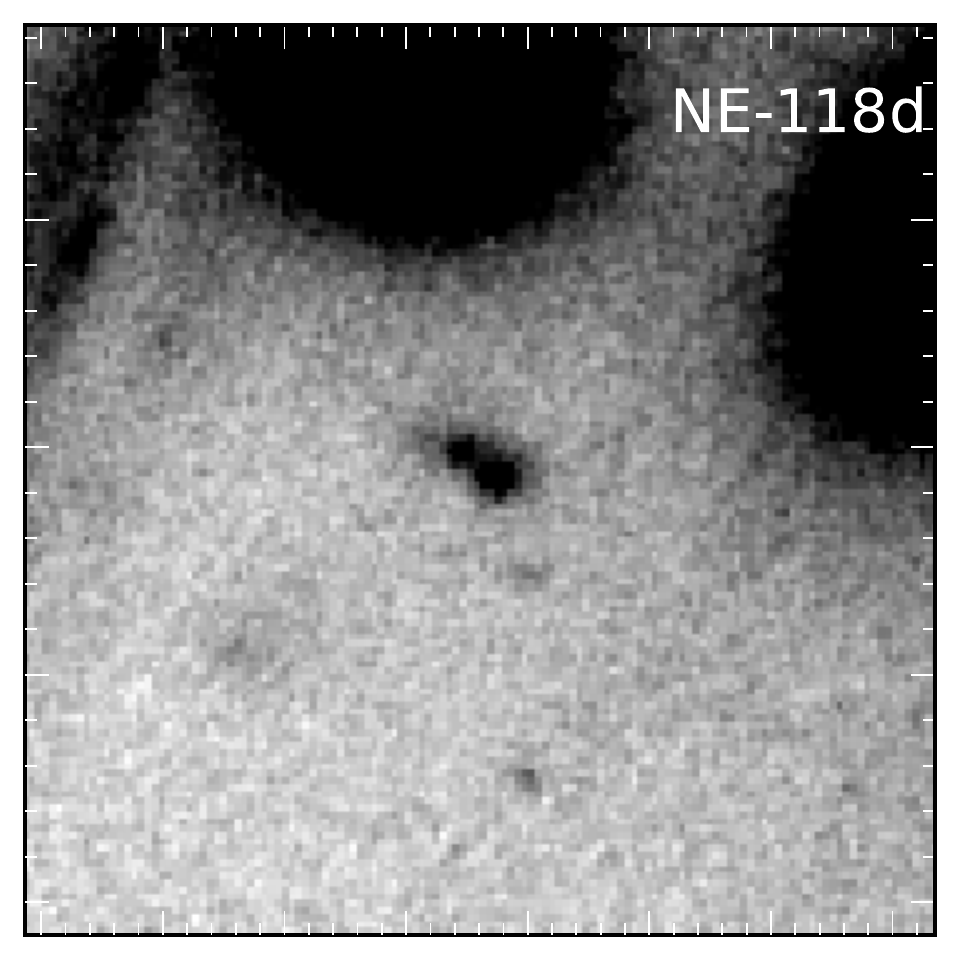} \hspace{2in}
 \caption{{\em HST} F814W stamps of all LAEs in the MUSE footprint. The ID of each
 LAE is plotted in the top right corner of each panel. Each stamp is 4\arcsec\ on each
 side, and a scalebar with a size of 1\arcsec\ is shown in the top left image. \label{fig:stamps_814}}
\end{figure*}

%% file: tab_lya.tex
\begin{table*}
\begin{center}
\begin{tabular}{llllll}
 {\bf ID} & {\bf z} &{\bf f$_{\Lya}$}$^{(a)}$ & {\bf Log(L$_{\Lya}$/(erg/s))}$^{(b)}$ & {\bf Log(N$_H$/cm$^{-2}$)}$^{(c)}$ & {\bf v$_{\rm exp}$}$^{(d)}$ \\ 
& & 10$^{-20}$ erg/s/cm$^{-2}$ & & & \\ \hline 
NE-91 & 2.9760 & 108$\pm$14 &  40.91$\pm$0.06 & --- & --- \\ 
SW-49 & 3.1166 & 1080$\pm$129 & 41.91$\pm$0.06 & 17.07$^{+0.26}_{-0.03}$ & 66.78$^{+7.07}_{-6.11}$ \\
SW-50 & 3.1169 & 984$\pm$18 & 41.92$\pm$0.01 & 17.53$^{+0.24}_{-0.23}$ & 123.63$^{+6.57}_{-5.85}$ \\
SW-68 & 3.1166 & 1942$\pm$16 & 42.22$\pm$0.01 & 20.01$^{+0.20}_{-0.25}$ & 463.18$^{+20.40}_{-48.58}$ \\
NE-93 & 3.1690 & 431$\pm$26 & 41.58$\pm$0.02 & 18.36$^{+1.16}_{-0.65}$ & 44.65$^{+27.91}_{-1.79}$ \\
SW-51 & 3.2271 & 36$\pm$2 & 41.24$\pm$0.01 & 18.53$^{+0.50}_{-0.71}$ & 97.99$^{+14.63}_{-17.76}$ \\
NE-94 & 3.2857 & 524$\pm$27 & 41.70$\pm$0.02 & 19.22$^{+0.37}_{-0.20}$ & 183.33$^{+23.37}_{-8.92}$ \\
NE-96 & 3.4514 & 228$\pm$15 & 41.39$\pm$0.03 & 18.59$^{+0.14}_{-0.25}$ & 56.68$^{+7.35}_{-8.85}$ \\
NE-97 & 3.7131 & 155$\pm$19 & 41.30$\pm$0.06 & 20.39$^{+0.09}_{-0.14}$ & 180.58$^{+13.00}_{-12.00}$  \\
SW-52 & 4.1130 & 106$\pm$4 & 41.24$\pm$0.02 & 17.12$^{+0.32}_{-0.17}$ & 98.10$^{+12.93}_{-10.29}$ \\
NE-98 & 5.0510 & 186$\pm$8 & 41.70$\pm$0.02 & 19.42$^{+0.22}_{-0.21}$ & 149.27$^{+11.83}_{-11.83}$ \\
NE-99 & 5.2373 & 113$\pm$11 & 41.47$\pm$0.05 & 17.99$^{+0.97}_{-0.56}$ & 107.87$^{+223.75}_{-127.68}$ \\
NE-100 & 5.8940 & 60$\pm$32 & 41.36$\pm$0.33 & --- & --- \\
SW-53 & 6.1074 & 2694$\pm$67 & 43.04$\pm$0.01 & 19.81$^{+0.07}_{-0.08}$ & 150.56$^{+3.69}_{-4.1}$ \\
\end{tabular}
\end{center}
\caption{Properties derived from \Lya spectroscopy, after averaging results for multiple images. The columns are $^{(a)}$ lens-corrected \Lya flux, $^{(b)}$ luminosity, $^{(c)}$ the hydrogen 
column density, and $^{(d)}$ expansion velocity.\label{tab:Lya}}
\end{table*}

%% file: tab_sed.tex
\begin{table*}
\begin{center}
\begin{tabular}{lccccccc}
{\bf ID} & {\bf Log(M$_\star/$M$_\odot$)} & {\bf Log(Age$_{\rm SSP}/yr)$} & {\bf E(B-V)} & {\bf Log(SFR$_{\rm SED}$/(M$_\odot$ yr$^{-1}$))} & {\bf Log(sSFR/yr)} & $\beta$ & M$_{\rm UV}$\\ \hline 
NE-91   & 8.02$^{+0.20}_{-0.19}$ & 8.26$^{+0.45}_{-0.67}$ & 0.03$^{+0.03}_{-0.02}$ & -0.38$^{+0.16}_{-1.32}$ & -8.56$^{+0.39}_{-1.10}$ & -2.42$\pm0.09$ & -17.77 \\
SW-49    & 6.91$^{+0.18}_{-0.08}$ & 6.70$^{+0.89}_{-0.06}$ & 0.045$^{+0.03}_{-0.015}$ & -0.16$^{+0.90}_{-0.30}$ & -7.00$^{+0.80}_{-0.30}$ & -2.79$\pm0.08$ & -17.09 \\
SW-50  & 8.23$^{+0.05}_{-0.06}$ & 6.28$^{+1.28}_{-0.63}$ & 0.30$^{+0.00}_{-0.20}$ &  1.88$^{+0.69}_{-1.99}$ & -6.34$^{+0.69}_{-2.82}$ & -1.95$\pm0.05$ & -18.05 \\
SW-68 & 7.78$^{+0.26}_{-0.16}$ & 7.12$^{+0.69}_{-0.48}$ & 0.10$^{+0.10}_{-0.05}$ &  0.36$^{+0.67}_{-0.51}$ & -7.39$^{+0.66}_{-0.90}$ & -2.09$\pm0.08$ & -18.18 \\
NE-93   & 6.24$^{+0.27}_{-0.03}$ & 6.49$^{+0.84}_{-0.03}$ & 0.035$^{+0.036}_{-0.005}$&-0.47$^{+0.91}_{-0.07}$& -6.61$^{+0.80}_{-0.03}$ & -2.85$\pm0.46$ & -15.63 \\
SW-51  & 6.81$^{+0.49}_{-0.34}$ & 7.59$^{+0.82}_{-0.65}$ & 0.05$^{+0.05}_{-0.03}$ & -0.96$^{+0.30}_{-0.27}$ & -7.87$^{+0.63}_{-0.69}$ & -2.31$\pm0.05$ & -15.83\\
NE-94   & 8.50$^{+0.16}_{-0.16}$ & 6.70$^{+0.66}_{-0.66}$ & 0.30$^{+0.05}_{-0.05}$ &  1.61$^{+0.93}_{-0.59}$ & -6.74$^{+0.64}_{-0.75}$ & -1.77$\pm0.08$ & -18.30 \\
NE-96   & 7.33$^{+0.41}_{-0.44}$ & 7.24$^{+0.87}_{-0.96}$ & 0.10$^{+0.15}_{-0.07}$ & -0.36$^{+1.34}_{-0.64}$ & -7.46$^{+1.12}_{-1.26}$ & -2.14$\pm0.22$ & -16.60\\
NE-97   & 6.35$^{+0.18}_{-0.17}$ & 6.40$^{+0.72}_{-0.75}$ & 0.03$^{+0.03}_{-0.02}$ & -0.18$^{+0.81}_{-0.69}$ & -6.47$^{+0.71}_{-0.85}$ & -2.90$\pm0.15$ & -15.98\\
SW-52 & 6.36$^{+0.38}_{-0.25}$ & 7.12$^{+0.53}_{-0.66}$ & 0.03$^{+0.03}_{-0.02}$ & -0.90$^{+0.61}_{-0.33}$ & -7.26$^{+0.74}_{-0.88}$ & -2.50$\pm0.28$ & -15.33 \\
NE-98   & 6.46$^{+0.10}_{-0.09}$ & 6.16$^{+0.36}_{-0.66}$ & 0.01$^{+0.02}_{-0.01}$ &  0.28$^{+0.66}_{-0.54}$ & -6.19$^{+0.68}_{-0.45}$ & -3.17$\pm0.02$ $^{a}$ & -16.19 \\
NE-99   & --- & --- & --- & --- & --- & --- & -14.42\\
NE-100   & --- & --- & --- & --- & --- & --- & $>$-14.70 \\
SW-53   & 7.47$^{+0.00}_{-0.00}$ & 6.04$^{+0.30}_{-0.54}$ & 0.00$^{+0.00}_{-0.00}$ &  1.37$^{+0.60}_{-0.30}$ & -6.10$^{+0.60}_{-0.30}$ & -3.17$\pm0.02$ $^{a}$& -18.97 \\
\end{tabular}
\end{center}
\caption{Stellar properties derived from SED modelling using {\sc LePhare} in combination with BC03 templates. The properties were derived after averaging the results of multiple images of the same source, if applicable. See Appendix \ref{sec:app_sed} for results of all individual LAEs. \\
$^{a}$ This is the maximal UV slope in our used templates, the photometric UV-slope is steeper and the small errors are therefore not representative but
a result of our method of calculating $\beta$.
 \label{tab:sed}}
 \end{table*}

%% file: AS1063_total_cat_v2.tex
  \bottomcaption{Redshifts NE of AS1063. Sources are labelled using their position with regard to the pointing in which they were found, i.e. IDs with SW- (NE-) are found in the south-western (north-eastern) half of the MUSE observations. We crossreferenced the sources with a multiple image detection in Table \ref{tab:multim} with the label MI-. Quality flags (QF) are {\em 4} very certain redshift ($\delta z<0.003$), {\em 3} certain redshift ($\delta z<0.001$), {\em 2} good redshift ($\delta z<0.01$), {\em 1} tentative redshift. The QF$>$90 are based on a single emission line, where {\em 93} illustrates that the profile of the line is asymmetrical, {\em 92} shows it is a doublet, and {\em 91} has only a single emission line detection, without any significant features in the profile. In the last column, we show if we detect emission lines (EL=Y) or not (EL=N). \label{tab:redshifts}} 
 \begin{supertabular}{cccccc}
 {\bf ID}& {\bf RA}& {\bf DEC} & $z$ & {\bf QF} & EL\\
 & (J2000) & J2000) & & & \\
 \hline 
NE-1 &  342.18700 & -44.52617 & 0.0000 & 2 & N \\ 
NE-2 &  342.19297 & -44.52816 & 0.0000 & 2 & N\\ 
SW-58 & 342.17277 & -44.54545 & 0.0000 & 3 & N \\ 
SW-35 & 342.18081 & -44.54647 & 0.1530 & 4 & Y \\ 
SW-36 & 342.17165 & -44.52896 & 0.1602 & 4 & Y \\ 
NE-3 &  342.19667 & -44.52260 & 0.2409 & 4 & Y\\ 
NE-4 &  342.18910 & -44.52116 & 0.2637 & 3 & Y\\ 
NE-5 &  342.18922 & -44.52118 & 0.2637 & 4 & Y\\ 
NE-6 &  342.18969 & -44.52006 & 0.2641 & 4 & Y\\ 
SW-1 &  342.16708 & -44.53469 & 0.3263 & 4 & N \\
NE-7 &  342.17937 & -44.52792 & 0.3275 & 4 & N\\ 
SW-73 & 342.16877 & -44.53399 & 0.3278 & 1 & N \\ 
NE-8 &  342.18726 & -44.52038 & 0.3279 & 1 & N \\ 
NE-95 & 342.19711 & -44.52467 & 0.3300 & 2 & N\\ 
NE-9 &  342.19795 & -44.52782 & 0.3320 & 2 & N \\ 
NE-10 & 342.18430 & -44.52807 & 0.3330 & 4 & N \\ 
SW-2 &  342.17691 & -44.53408 & 0.3343 & 4 & N \\ 
NE-11 & 342.18657 & -44.53028 & 0.3347 & 3 & Y \\ 
SW-3 &  342.18579 & -44.53454 & 0.3348 & 4 & N \\ 
SW-4 &  342.17449 & -44.54621 & 0.3350 & 3 & N \\ 
SW-64 & 342.18556 & -44.53859 & 0.3350 & 1 & N \\ 
SW-5 &  342.17085 & -44.53587 & 0.3360 & 4 & Y \\ 
SW-6 &  342.17546 & -44.53542 & 0.3365 & 4 & Y \\ 
NE-12 & 342.19189 & -44.52967 & 0.3366 & 3 & N \\ 
NE-13 & 342.19207 & -44.52283 & 0.3368 & 1 & N \\ 
SW-9 &  342.17704 & -44.53694 & 0.3368 & 4 & Y \\ 
SW-7 &  342.17158 & -44.53948 & 0.3370 & 3 & N \\ 
NE-14 & 342.19366 & -44.51850 & 0.3378 & 4 & N \\ 
NE-15 & 342.18901 & -44.52470 & 0.3382 & 4 & N \\ 
SW-8 &  342.18694 & -44.53536 & 0.3383 & 4 & N \\ 
SW-60 & 342.17993 & -44.53560 & 0.3384 & 2 & N \\ 
NE-16 & 342.18419 & -44.52030 & 0.3386 & 3 & N \\ 
NE-17 & 342.18812 & -44.53281 & 0.3387 & 3 & N \\ 
NE-18 & 342.18663 & -44.52248 & 0.3388 & 4 & N \\ 
SW-71 & 342.18116 & -44.53788 & 0.3388 & 2 & N \\ 
NE-19 & 342.17960 & -44.52307 & 0.3390 & 4 & Y \\ 
NE-20 & 342.19327 & -44.52413 & 0.3390 & 2 & N \\ 
SW-61 & 342.17462 & -44.53699 & 0.3390 & 2 & N \\ 
NE-21 & 342.18917 & -44.52368 & 0.3397 & 2 & N \\ 
SW-72 & 342.17526 & -44.53141 & 0.3398 & 1 & N \\ 
NE-22 & 342.19589 & -44.52368 & 0.3400 & 1 & N  \\ 
NE-23 & 342.19547 & -44.53244 & 0.3401 & 3& N  \\ 
NE-24 & 342.18295 & -44.52495 & 0.3406 & 3 & N \\ 
NE-25 & 342.19517 & -44.52925 & 0.3410 & 1 & N \\ 
SW-10 & 342.17490 & -44.53412 & 0.3410 & 4 & N \\ 
SW-11 & 342.16657 & -44.53483 & 0.3420 & 4 & N \\ 
NE-26 & 342.19328 & -44.51782 & 0.3422 & 4 & N \\ 
NE-27 & 342.19269 & -44.51492 & 0.3424 & 3 & N \\ 
NE-28 & 342.19421 & -44.52209 & 0.3424 & 4 & Y \\ 
NE-29 & 342.18844 & -44.51775 & 0.3427 & 4 & Y\\ 
NE-30 & 342.18902 & -44.51923 & 0.3427 & 4 & Y\\ 
NE-31 & 342.19019 & -44.51652 & 0.3428 & 4 & Y \\ 
NE-32 & 342.19293 & -44.52207 & 0.3431 & 4 & Y\\ 
NE-33 & 342.20085 & -44.52723 & 0.3431 & 2 & N \\ 
SW-12 & 342.18205 & -44.54034 & 0.3435 & 4 & N \\ 
NE-34 & 342.20017 & -44.52722 & 0.3436 & 3 & N \\ 
NE-35 & 342.18539 & -44.51864 & 0.3440 & 4 & N \\ 
NE-36 & 342.18571 & -44.52602 & 0.3440 & 3 & N \\ 
SW-13 & 342.18940 & -44.53689 & 0.3440 & 4 & N \\ 
NE-37 & 342.17724 & -44.52502 & 0.3446 & 2 & N \\ 
NE-38 & 342.19270 & -44.51976 & 0.3447 & 3 & N \\ 
NE-39 & 342.19333 & -44.52642 & 0.3448 & 4 & N \\ 
SW-15 & 342.18685 & -44.53389 & 0.3450 & 3 & N \\ 
NE-40 & 342.20125 & -44.52404 & 0.3453 & 2 & N \\ 
NE-41 & 342.19123 & -44.53513 & 0.3454 & 3 & N \\ 
NE-42 & 342.19550 & -44.52599 & 0.3455 & 4 & N \\ 
SW-14 & 342.17447 & -44.52899 & 0.3457 & 4 & N \\ 
NE-43 & 342.20122 & -44.52067 & 0.3459 & 3 & N \\ 
NE-44 & 342.19577 & -44.52415 & 0.3460 & 1 & N \\ 
NE-45 & 342.18453 & -44.52931 & 0.3463 & 3 & N \\ 
NE-46 & 342.17794 & -44.52407 & 0.3464 & 4 & N \\ 
NE-47 & 342.18111 & -44.52924 & 0.3464 & 4 & N \\ 
SW-17 & 342.17692 & -44.53744 & 0.3465 & 3 & N \\ 
NE-48 & 342.19001 & -44.53450 & 0.3470 & 1 & N \\ 
NE-49 & 342.19110 & -44.53295 & 0.3470 & 1 & N \\ 
SW-16 & 342.17648 & -44.53363 & 0.3470 & 4 & N \\ 
NE-50 & 342.18569 & -44.53051 & 0.3471 & 3 & N \\ 
NE-51 & 342.18663 & -44.53109 & 0.3475 & 4 & N \\ 
SW-21 & 342.17178 & -44.54054 & 0.3475 & 4 & N \\ 
SW-20 & 342.18898 & -44.54037 & 0.3476 & 4 & N \\ 
NE-52 & 342.18368 & -44.53056 & 0.3478 & 4 & N \\ 
NE-53 & 342.18916 & -44.52954 & 0.3479 & 4 & N \\ 
NE-54 & 342.18293 & -44.53046 & 0.3480 & 4 & N \\ 
NE-55 & 342.18857 & -44.52671 & 0.3480 & 2 & N \\ 
NE-56 & 342.19656 & -44.51974 & 0.3480 & 1 & N \\ 
NE-57 & 342.18256 & -44.52688 & 0.3484 & 4 & N \\ 
NE-58 & 342.19528 & -44.53483 & 0.3485 & 2 & N \\ 
SW-22 & 342.17903 & -44.53276 & 0.3485 & 3 & N \\ 
SW-23 & 342.16985 & -44.53555 & 0.3485 & 4 & N \\ 
NE-59 & 342.19159 & -44.53335 & 0.3486 & 4 & N \\ 
SW-18 & 342.18303 & -44.53099 & 0.3488 & 3 & N \\ 
SW-25 & 342.18671 & -44.54074 & 0.3488 & 4 & N \\ 
NE-60 & 342.18669 & -44.52061 & 0.3489 & 2 & N \\ 
SW-19 & 342.18550 & -44.53305 & 0.3489 & 4 & N \\ 
NE-61 & 342.18564 & -44.53124 & 0.3490 & 3 & N \\ 
SW-63 & 342.18488 & -44.54065 & 0.3491 & 2 & N \\ 
NE-62 & 342.18815 & -44.52973 & 0.3493 & 4 & N \\ 
SW-24 & 342.16826 & -44.53658 & 0.3493 & 4 & N \\ 
NE-63 & 342.18813 & -44.52597 & 0.3494 & 4 & N \\ 
SW-55 & 342.16909 & -44.54040 & 0.3494 & 2 & N \\ 
SW-56 & 342.17217 & -44.54314 & 0.3494 & 2 & N \\ 
NE-64 & 342.20508 & -44.52591 & 0.3495 & 2 & N \\ 
NE-65 & 342.18907 & -44.52645 & 0.3498 & 2 & N \\ 
SW-26 & 342.18307 & -44.53307 & 0.3498 & 4 & N \\ 
SW-28 & 342.18196 & -44.53857 & 0.3502 & 3 & N \\ 
SW-62 & 342.18197 & -44.53858 & 0.3502 & 3 & N \\ 
NE-66 & 342.20420 & -44.52523 & 0.3503 & 4 & N \\ 
NE-67 & 342.18406 & -44.52694 & 0.3504 & 4 & N \\ 
SW-27 & 342.17793 & -44.53239 & 0.3505 & 4 & N \\ 
NE-68 & 342.17961 & -44.52858 & 0.3510 & 2 & N \\ 
NE-69 & 342.19728 & -44.52325 & 0.3510 & 4 & Y \\ 
NE-70 & 342.19534 & -44.53490 & 0.3514 & 2 & N \\ 
NE-71 & 342.17891 & -44.52472 & 0.3515 & 3 & N \\ 
NE-72 & 342.19462 & -44.53228 & 0.3520 & 2 & N \\ 
SW-30 & 342.17872 & -44.54657 & 0.3525 & 3 & N \\ 
SW-29 & 342.18452 & -44.54318 & 0.3526 & 4 & N \\ 
NE-73 & 342.18353 & -44.53011 & 0.3530 & 4 & N \\ 
NE-74 & 342.19002 & -44.52413 & 0.3530 & 2 & N \\ 
NE-75 & 342.20030 & -44.52519 & 0.3530 & 4 & N \\ 
NE-76 & 342.19137 & -44.53433 & 0.3532 & 4 & N \\ 
SW-57 & 342.18324 & -44.54386 & 0.3532 & 2 & N \\ 
SW-77 & 342.18533 & -44.53412 & 0.3532 & 3 & N \\ 
NE-77 & 342.17911 & -44.52866 & 0.3533 & 1 & N \\ 
SW-31 & 342.18436 & -44.53617 & 0.3535 & 3 & N \\ 
NE-78 & 342.18674 & -44.52786 & 0.3550 & 4 & N \\ 
SW-32 & 342.16265 & -44.53809 & 0.3550 & 3 & N \\ 
SW-33 & 342.16044 & -44.53896 & 0.3553 & 4 & N \\ 
NE-79 & 342.19670 & -44.52909 & 0.3572 & 3 & N \\ 
SW-34 & 342.17366 & -44.53278 & 0.3588 & 4 & N \\ 
NE-80 & 342.19823 & -44.52741 & 0.3606 & 3 & N \\ 
NE-81 & 342.20442 & -44.52459 & 0.3799 & 4 & Y \\ 
NE-82 & 342.19539 & -44.51676 & 0.3940 & 1 & N \\ 
NE-83 & 342.20296 & -44.52716 & 0.4382 & 4 & Y \\ 
NE-84 & 342.20568 & -44.52325 & 0.4386 & 4 & Y \\ 
SW-37 & 342.16658 & -44.54021 & 0.4582 & 4 & Y \\ 
NE-85 & 342.18576 & -44.52405 & 0.4845 & 4 & Y \\ 
SW-39 & 342.18870 & -44.53788 & 0.6108 & 4 & Y \\ 
SW-38 & 342.17403 & -44.53244 & 0.6111 & 4 & Y \\ 
SW-40 & 342.18442 & -44.53957 & 0.6518 & 4 & Y \\ 
SW-41 & 342.17925 & -44.54219 & 0.6980 & 4 & Y \\ 
NE-86 & 342.20357 & -44.52496 & 0.7048 & 3 & Y \\ 
SW-42 & 342.18014 & -44.54407 & 0.7145 & 4 & Y \\ 
NE-87a/ & 342.18429 & -44.52529 & 0.7287 & 4 & Y \\ 
MI-87a\\
NE-87b/ & 342.18894 & -44.52864 & 0.7287 & 4 & Y \\ 
MI-87b\\
NE-87c/ & 342.19010 & -44.53010 & 0.7287 & 4 & Y \\ 
MI-87c\\
SW-43a/  & 342.17209 & -44.53052 & 1.0350 & 4 & Y \\ 
MI-7c\\
SW-43b/  & 342.17597 & -44.53615 & 1.0350 & 4 & Y \\ 
MI-7b\\
SW-43c/  & 342.18069 & -44.53866 & 1.0350 & 4 & Y \\ 
MI-7a\\
NE-88a/ & 342.18642 & -44.52116 & 1.2277 & 4 & Y \\ 
MI-2c\\
NE-88b/ & 342.19588 & -44.52895 & 1.2278 & 4 & Y \\ 
MI-2a\\
NE-88c/ & 342.19450 & -44.52698 & 1.2279 & 4 & Y \\ 
MI-2b\\
NE-88d/ & 342.19520 & -44.52786 & 1.2279 & 4 & Y \\ 
MI-2d\\
NE-102a/ & 342.19369 & -44.53014 & 1.2583 & 91 & Y \\ 
MI-13a\\
NE-102b/ & 342.19331 & -44.52942 & 1.2583 & 91 & Y \\
MI-13b\\
NE-89/ & 342.19271 & -44.53119 & 1.2592 & 4 & Y \\ 
MI-3a\\
NE-89/ & 342.19213 & -44.52983 & 1.2593 & 4 & Y \\ 
MI-3b\\
SW-44 & 342.17253 & -44.54128 & 1.2690 & 3 & Y \\ 
SW-45 & 342.17552 & -44.54558 & 1.2690 & 92 & Y \\ 
NE-90a/ & 342.17921 & -44.52359 & 1.3972 & 92 & Y \\ 
MI-4c\\
NE-90b/ & 342.18783 & -44.52731 & 1.3972 & 92 & Y \\ 
MI-4b\\
NE-90c/ & 342.19317 & -44.53653 & 1.3972 & 92 & Y \\ 
MI-4a\\
SW-46a/  & 342.18843 & -44.53997 & 1.4285 & 4 & Y \\ 
MI-6a\\
SW-46b/  & 342.17583 & -44.53258 & 1.4285 & 4 & Y \\ 
MI-6b\\
SW-46c/  & 342.17409 & -44.52844 & 1.4285 & 4 & Y \\ 
MI-6c\\
SW-47 & 342.17699 & -44.54633 & 1.4770 & 92 & Y \\ 
SW-69a/  & 342.18186 & -44.54050 & 1.8370 & 92 & Y \\ 
MI-8a\\
SW-69b/  & 342.17424 & -44.53711 & 1.8370 & 92 & Y \\ 
MI-8b\\
SW-48 & 342.16135 & -44.53835 & 2.5770 & 91 & Y \\ 
NE-91/ & 342.19238 & -44.52505 & 2.9760 & 91 & Y\\ 
MI-91a \\
NE-92 & 342.19171 & -44.53052 & 3.0600 & 1 & N\\ 
SW-50 & 342.16225 & -44.53829 & 3.1160 & 4 & Y \\ 
SW-68a/  & 342.18745 & -44.53869 & 3.1166 & 4 & Y \\ 
MI-20a\\
SW-68b/  & 342.17886 & -44.53587 & 3.1166 & 4 & Y \\ 
MI-20b\\
SW-49a\  & 342.17505 & -44.54102 & 3.1169 & 4 & Y \\ 
MI-11a\\
SW-49b/  & 342.17318 & -44.54000 & 3.1169 & 4 & Y \\ 
MI-11b\\
NE-93a/ & 342.18283 & -44.52028 & 3.169 & 92 & Y \\ 
MI-93a\\
NE-93b/ & 342.19196 & -44.52409 & 3.169 & 92 & Y \\ 
MI-93b\\
SW-51 & 342.17402 & -44.54124 & 3.2275 & 4 & Y \\ 
NE-94a/ & 342.18935 & -44.51871 & 3.2857 & 92 & Y \\ 
MI-94a\\
NE-94b/ & 342.19615 & -44.52292 & 3.2857 & 92 & Y \\ 
MI-94b\\
NE-96 & 342.19709 & -44.52483 & 3.4514 & 92 & Y \\ 
SW-70a/  & 342.18586 & -44.53883 & 3.6063 & 2 & N \\ 
MI-21a\\
SW-70b / & 342.17892 & -44.53668 & 3.6067 & 2 & N \\ 
MI-21b\\
NE-97 & 342.19100 & -44.52679 & 3.7131 & 93 & Y \\ 
SW-52a/  & 342.17910 & -44.53863 & 4.1130 & 92 & Y \\ 
MI-18a\\
SW-52b/  & 342.18164 & -44.53936 & 4.1130 & 92 & Y \\ 
MI-18b\\
NE-98a/ & 342.19015 & -44.53094 & 5.0510 & 92 & Y \\ 
MI-98a\\
NE-98b/ & 342.19085 & -44.53566 & 5.0510 & 92 & Y \\ 
MI-98b\\
NE-99a/ & 342.18378 & -44.52122 & 5.2373 & 92 & Y \\ 
MI-99a\\
NE-99b/ & 342.18874 & -44.52276 & 5.2373 & 92 & Y \\ 
MI-99b\\
NE-100/ & 342.19701 & -44.52212 & 5.894 & 91 & Y \\ 
MI-100b\\
SW-53a/& 342.18104 & -44.53460 & 6.1074 & 92 & Y \\ 
MI-14b\\
SW-53b/  & 342.19089 & -44.53746 & 6.1074 & 92 & Y \\ 
MI-14a\\
NE-101c & 342.18402 & -44.53159 & 6.1074 & 4 & Y \\ 
MI-14e\\
NE-101d/ & 342.18910 & -44.53001 & 6.1074 & 4 & Y \\ 
MI-14c\\
SW-54  & 342.18407 & -44.53532 & 6.1074 & 91 & Y \\ 
\end{supertabular}

%% file: SED_figures.tex
\begin{figure*}
\begin{center}

\includegraphics[width=0.425\textwidth]{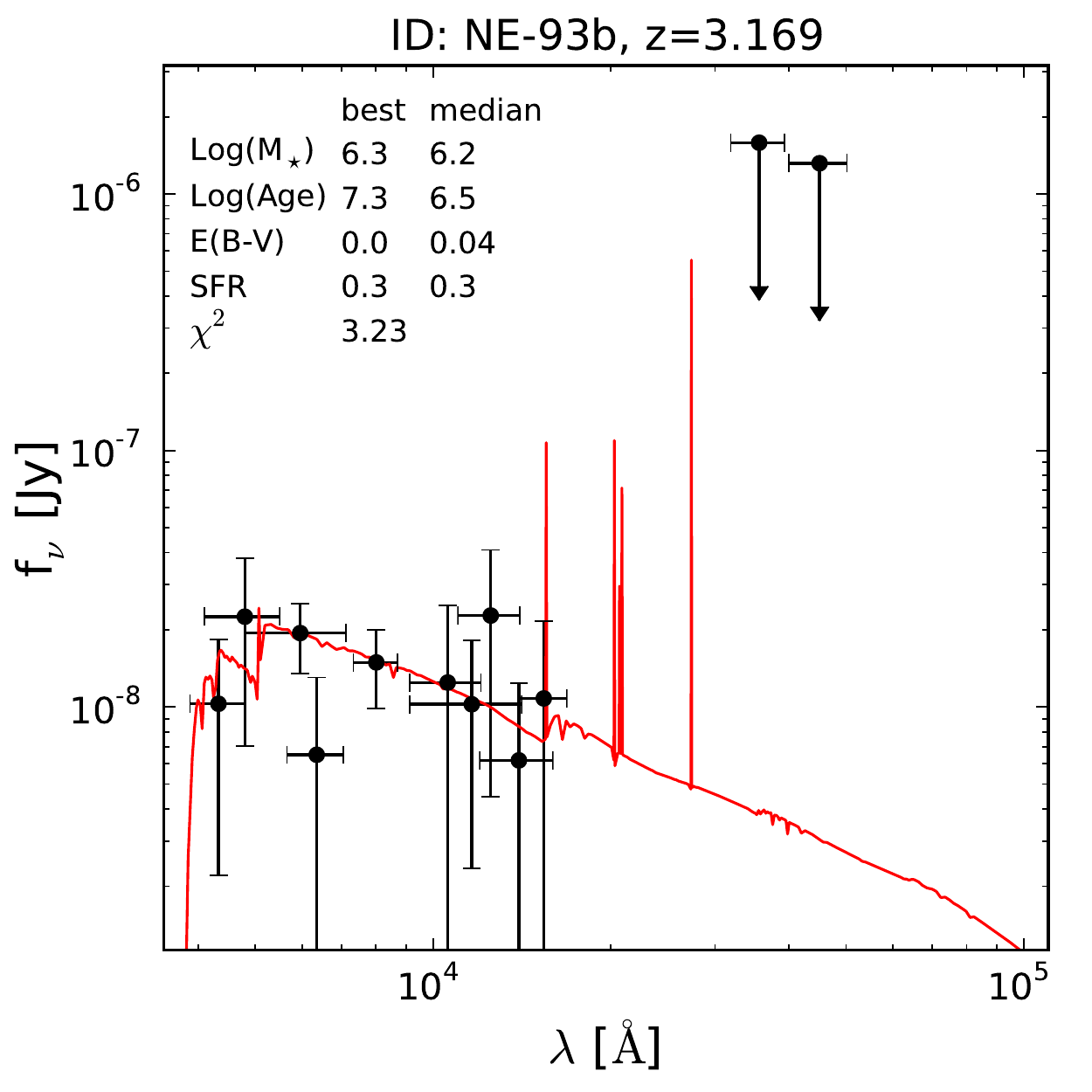}
\includegraphics[width=0.425\textwidth]{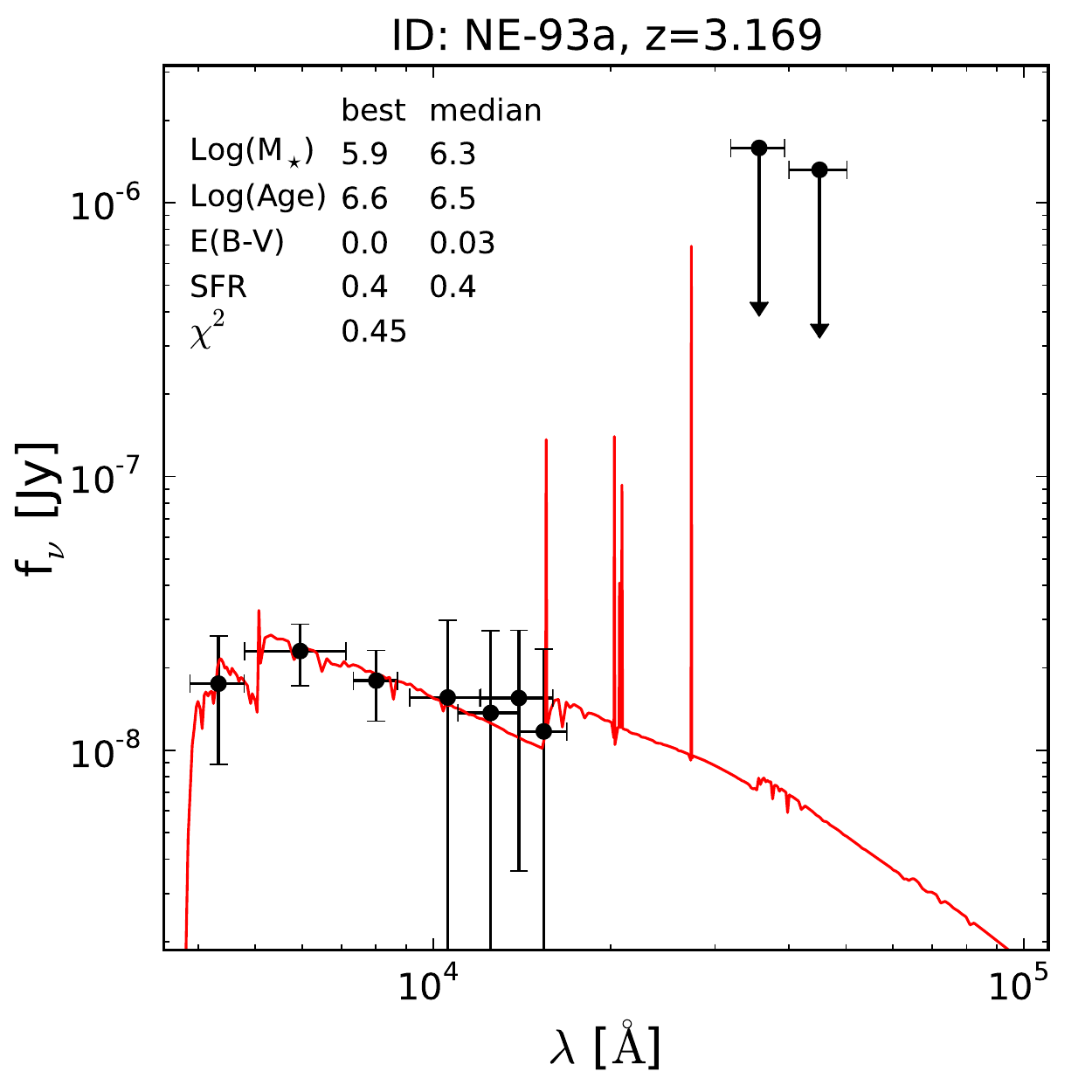}
\includegraphics[width=0.425\textwidth]{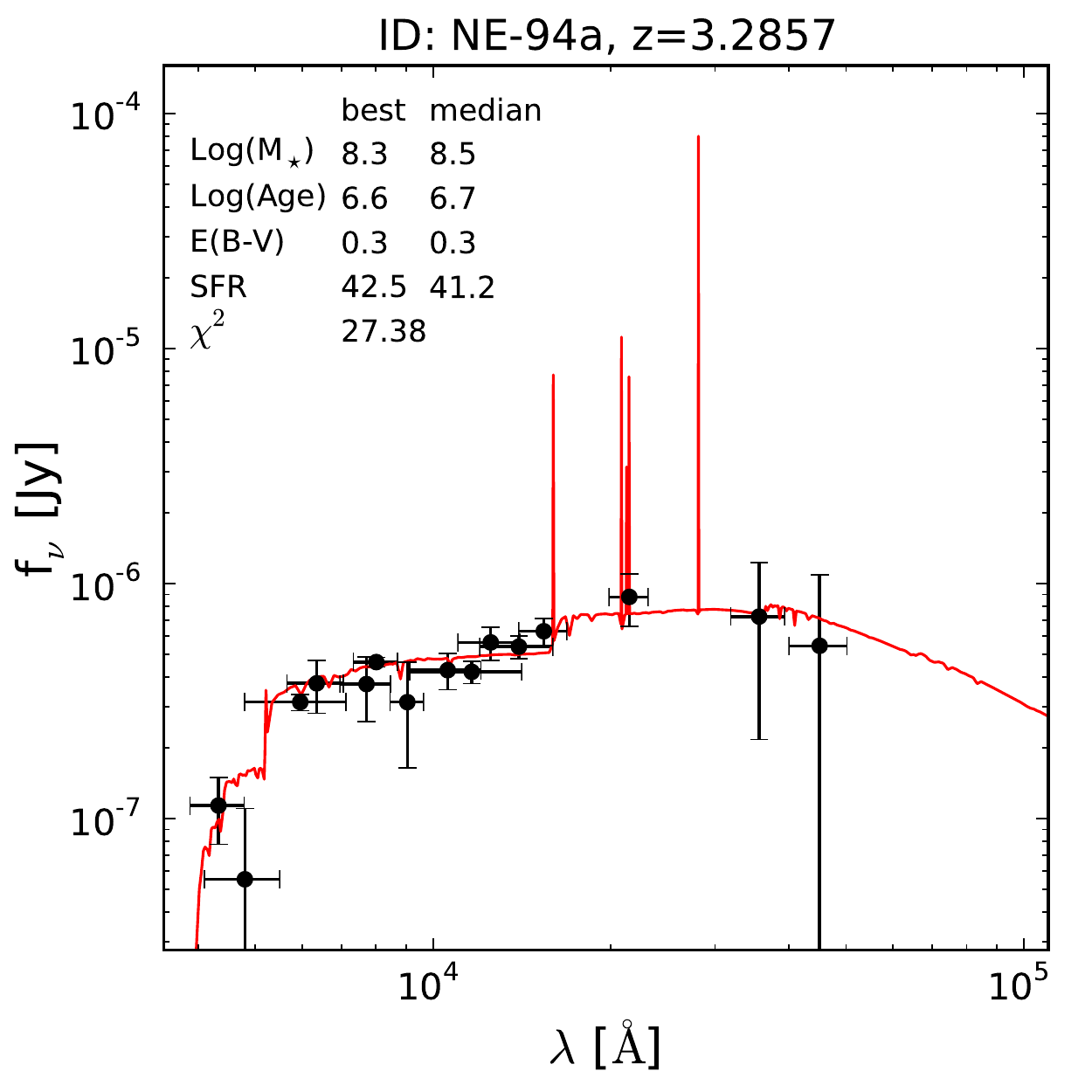}
\includegraphics[width=0.425\textwidth]{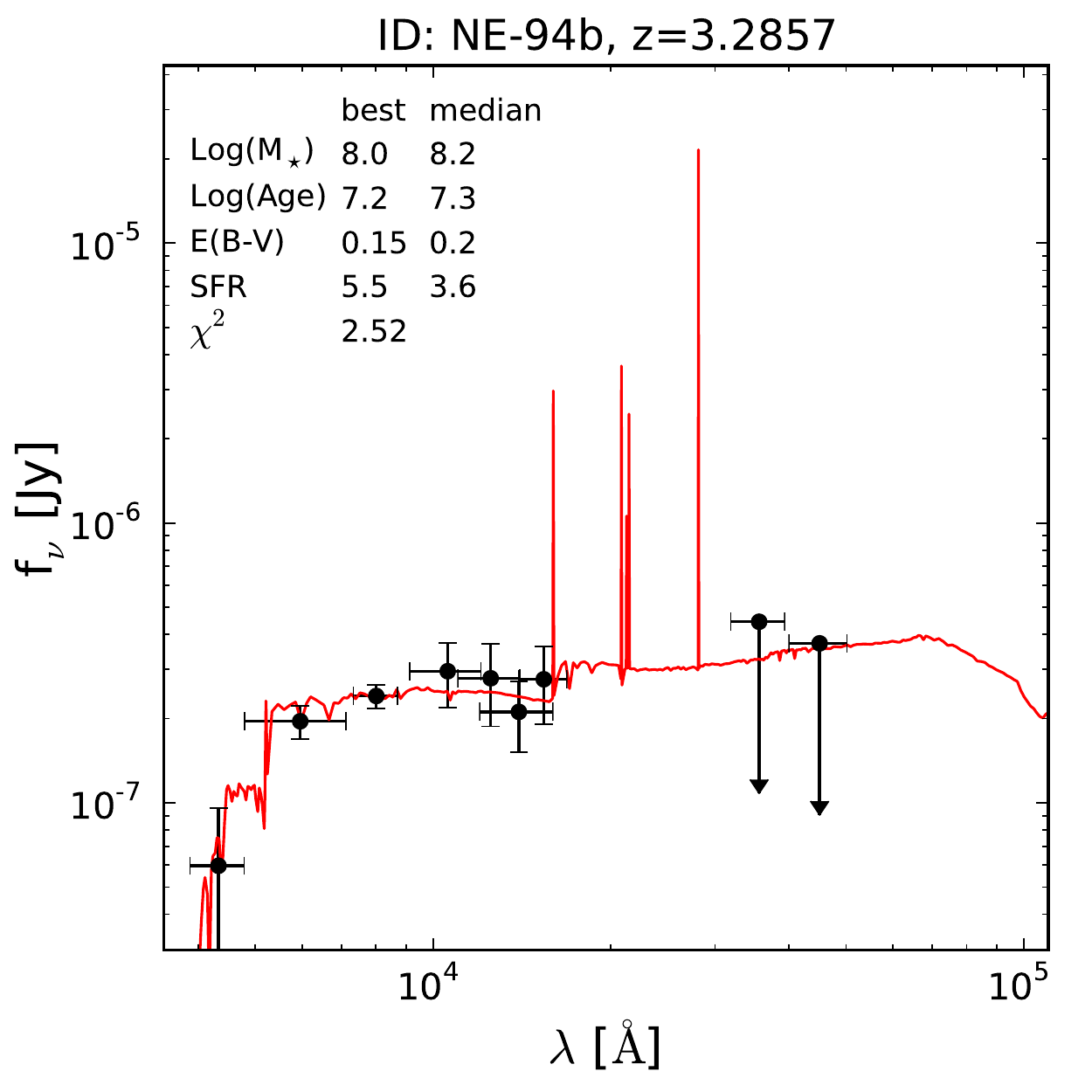}
\includegraphics[width=0.425\textwidth]{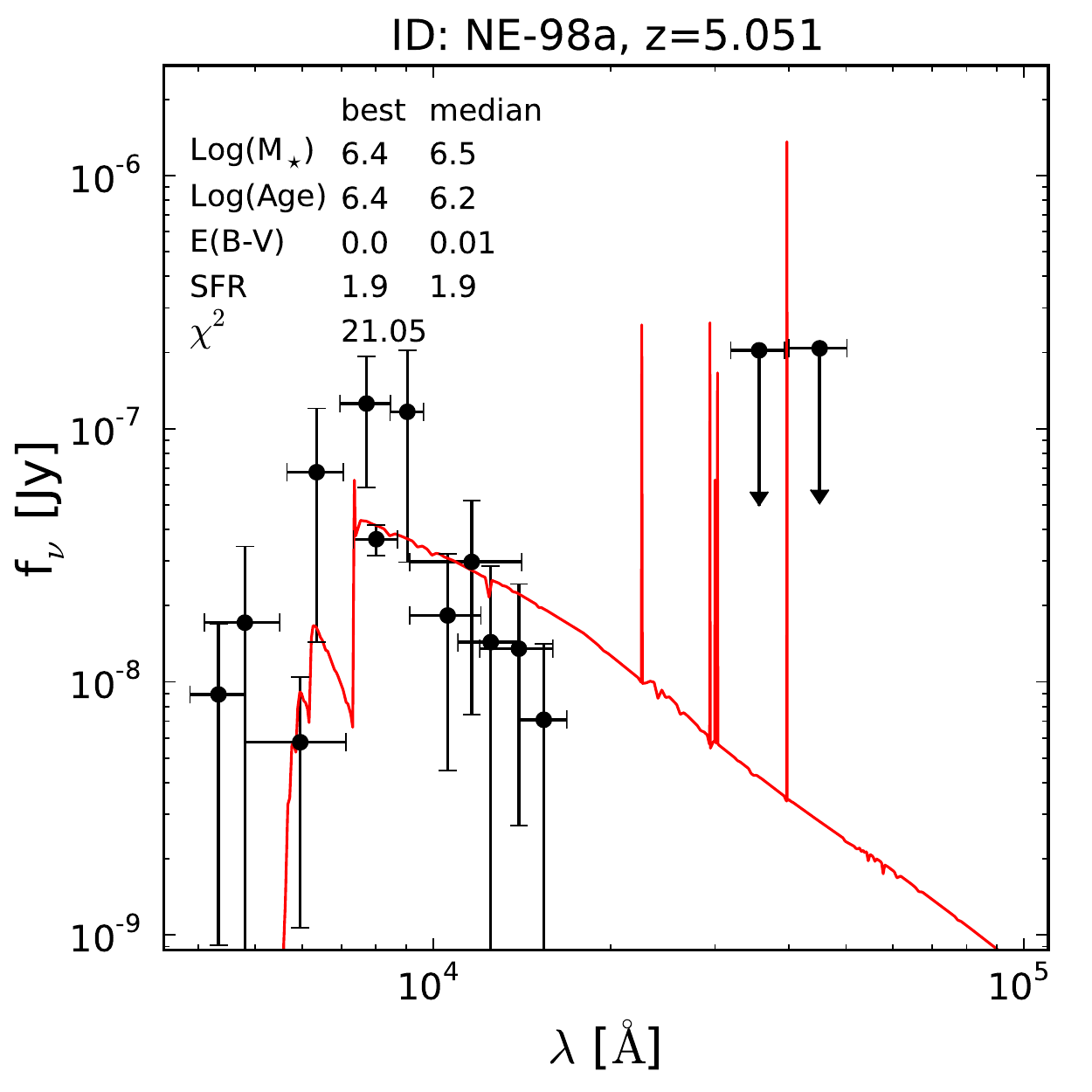}
\includegraphics[width=0.425\textwidth]{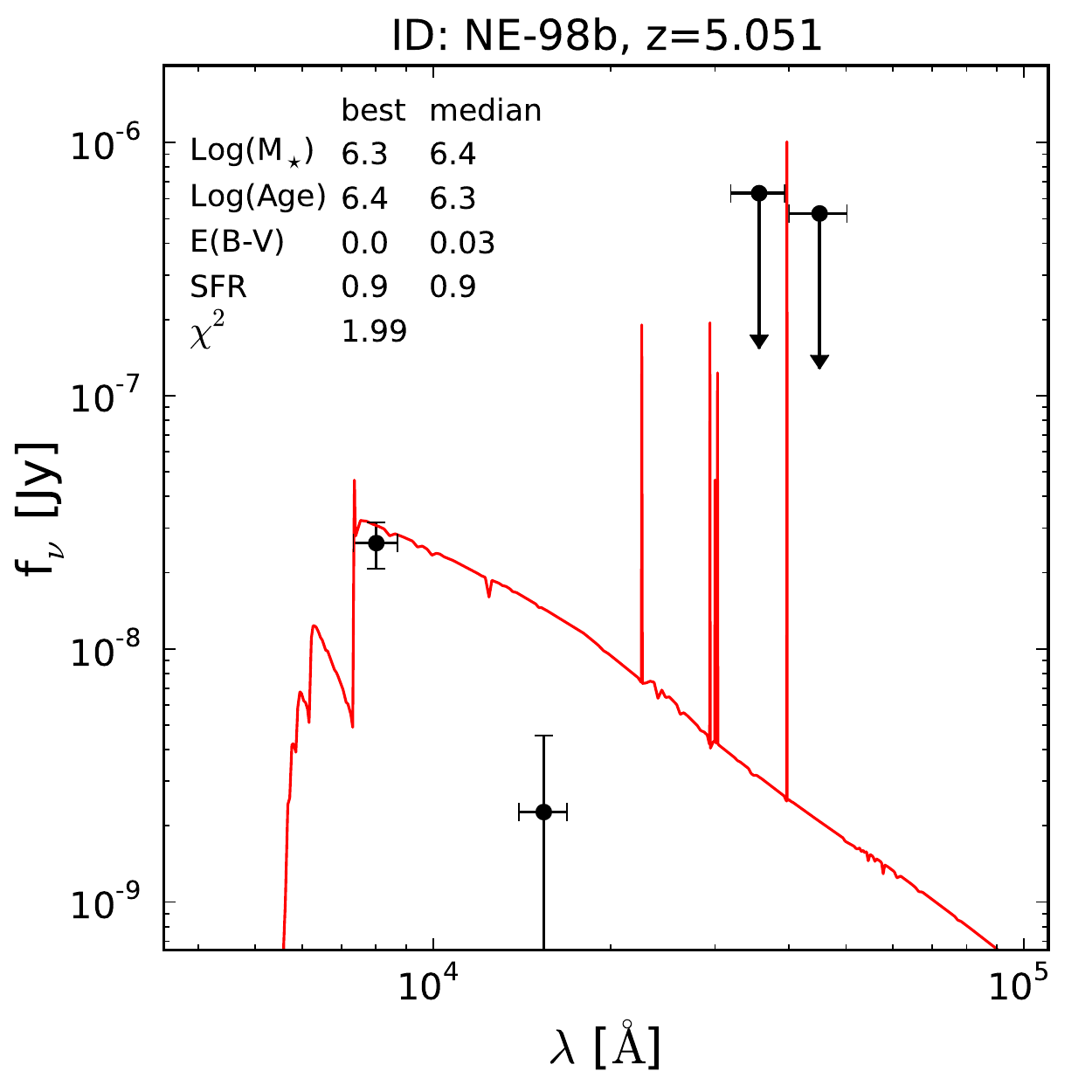}
\caption{\label{fig:seds_p1}}
\end{center}
\end{figure*}

\begin{figure*}
\begin{center}
\ContinuedFloat
\includegraphics[width=0.425\textwidth]{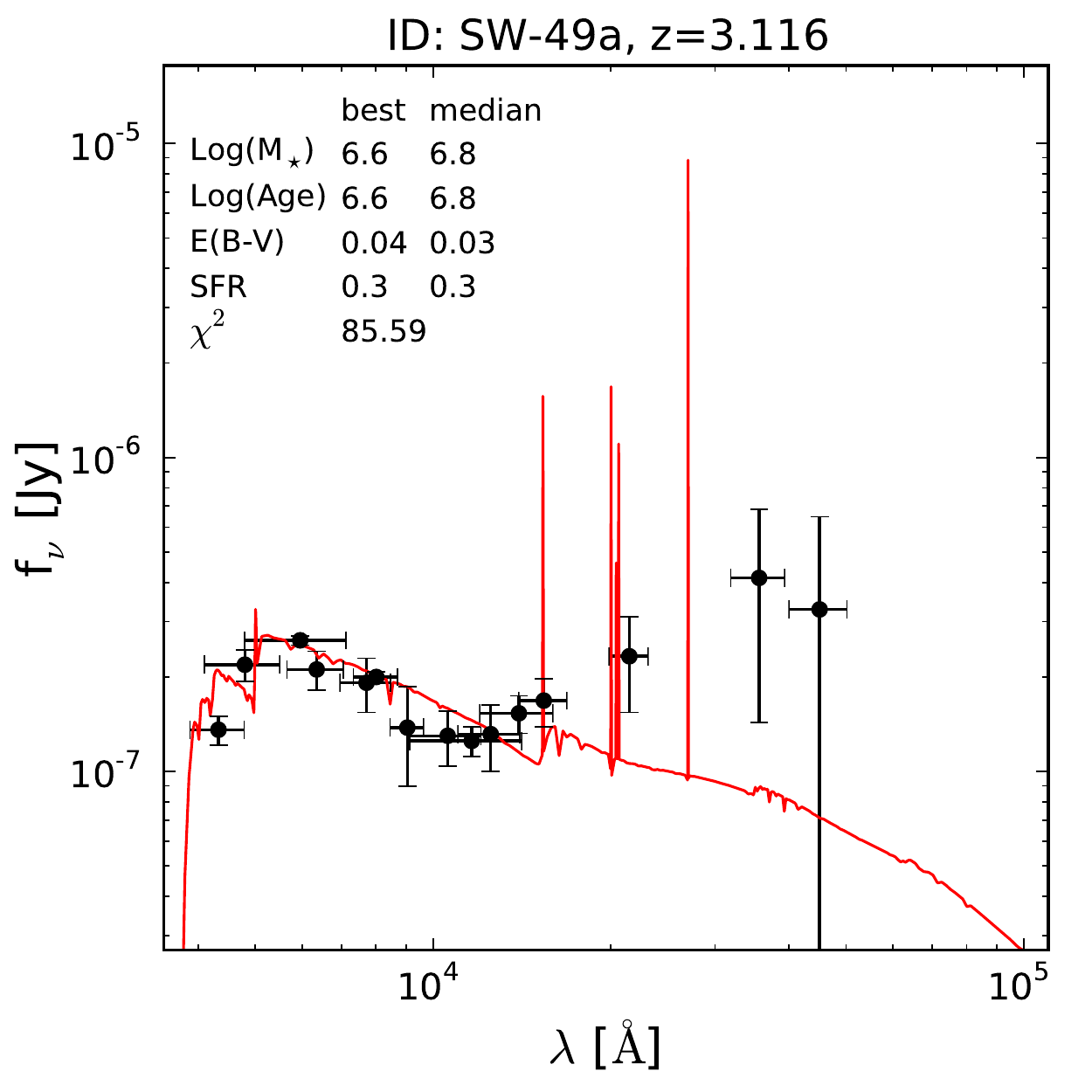}
\includegraphics[width=0.425\textwidth]{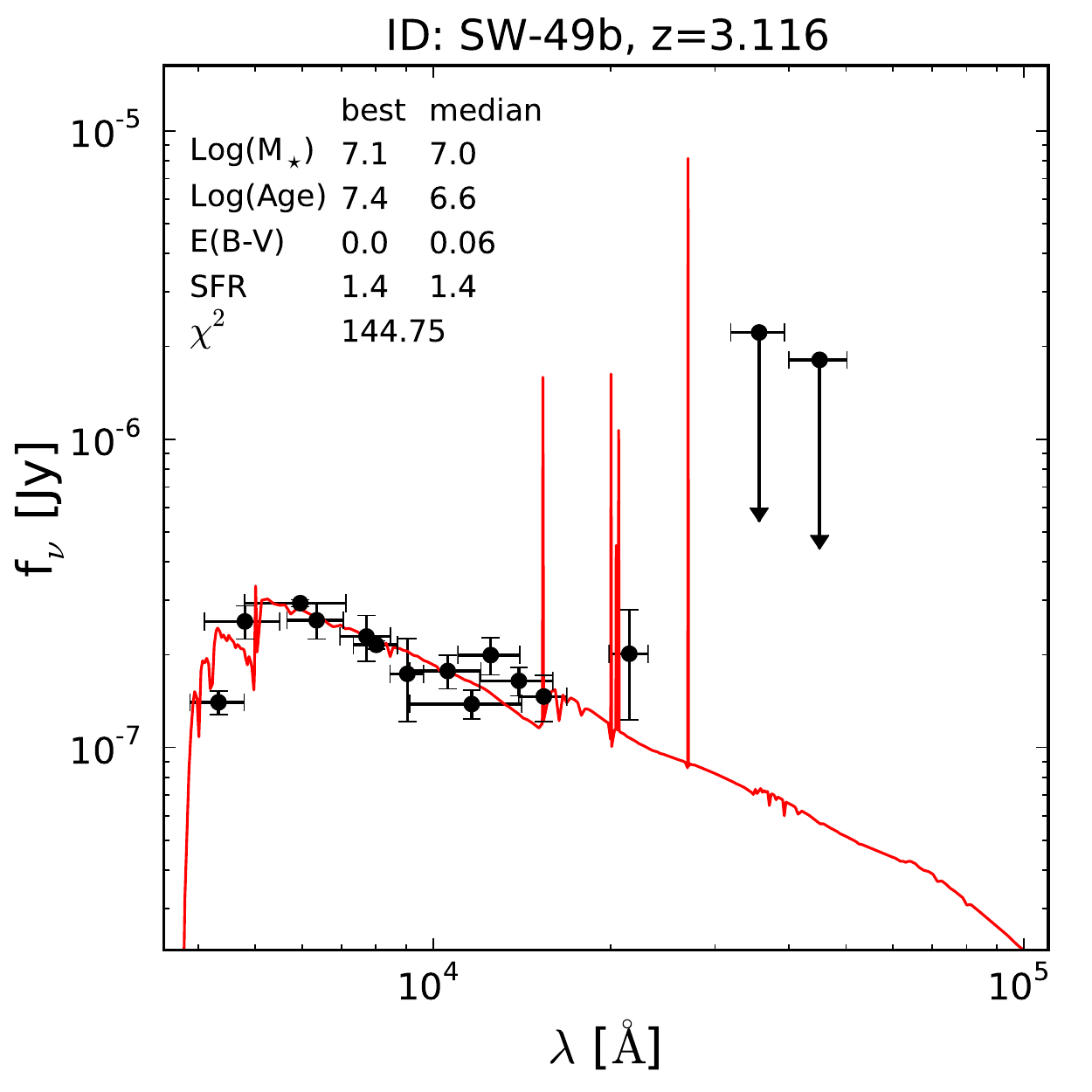}
\includegraphics[width=0.425\textwidth]{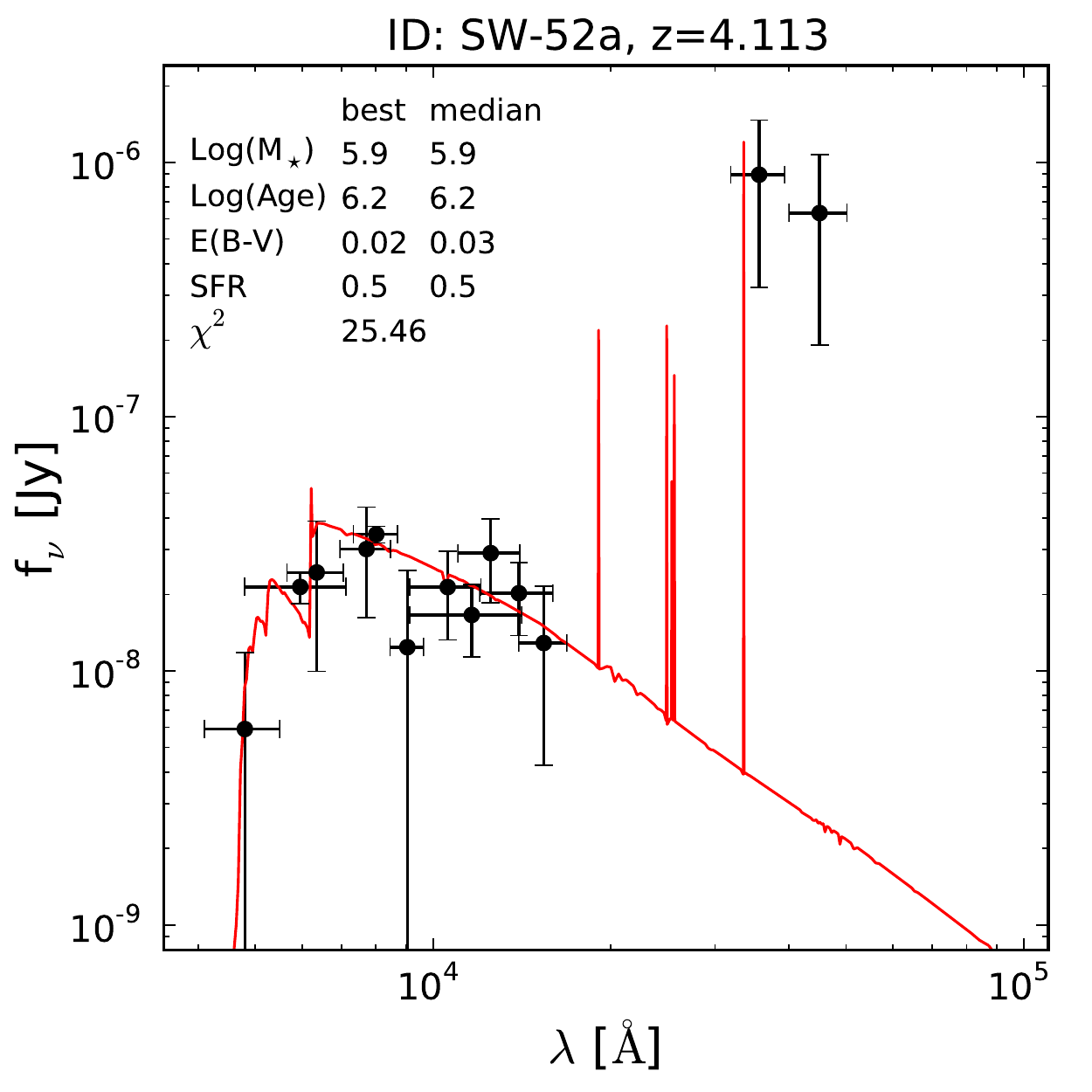}
\includegraphics[width=0.425\textwidth]{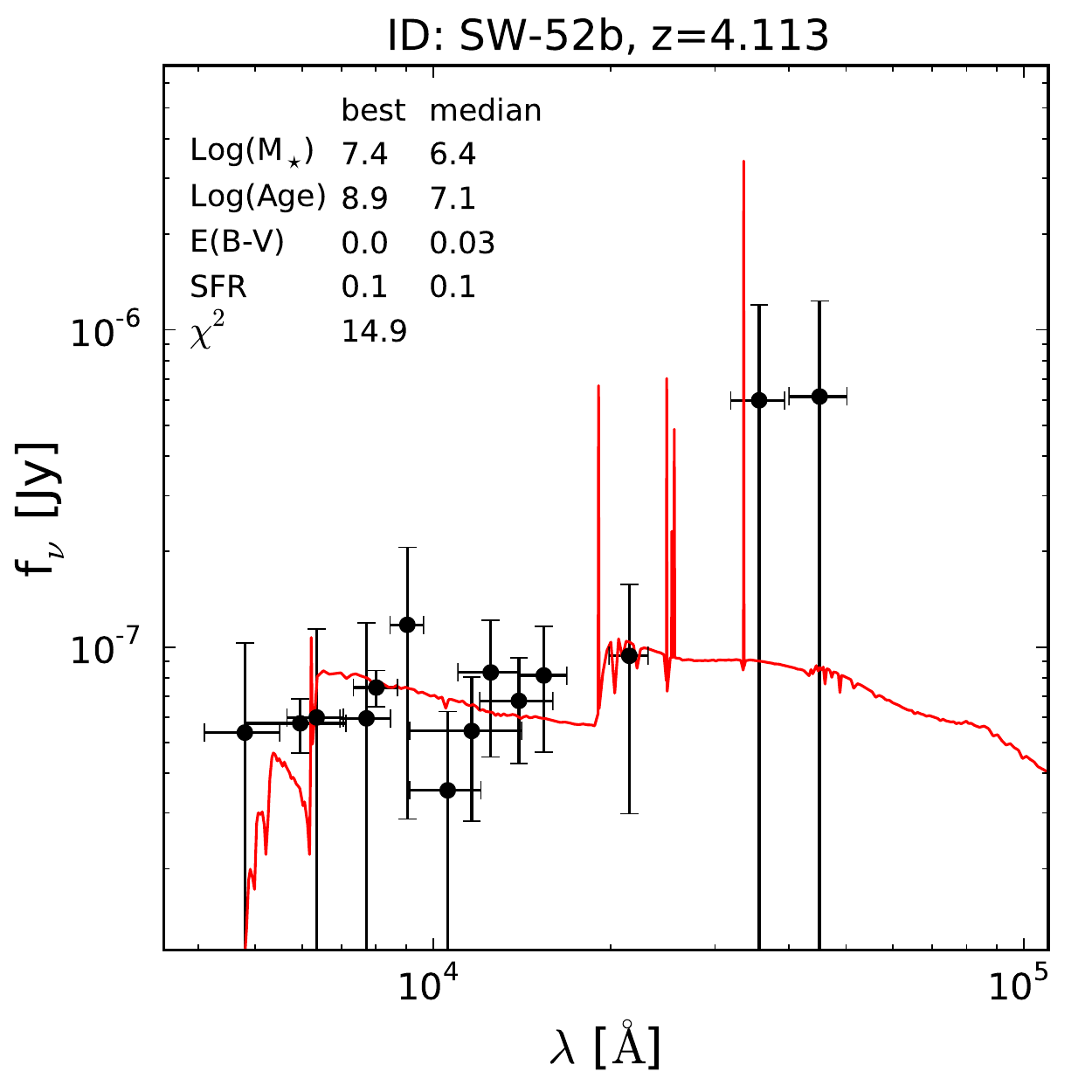}
\includegraphics[width=0.425\textwidth]{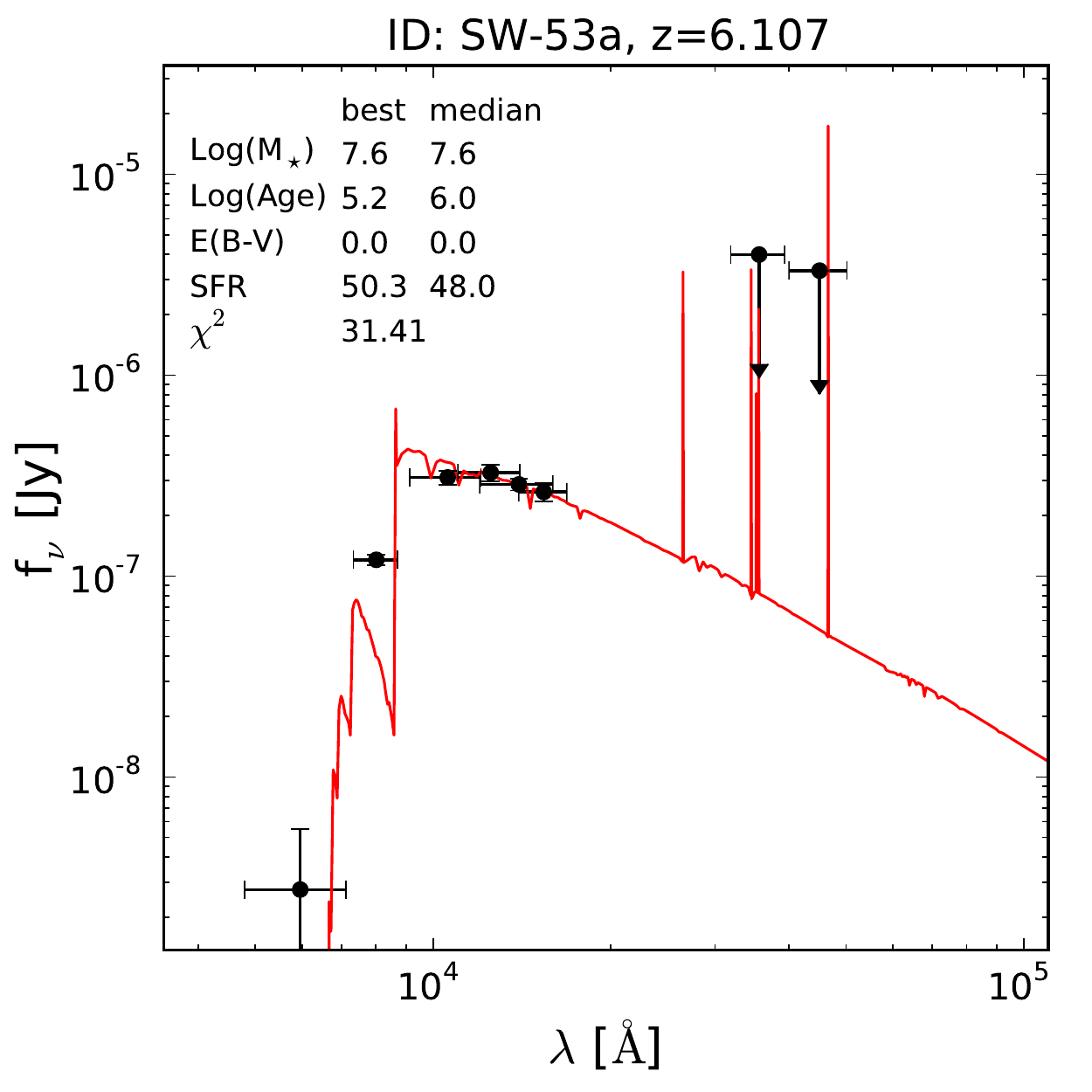}
\includegraphics[width=0.425\textwidth]{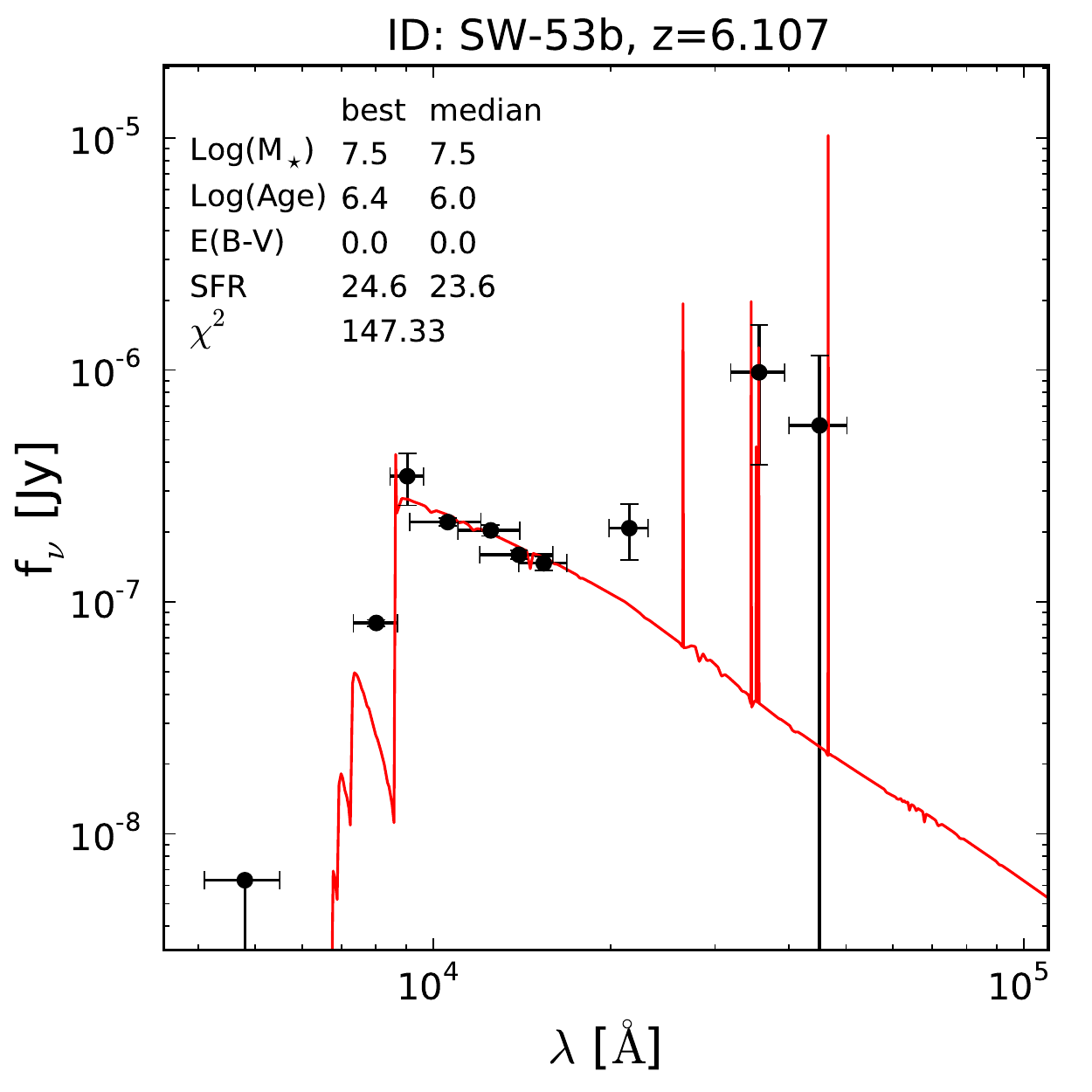}
\caption{\label{fig:seds_p2}}
\end{center}
\end{figure*}

\begin{figure*}
\begin{center}
\ContinuedFloat
\includegraphics[width=0.425\textwidth]{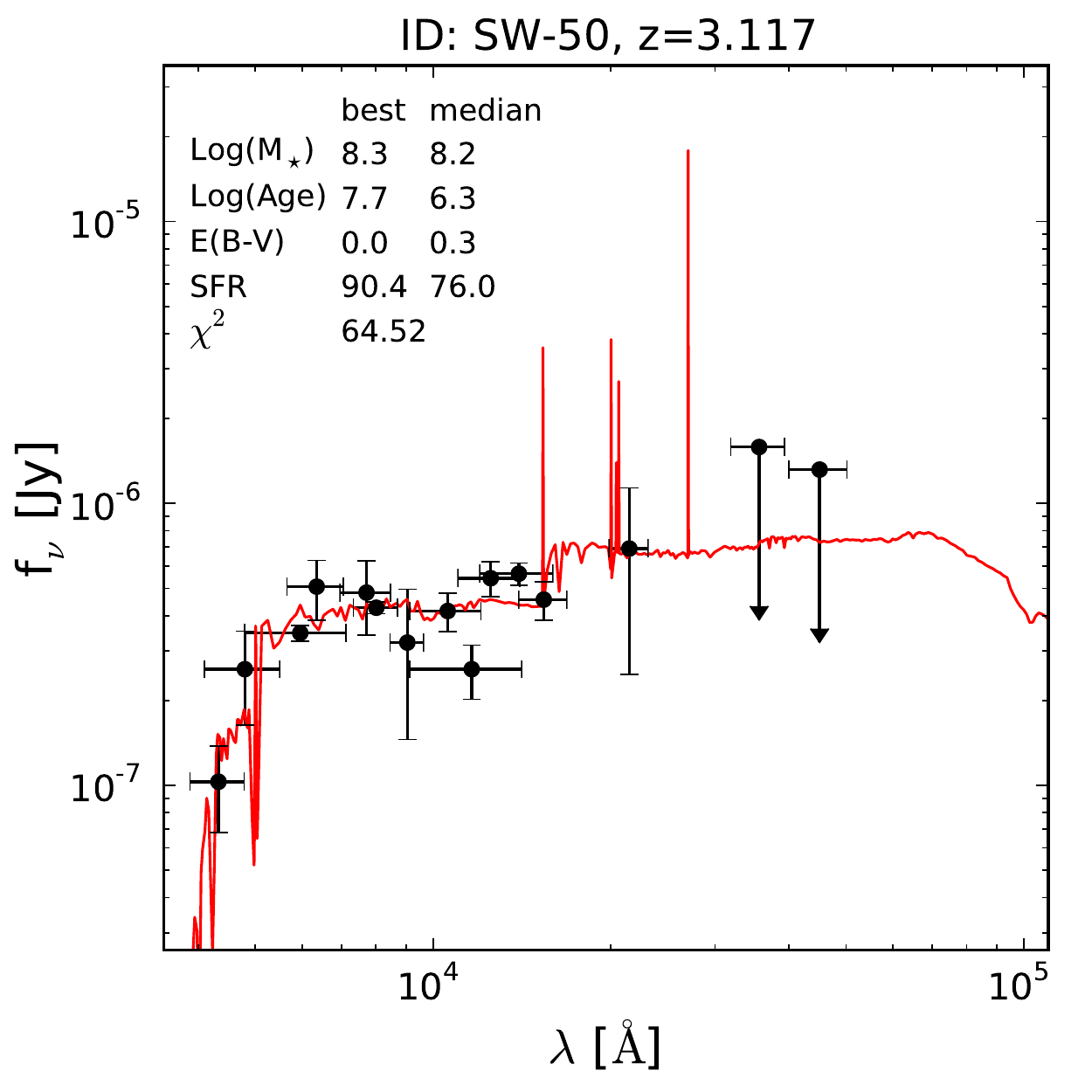}
\includegraphics[width=0.425\textwidth]{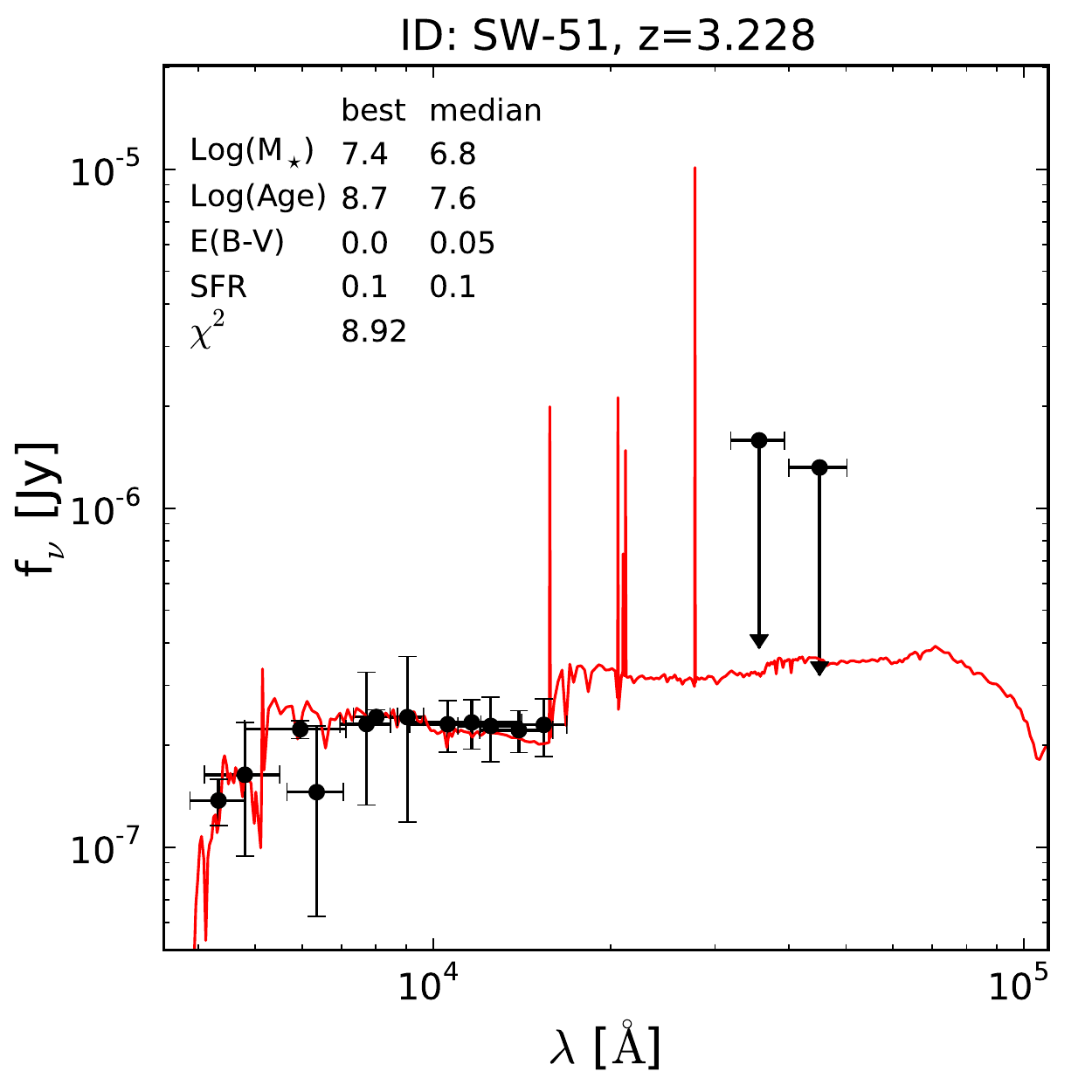}
\includegraphics[width=0.425\textwidth]{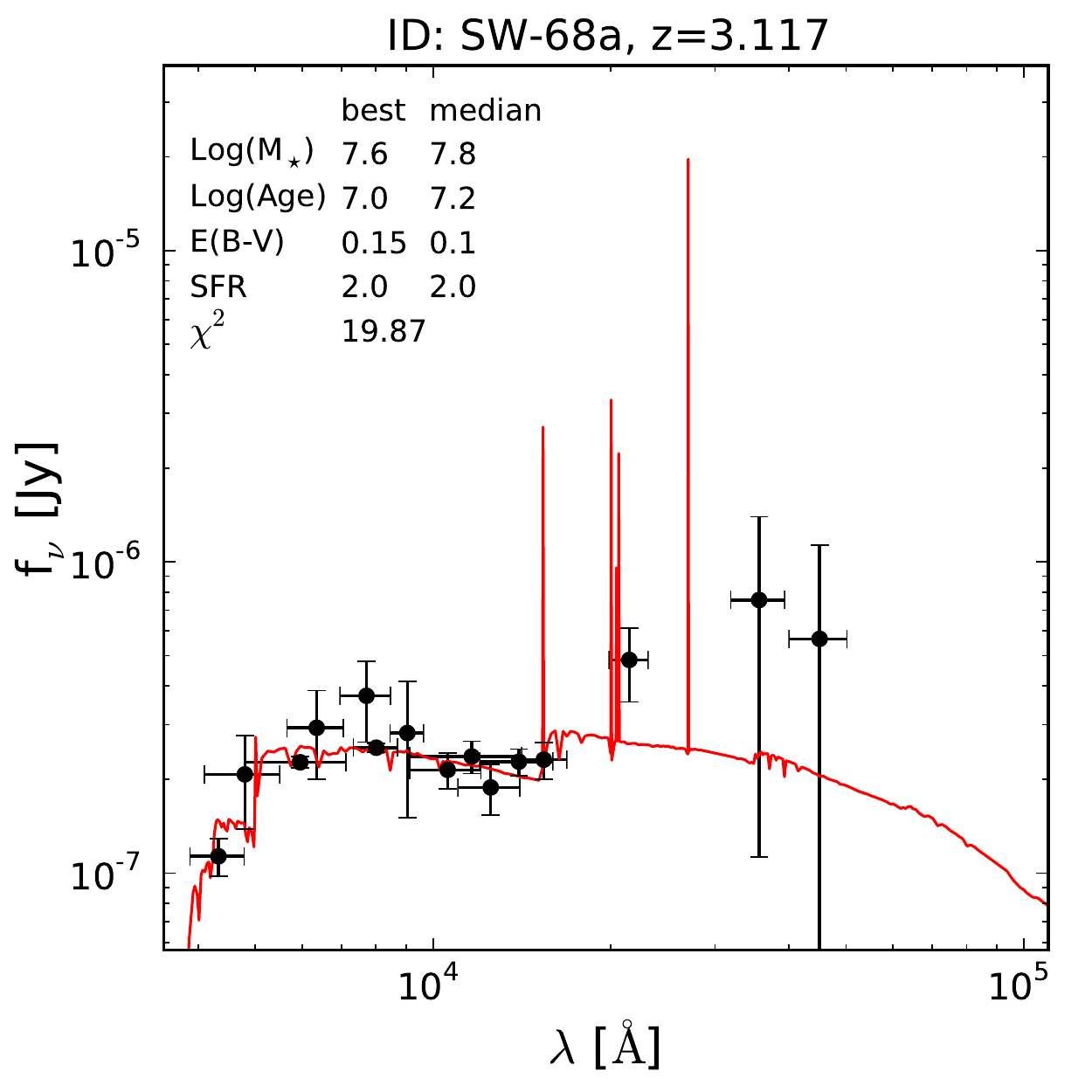}
\includegraphics[width=0.425\textwidth]{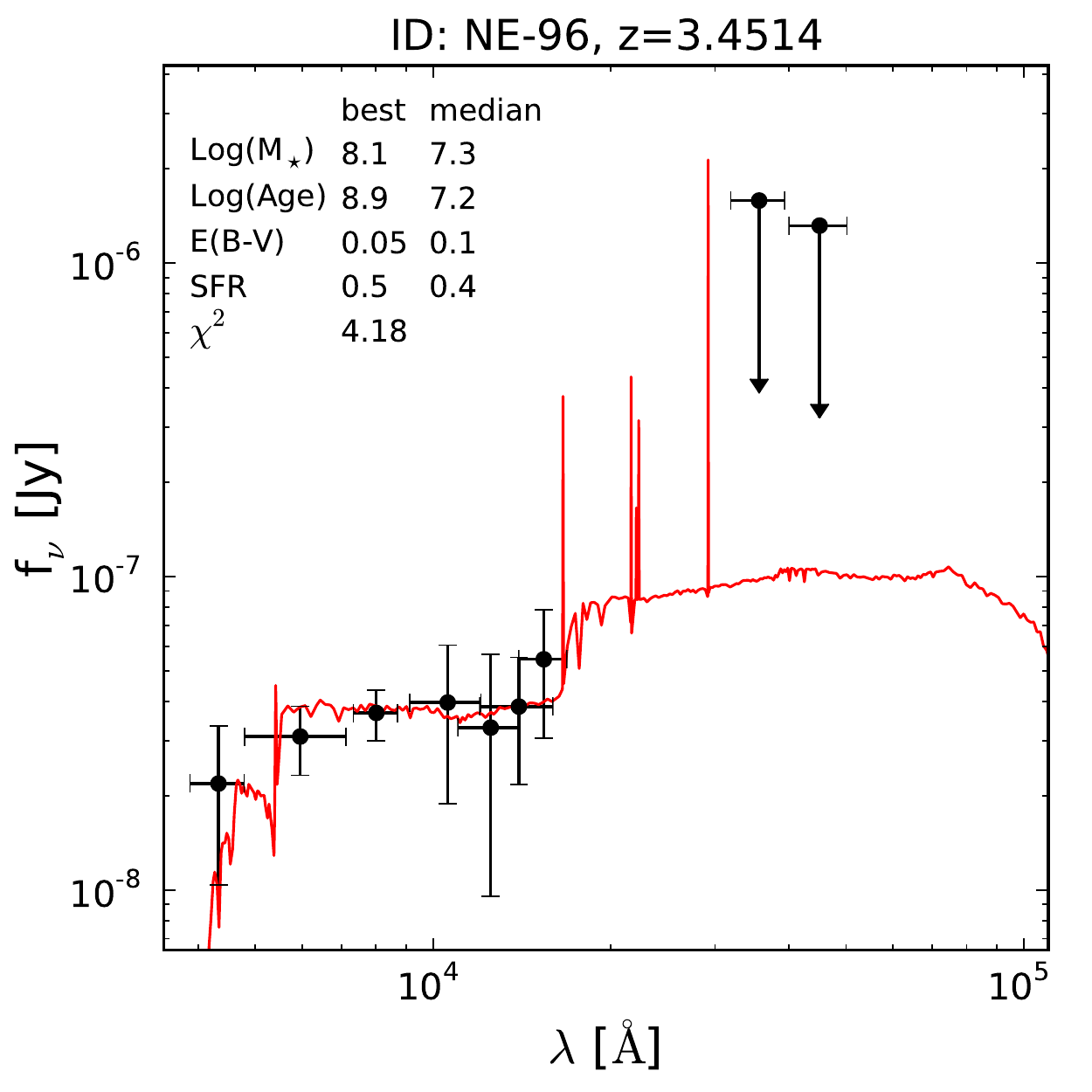}
\includegraphics[width=0.425\textwidth]{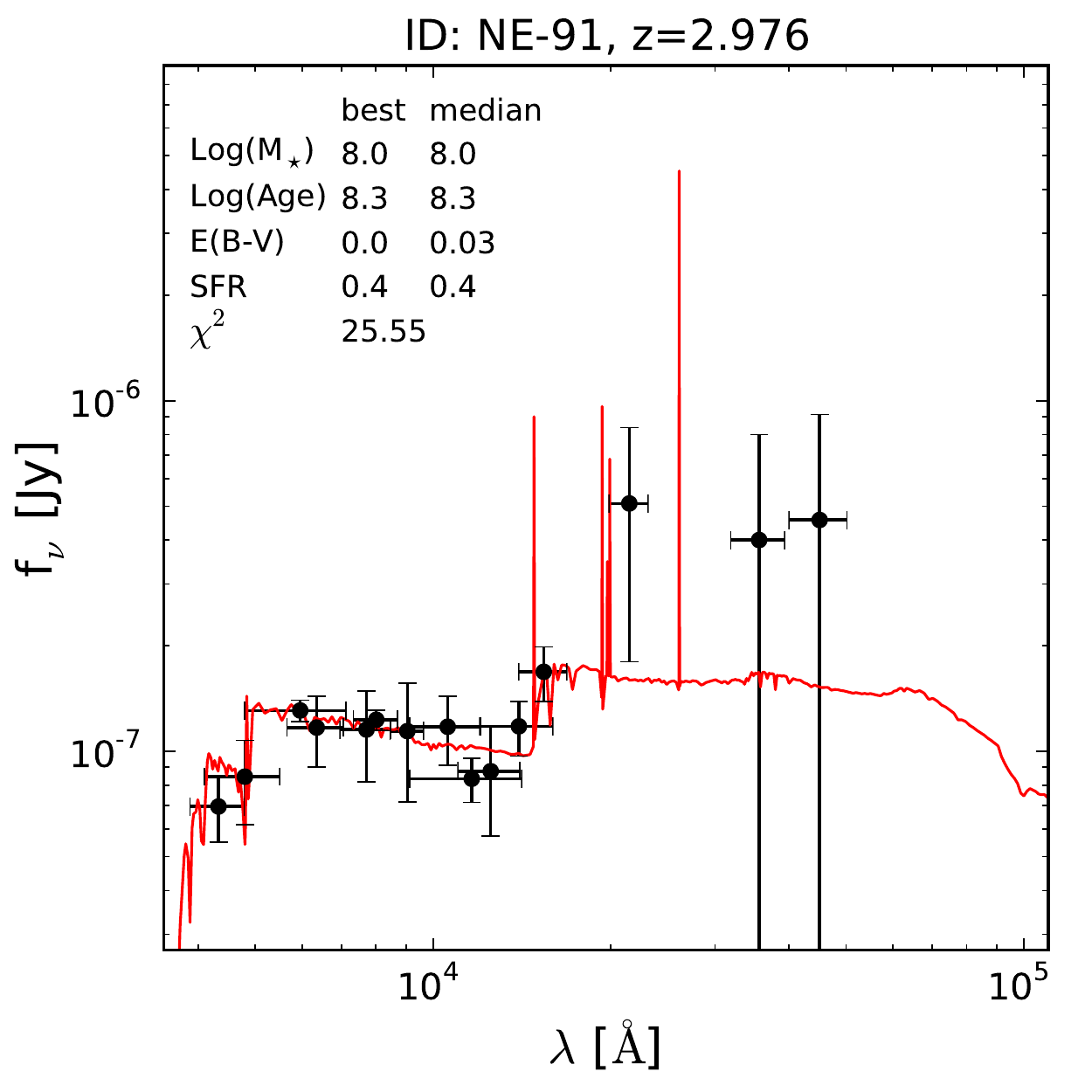}
\includegraphics[width=0.425\textwidth]{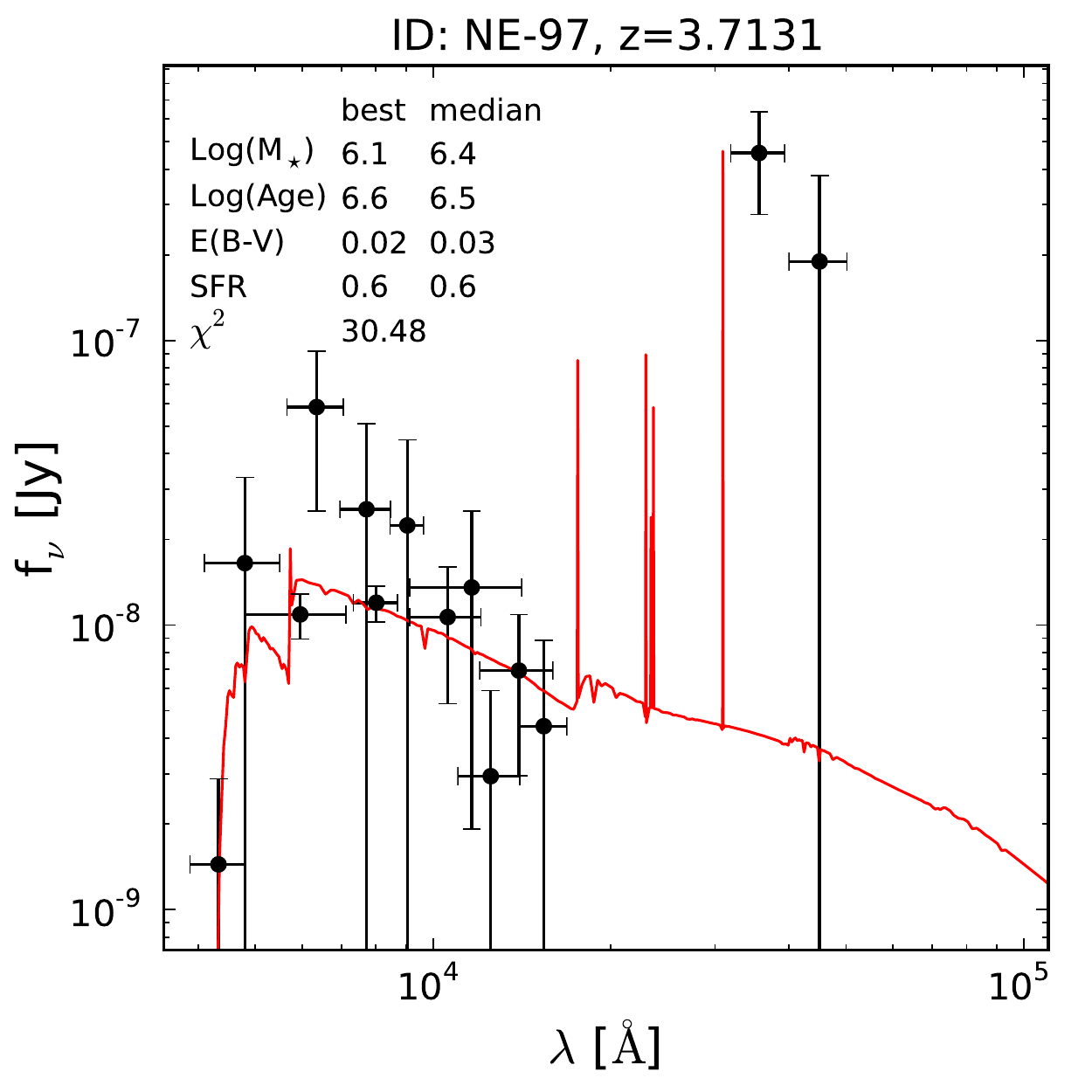}
\caption{Sed fitting results of individual LAEs. The red lines 
correspond to the best-fitting SED-model and the black circles are the
photometric datapoints with 2$\sigma$-errorbars. The details of the best
fitting model and the median marginalized values for each galaxy are 
shown wiht text in the upper left corners.\label{fig:ind_seds}}
\end{center}
\end{figure*}


%% file: tab_sed_ind.tex
\begin{sidewaystable*}
\begin{center}
\caption{SED fitting results of every individual LAE image. \label{tab:sed_ind}}
\begin{tabular}{llllllllll}
{\bf ID} & {\bf $\mu$} & {\bf Log(M$_\star$/M$_\odot$)} & {\bf Log(Age$_{SSP}/yr)$} & {\bf E(B-V)} & {\bf Log(SFR/(M$_\odot yr^{-1}$))} & {\bf Log(sSFR/yr)} & $\beta$ & {\rm f$_{esc,\Lya}$} & M$_{F814W}$ \\ 
& & & & & & & & \\ \hline 
NE-91 & 4.41$\pm0.11$  & 8.04$^{+0.19}_{-0.20}$ & 8.26$^{+0.45}_{-0.67}$ & 0.03$^{+0.03}_{-0.02}$ & -0.38$^{+0.15}_{-1.30}$ & -8.56$^{+0.39}_{-0.70}$ & -2.42$\pm0.09$ & 0.18 & 26.176 \\
SW-49a & 16.50$\pm1.00$  & 6.79$^{+0.27}_{-0.17}$ & 6.82$^{+0.83}_{-0.24}$ & 0.03$^{+0.03}_{-0.03}$ & -0.52$^{+0.63}_{-0.52}$ & -7.42$^{+0.78}_{-0.45}$ & -2.79$\pm0.10$ & 2.10 & 25.648\\
SW-49b & 15.80$\pm0.92$  & 7.01$^{+0.09}_{-0.17}$ & 6.64$^{+0.88}_{-0.84}$ & 0.06$^{+0.02}_{-0.05}$ & 0.15$^{+0.98}_{-0.72}$ & -6.70$^{+0.76}_{-0.98}$ & -2.79$\pm0.05$ & 0.68 & 25.568\\
SW-50 & 11.20$\pm0.76$  & 8.23$^{+0.05}_{-0.06}$ & 6.28$^{+1.28}_{-0.63}$ & 0.30$^{+0.00}_{-0.20}$ & 1.88$^{+0.69}_{-2.76}$ & -6.34$^{+0.69}_{-2.12}$ & -1.95$\pm0.05$ & 0.01 & 24.825\\
SW-68a & 6.09$\pm0.12$  & 7.78$^{+0.28}_{-0.15}$ & 7.18$^{+0.78}_{-0.42}$ & 0.10$^{+0.10}_{-0.05}$ & 0.30$^{+0.48}_{-0.52}$ & -7.50$^{+0.63}_{-0.83}$ & -2.09$\pm0.08$ & 0.56 & 25.392\\
NE-93a & 4.95$\pm0.12$  & 6.28$^{+0.25}_{-0.17}$ & 6.46$^{+0.84}_{-0.66}$ & 0.03$^{+0.03}_{-0.02}$ & -0.39$^{+0.89}_{-0.61}$ & -6.58$^{+0.78}_{-0.88}$ & -2.85$\pm0.50$ & 0.63 & 28.264\\
NE-93b & 5.08$\pm0.11$  & 6.20$^{+0.29}_{-0.19}$ & 6.52$^{+0.84}_{-0.72}$ & 0.04$^{+0.04}_{-0.03}$ & -0.55$^{+0.93}_{-0.58}$ & -6.64$^{+0.82}_{-0.89}$ & -2.85$\pm0.41$ & 1.23 & 28.466 \\
SW-51 & 55.60$\pm8.95$  & 6.81$^{+0.49}_{-0.34}$ & 7.59$^{+0.82}_{-0.65}$ & 0.05$^{+0.05}_{-0.03}$ & -0.96$^{+0.30}_{-0.27}$ & -7.87$^{+0.63}_{-0.69}$ & -2.31$\pm0.05$ & 0.27 & 25.440 \\
NE-94a & 7.09$\pm0.15$  & 8.50$^{+0.16}_{-0.16}$ & 6.70$^{+0.66}_{-0.66}$ & 0.30$^{+0.05}_{-0.05}$ & 1.61$^{+0.93}_{-0.59}$ & -6.74$^{+0.64}_{-0.75}$ & -1.76$\pm0.08$ & 0.03 & 24.734 \\
NE-94b & 8.12$\pm0.21$  & 8.15$^{+0.59}_{-0.32}$ & 7.30$^{+0.96}_{-1.20}$ & 0.20$^{+0.10}_{-0.10}$ & 0.55$^{+1.39}_{-0.65}$ & -7.42$^{+1.26}_{-1.64}$ & -1.85$\pm0.18$ & 0.07 & 25.445 \\
NE-96 & 4.78$\pm0.13$  & 7.33$^{+0.40}_{-0.44}$ & 7.24$^{+0.87}_{-0.96}$ & 0.10$^{+0.15}_{-0.07}$ & -0.36$^{+1.34}_{-0.64}$ & -7.48$^{+1.14}_{-1.24}$ & -2.14$\pm0.22$ & 0.52 & 27.487\\
NE-97 & 2.98$\pm0.06$  & 6.35$^{+0.20}_{-0.17}$ & 6.46$^{+0.72}_{-0.66}$ & 0.03$^{+0.03}_{-0.02}$ & -0.23$^{+0.84}_{-0.66}$ & -6.52$^{+0.72}_{-0.85}$ & -2.90$\pm0.15$ & 0.27 & 28.704\\
SW-52a & 23.6$\pm1.47$  & 5.93$^{+0.10}_{-0.13}$ & 6.22$^{+0.48}_{-0.72}$ & 0.03$^{+0.02}_{-0.02}$ & -0.32$^{+0.68}_{-0.74}$ & -6.28$^{+0.71}_{-0.60}$ & -3.05$\pm0.10$ & 0.32 & 27.557\\
SW-52b & 27.9$\pm1.56$  & 6.40$^{+0.39}_{-0.26}$ & 7.12$^{+0.60}_{-0.60}$ & 0.03$^{+0.03}_{-0.02}$ & -0.91$^{+0.56}_{-0.33}$ & -7.30$^{+0.73}_{-0.86}$ & -2.50$\pm0.28$ & 2.46 & 26.719\\
NE-98a & 8.84$\pm0.35$  & 6.46$^{+0.10}_{-0.09}$ & 6.16$^{+0.36}_{-0.66}$ & 0.01$^{+0.02}_{-0.01}$ & 0.28$^{+0.66}_{-0.54}$ & -6.19$^{+0.68}_{-0.45}$ & -3.17$\pm0.02$ & 0.24 & 27.491\\
NE-98b & 8.95$\pm0.24$  & --- & --- & --- & --- & --- & --- & --- & 27.855\\ 
NE-99a & 9.91$\pm0.33$  & --- & --- & --- & --- & --- & --- & --- & 28.854\\
NE-99b & 10.10$\pm0.29$  & --- & --- & --- & --- & --- & --- & --- & 29.584\\
NE-100 & 7.87$\pm0.21$  & --- & --- & --- & --- & --- & --- & --- & ---\\
NE-101d & 6.53$\pm0.51$  & 8.13$^{+0.23}_{-0.13}$ & 6.34$^{+0.72}_{-0.69}$ & 0.02$^{+0.03}_{-0.02}$ & 1.72$^{+0.76}_{-0.71}$ & -6.40$^{+0.75}_{-0.84}$ & -3.18$\pm0.00$ & 0.17 & ---\\
SW-53a & 5.56$\pm0.33$  & 7.59$^{+0.08}_{-0.02}$ & 5.95$^{+0.33}_{-0.60}$ & 0.00$^{+0.01}_{-0.00}$ & 1.68$^{+0.56}_{-0.41}$ & -5.95$^{+0.60}_{-0.39}$ & -3.02$\pm0.05$ & 0.09 & 26.198\\
SW-53b & 5.97$\pm0.12$  & 7.47$^{+0.00}_{-0.00}$ & 6.04$^{+0.30}_{-0.54}$ & 0.00$^{+0.00}_{-0.00}$ & 1.37$^{+0.60}_{-0.30}$ & -6.10$^{+0.60}_{-0.30}$ & -3.17$\pm0.02$ & 0.43 & 26.627\\
SW-70a\tablefootmark{a} & 8.43$\pm0.248$  & 8.58$^{+0.08}_{-0.01}$ & 6.10$^{+0.30}_{-0.60}$ & 0.25$^{+0.00}_{-0.00}$ & 2.45$^{+0.63}_{-0.50}$ & -6.16$^{+0.66}_{-0.43}$ & -2.00$\pm0.02$ & 0.00 & 23.233\\
SW-70b\tablefootmark{a} & 5.79$\pm0.146$  & 8.84$^{+0.31}_{-0.10}$ & 7.00$^{+0.42}_{-0.18}$ & 0.20$^{+0.00}_{-0.00}$ & 1.81$^{+0.07}_{-0.26}$ & -7.03$^{+0.11}_{-1.29}$ & -1.83$\pm0.12$ & 0.01 & 24.018\\
\end{tabular}
\tablefoot{Results of SED modelling of all individual images of LAEs. The stellar masses and SFRs have been corrected for lensing, using the magnification factor given in the second column. The last column shows the observed magnitude in the F814W {\em HST} band without correcting for magnification. LAEs NE-98b, NE-99a, NE-99b, and NE-100 did not have enough photometric detections to perform a sensible modelling.\\ \tablefoottext{a}{Lyman break galaxy, no significant \Lya emission.}}
\end{center}
\end{sidewaystable*}

%% file: tab_lya_ind.tex
\begin{sidewaystable*}
\begin{center}
\begin{tabular}{lllllll}
{\bf ID} & {\bf z} & {\bf f$_\Lya$} & {\bf Log(L$_\Lya$/(erg/s))} & {\bf Log(N$_H$/cm$^{-2}$)} & {\bf v$_{\rm exp}$} & {\bf SFR$_{\Lya}$} \\ 
& & 10$^{-20}$ erg/s/cm$^{-2}$ & & & km s$^{-1}$ & M$_\odot$ yr$^{-1}$ \\ \hline 
NE-91 & 2.9760 & 4.7$\pm$0.6 & 40.9$\pm$0.0611 &--- &---  & 0.0741 \\
SW-49a & 3.1169 & 179.0$\pm$21.3 & 42.0$\pm$0.0551 & 17.1$^{+0.2}_{0.1}$ & 72.9$^{+1.6}_{-3.5}$  & 0.83 \\
SW-49b & 3.1169 & 198.0$\pm$33.1 & 42.0$\pm$0.0795 & 17$^{+0.2}_{0.1}$ & 60.7$^{+3.1}_{-3.7}$  & 0.965 \\
SW-50 & 3.1169 & 110.0$\pm$2.0 & 41.9$\pm$0.00803 & 17.5$^{+0.3}_{0.2}$ & 124$^{+6}_{-6}$  & 0.757 \\
SW-68a & 3.1166 & 118.0$\pm$1.0 & 42.2$\pm$0.00364 & 20$^{+0.2}_{0.2}$ & 463$^{+21}_{-48}$  & 1.49 \\
NE-93a & 3.1696 & 15.8$\pm$1.6 & 41.4$\pm$0.0469 & 19$^{+0.4}_{0.5}$ & 42.9$^{+25.3}_{-16.1}$  & 0.255 \\
NE-93b & 3.1701 & 21.9$\pm$1.3 & 41.6$\pm$0.0265 & 17.7$^{+0.9}_{0.6}$ & 46.4$^{+11.6}_{-11.3}$  & 0.345 \\
SW-51 & 3.2271 & 19.8$\pm$0.9 & 40.5$\pm$0.0206 & 18.5$^{+0.5}_{0.7}$ & 98$^{+15}_{-17.8}$  & 0.0297 \\
NE-94a & 3.2848 & 37.2$\pm$1.9 & 41.7$\pm$0.0228 & 19.4$^{+0.3}_{0.3}$ & 192$^{+19}_{-25}$  & 0.458 \\
NE-94b & 3.2854 & 44.4$\pm$2.2 & 41.7$\pm$0.0219 & 19$^{+0.2}_{0.2}$ & 174$^{+11}_{-9}$  & 0.477 \\
NE-96 & 3.4514 & 10.9$\pm$0.7 & 41.4$\pm$0.0291 & 18.6$^{+0.1}_{0.3}$ & 56.7$^{+7.3}_{-8.9}$  & 0.224 \\
NE-97 & 3.7131 & 4.6$\pm$0.6 & 41.3$\pm$0.0581 & 20.4$^{+0.1}_{0.2}$ & 181$^{+13}_{-12}$  & 0.181 \\
SW-52a & 4.1130 & 24.9$\pm$0.9 & 41.2$\pm$0.0165 & 17.3$^{+0.2}_{0.2}$ & 87.8$^{+6}_{-7.9}$  & 0.157 \\
SW-52b & 4.1128 & 29.8$\pm$1.1 & 41.2$\pm$0.0167 & 17$^{+0.1}_{0.1}$ & 108$^{+5}_{-2}$  & 0.159 \\
NE-98a & 5.0514 & 16.5$\pm$0.7 & 41.7$\pm$0.0198 & 19.4$^{+0.2}_{0.2}$ & 149$^{+12}_{-12}$  & 0.451 \\
NE-98b & 5.0549 & 4.7$\pm$0.7 & 41.1$\pm$0.0693 & 18.2$^{+0.4}_{0.4}$ & -4.93$^{+2.64}_{-3.36}$  & 0.127 \\
NE-99a & 5.2399 & 10.5$\pm$1.2 & 41.5$\pm$0.0506 & 18.5$^{+0.8}_{0.6}$ & 236$^{+183}_{-107}$  & 0.28 \\
NE-99b & 5.2415 & 12.1$\pm$1.1 & 41.5$\pm$0.0423 & 17.4$^{+0.4}_{0.3}$ & -19.8$^{+5.1}_{-3.8}$  & 0.315 \\
NE-100 & 5.8940 & 4.7$\pm$2.5 & 41.4$\pm$0.329 &--- &---  & 0.207 \\
NE-101d & 6.1074 & 150.0$\pm$1.9 & 43$\pm$0.00546 &--- &---  & 8.67 \\
SW-53a & 6.1074 & 63.4$\pm$2.7 & 42.7$\pm$0.0189 & 19.8$^{+0.1}_{-0.1}$ & 151$^{+3}_{5}$  & 4.31 \\
SW-53b & 6.1074 & 161.0$\pm$4.0 & 43$\pm$0.0109 &--- &---  & 10.2 \\
SW-70a & 3.6065 & --- & 41.2$\pm$2.0 &--- &---  & 0.129 \\
SW-70b & 3.6065 & --- & 41.3$\pm$2.0 &--- &---  & 0.188 \\
\end{tabular}
\caption{Results from the \Lya line modelling simulations of all images of LAEs. We report in column 2 the
intrinsic redshift obtained from modelling the \Lya profile.
 Column 3 gives the
observed \Lya flux without correcting for lensing, while we corrected $L_{\rm Lya}$ and SFR$_{\Lya}$ 
for the effect of gravitational magnification using the magnification factors given in Table \ref{tab:sed_ind}.\label{tab:Lya_ind}}
\end{center}
\end{sidewaystable*}

%% file: Lya_triangles.tex
\begin{figure*}
\begin{center}
\includegraphics[width=\textwidth]{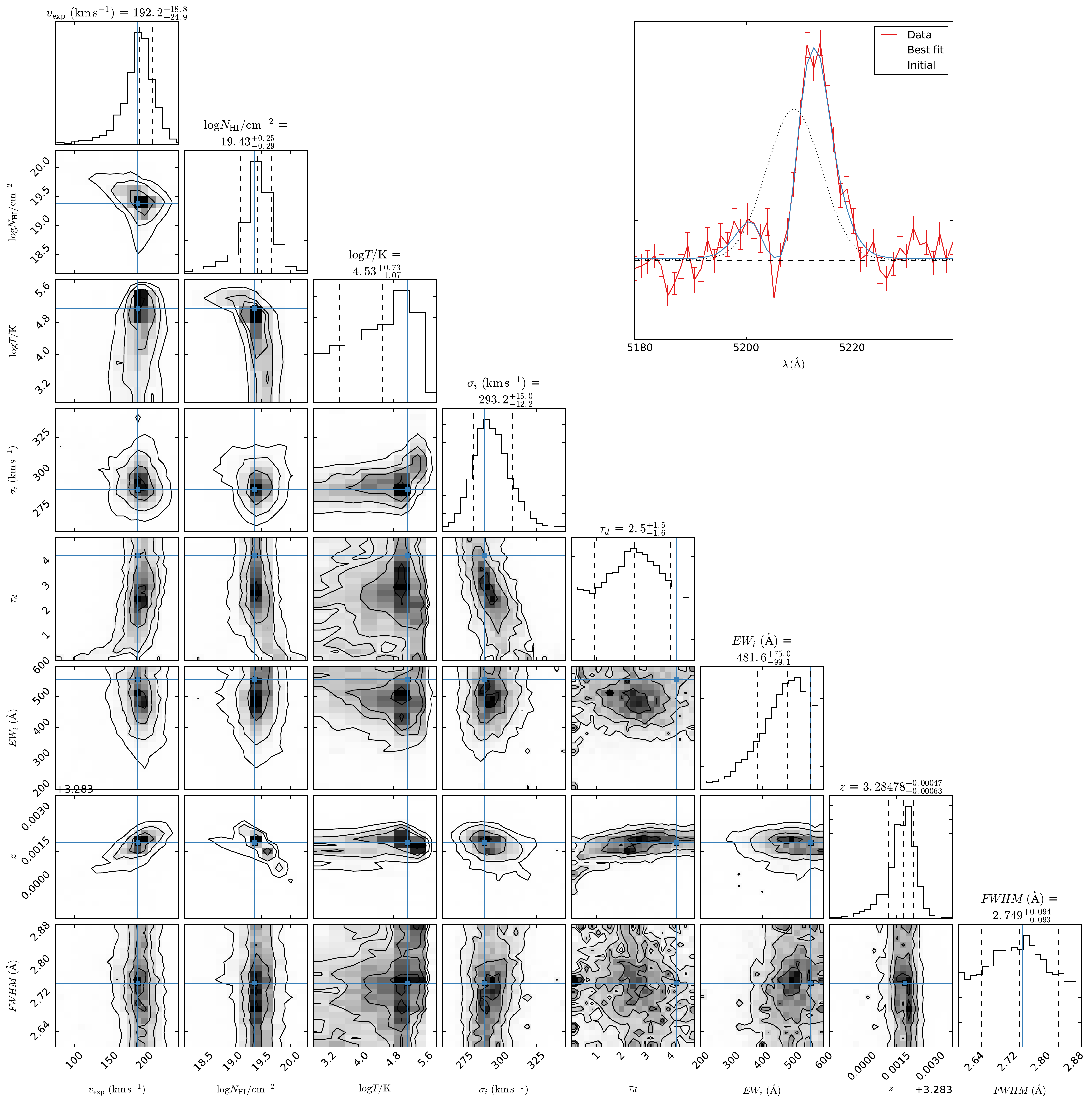}
\caption{Lya modelling results for LAE NE-94a. We show in the top
right panel the spectrum zoomed in on \Lya overplotted with the
best-fit model. We show the likelihood of each model parameter 
against all other model parameters in the other panels. 
The parameters plotted are (from left/top to right/bottom):
expansion velocity ($v_{\rm exp}$), hydrogen column density
($n_{\rm H}$), temperature ({\em T}), intrinsic spread of
velocities ($\sigma_i$), optical depth of dust ($\tau_d$),
intrinsic EW ($EW_i$), the systemic redshift ({\em z}), and
the FWHM of the observations. The best-fit parameters are indicated
by the blue lines in each panel, while the predicted parameter
ranges are shown at the top of each column.
\label{fig:lya_112}}
\end{center}
\end{figure*}

\begin{figure*}
\begin{center}
\includegraphics[width=\textwidth]{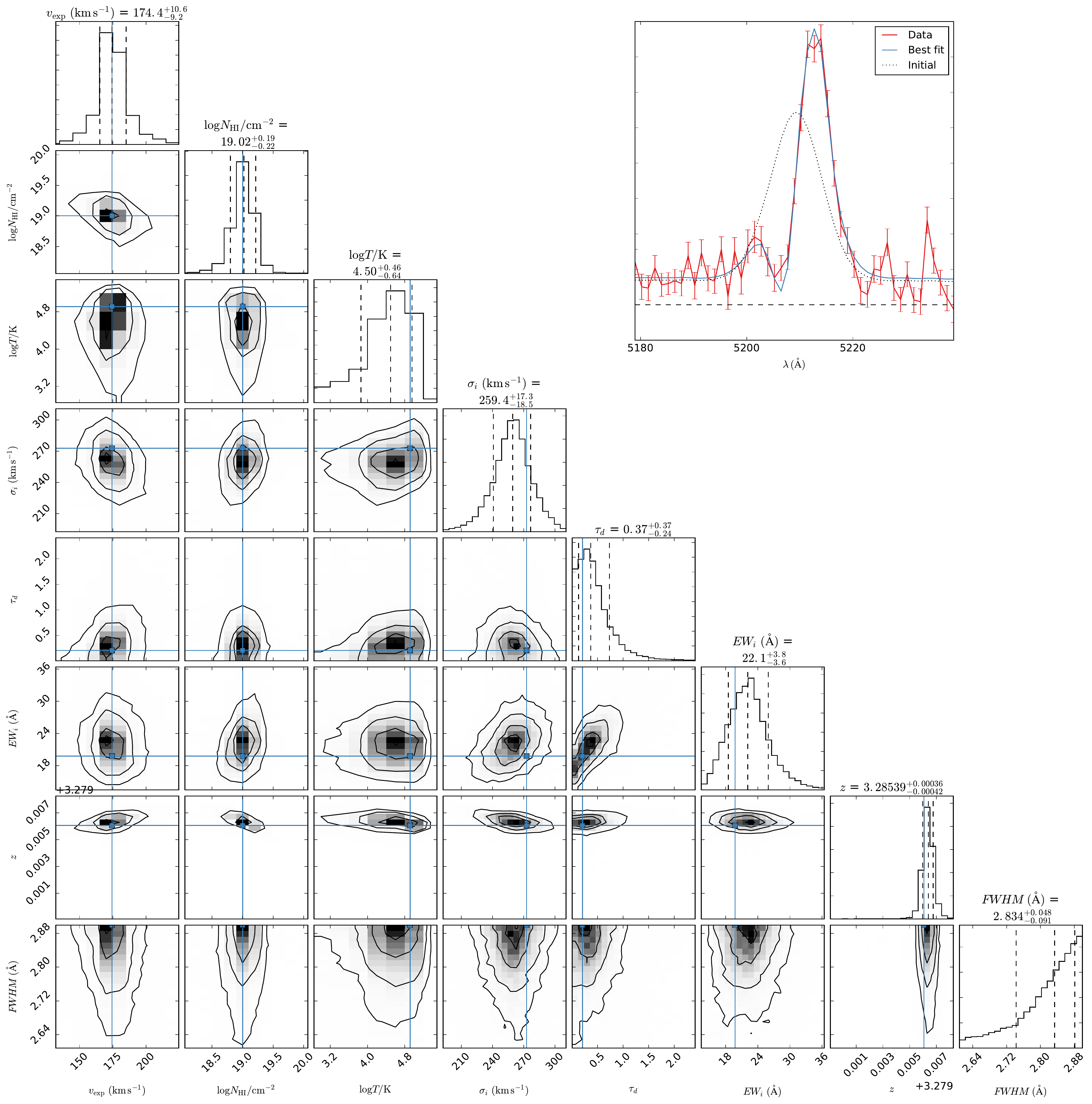}
\caption{Lya modelling results for NE-94b. The lines and legend
are the same as in Fig. \ref{fig:lya_112}. \label{fig:lya_1121}}
\end{center}
\end{figure*}

\clearpage

\begin{figure*}
\begin{center}
\includegraphics[width=\textwidth]{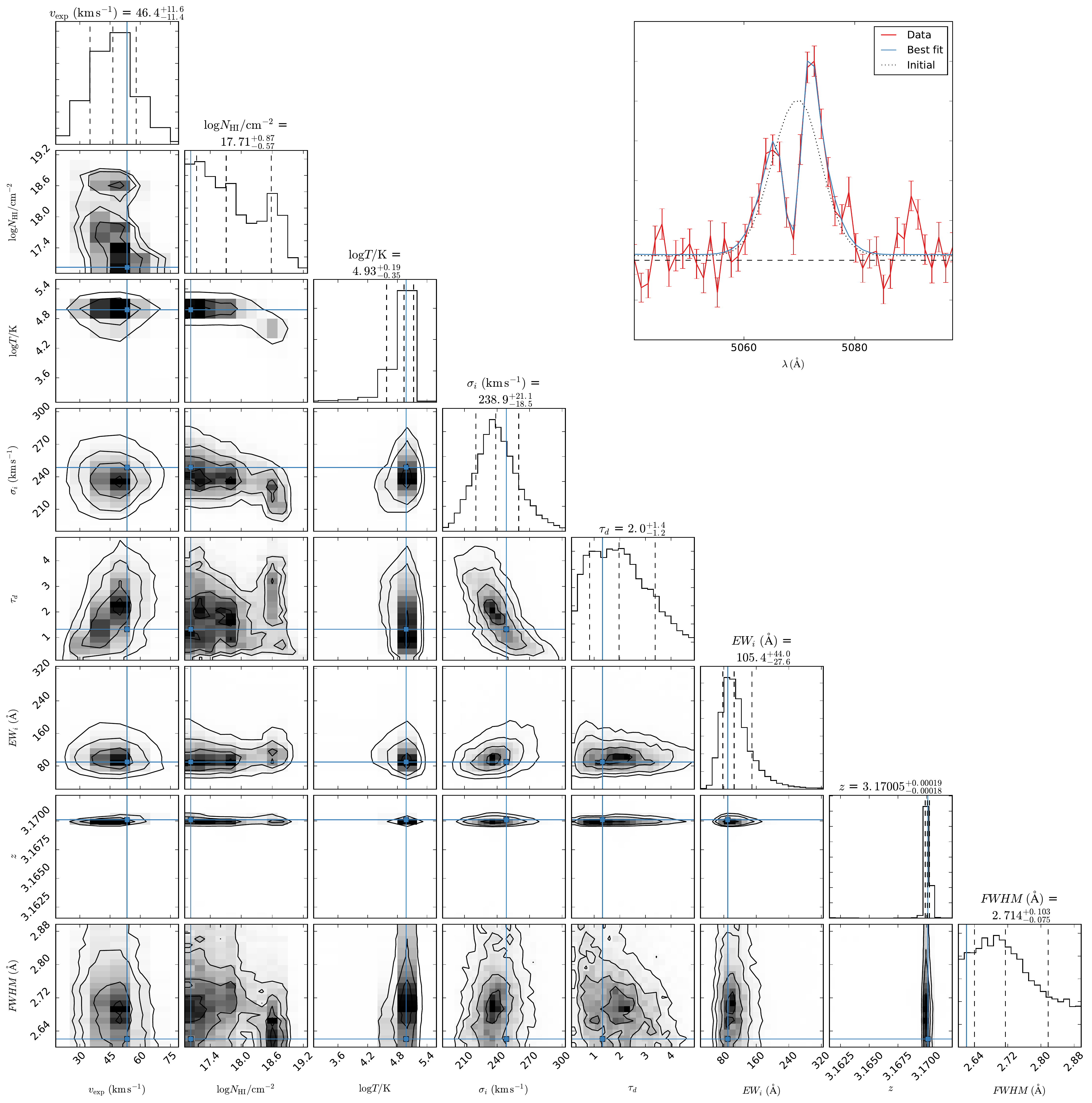}
\caption{Lya modelling results for NE-93b. The lines and legend
are the same as in Fig. \ref{fig:lya_112}.\label{fig:lya_113}}
\end{center}
\end{figure*}

\begin{figure*}
\begin{center}

\includegraphics[width=\textwidth]{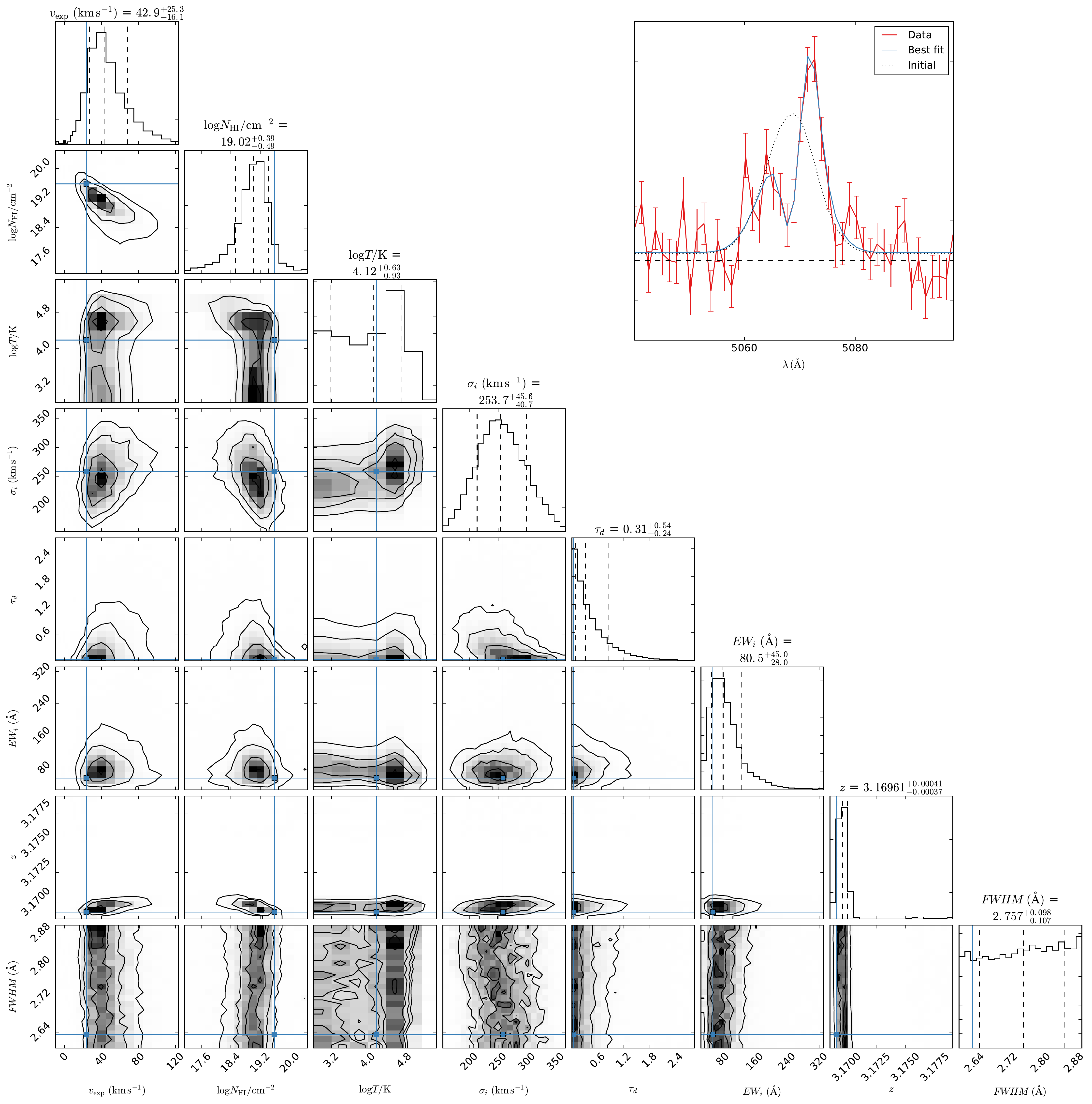}
\caption{Lya modelling results for NE-93a. The lines and legend
are the same as in Fig. \ref{fig:lya_112}.\label{fig:lya_1131}}
\end{center}
\end{figure*}

\begin{figure*}
\begin{center}

\includegraphics[width=\textwidth]{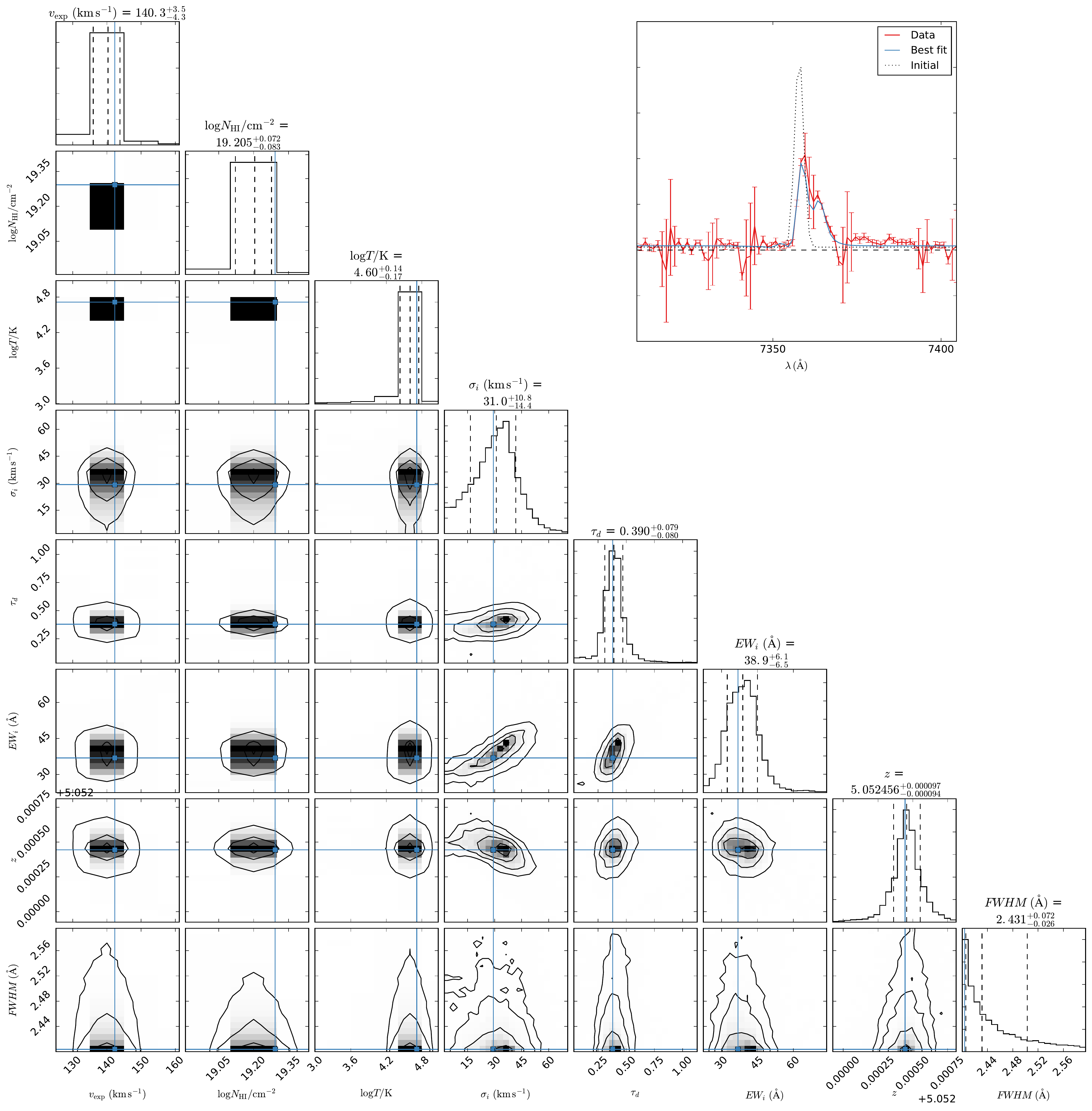}
\caption{Lya modelling results for NE-98a. The lines and legend
are the same as in Fig. \ref{fig:lya_112}.\label{fig:lya_114}}
\end{center}
\end{figure*}

\clearpage

\begin{figure*}
\begin{center}

\includegraphics[width=\textwidth]{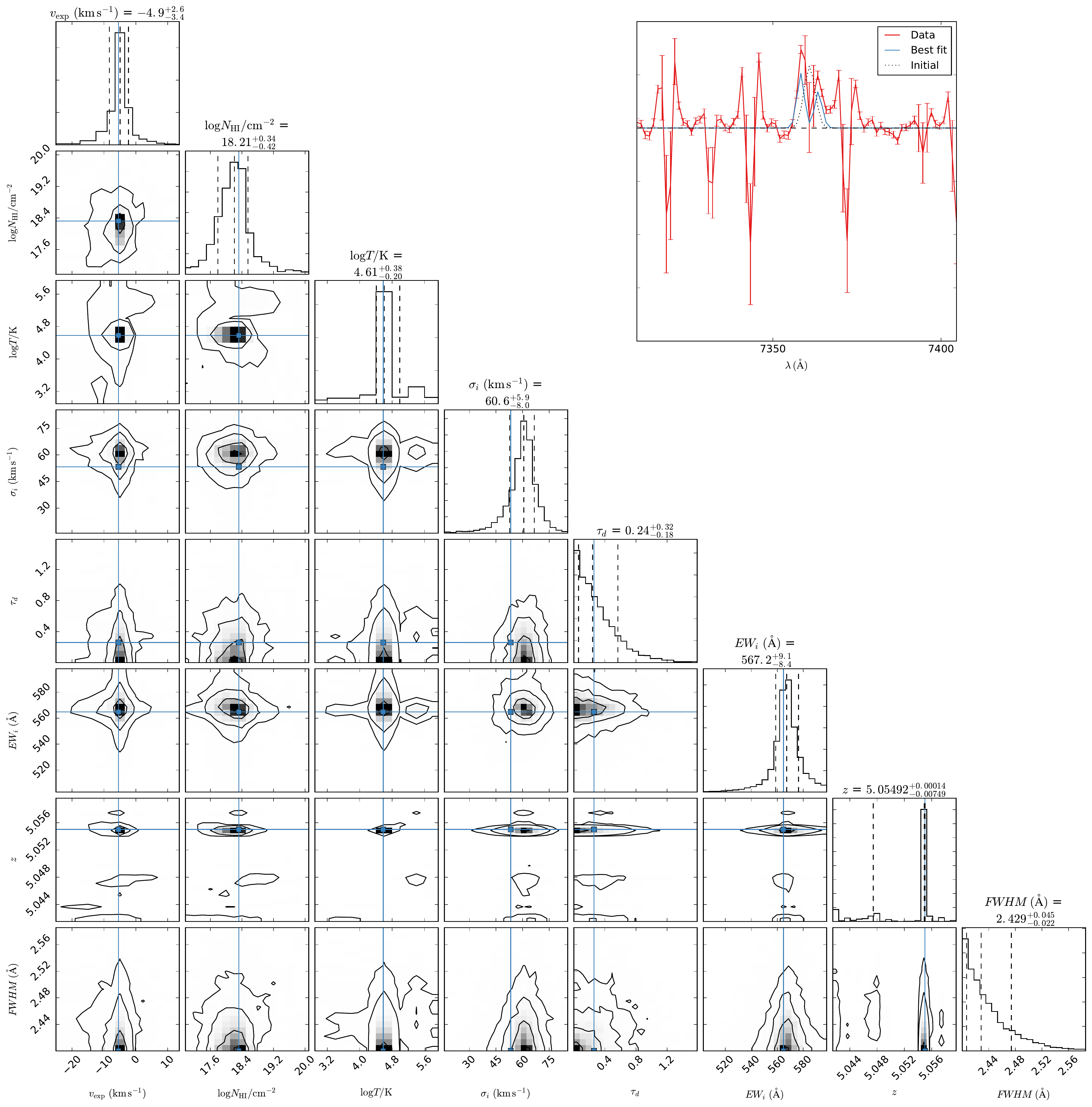}
\caption{Lya modelling results for NE-98b. The lines and legend
are the same as in Fig. \ref{fig:lya_112}.\label{fig:lya_1141}}
\end{center}
\end{figure*}

\begin{figure*}
\begin{center}

\includegraphics[width=\textwidth]{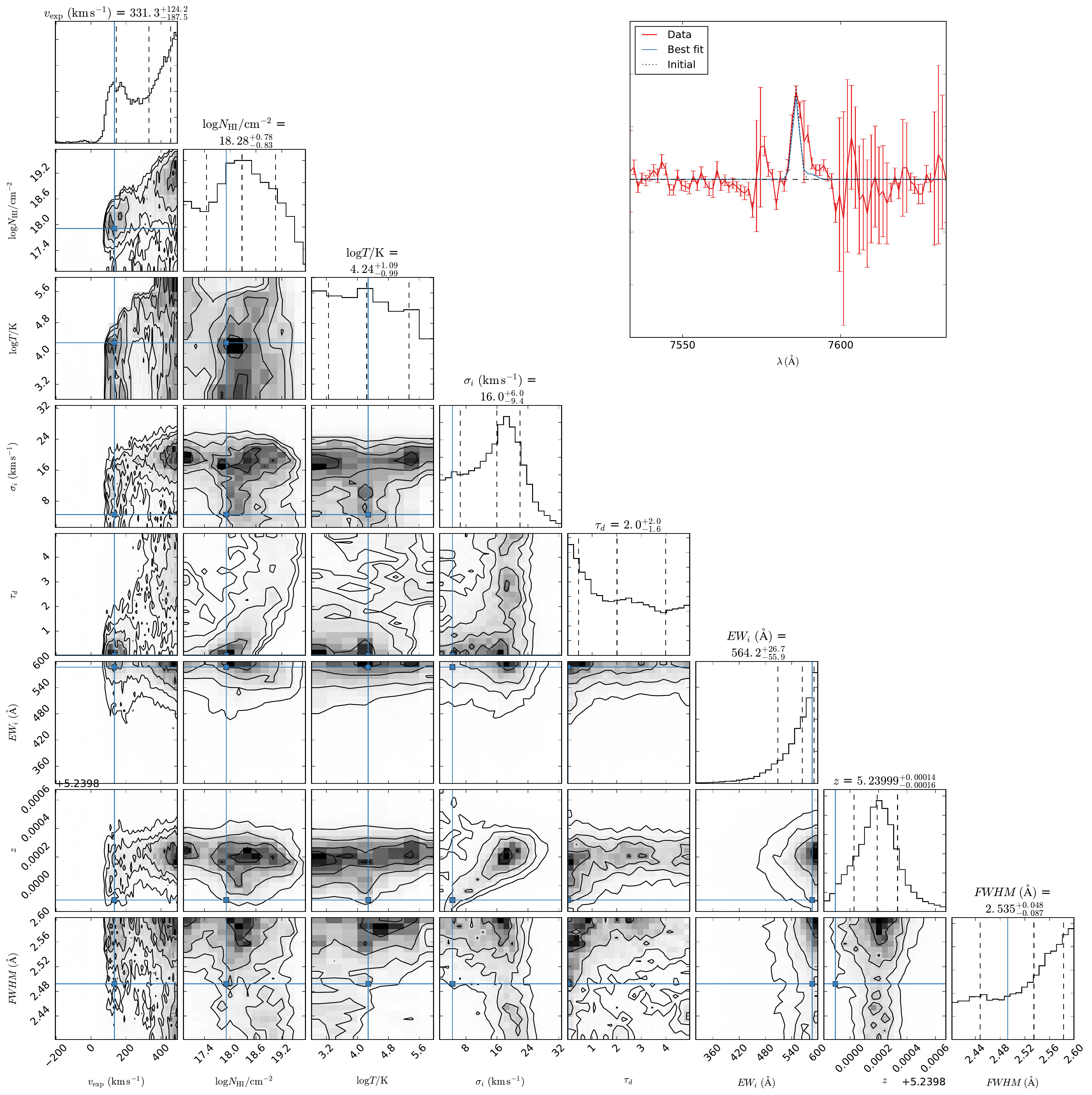}
\caption{Lya modelling results for NE-99a.The lines and legend
are the same as in Fig. \ref{fig:lya_112}.\label{fig:lya_1161}}
\end{center}
\end{figure*}

\clearpage

\begin{figure*}
\begin{center}

\includegraphics[width=\textwidth]{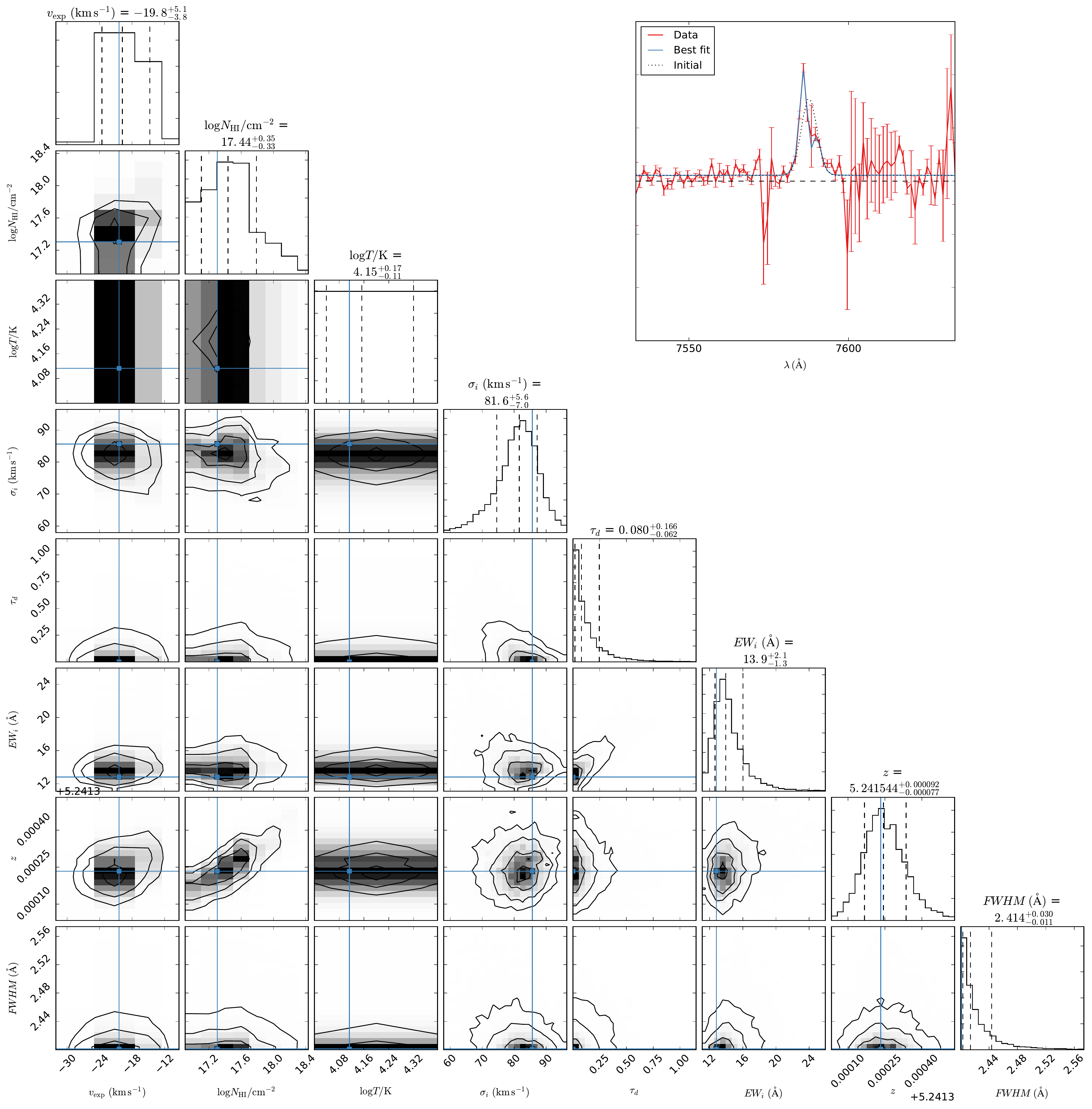}
\caption{Lya modelling results for NE-99b. The lines and legend
are the same as in Fig. \ref{fig:lya_112}.\label{fig:lya_1160}}
\end{center}
\end{figure*}

\begin{figure*}
\begin{center}
\includegraphics[width=\textwidth]{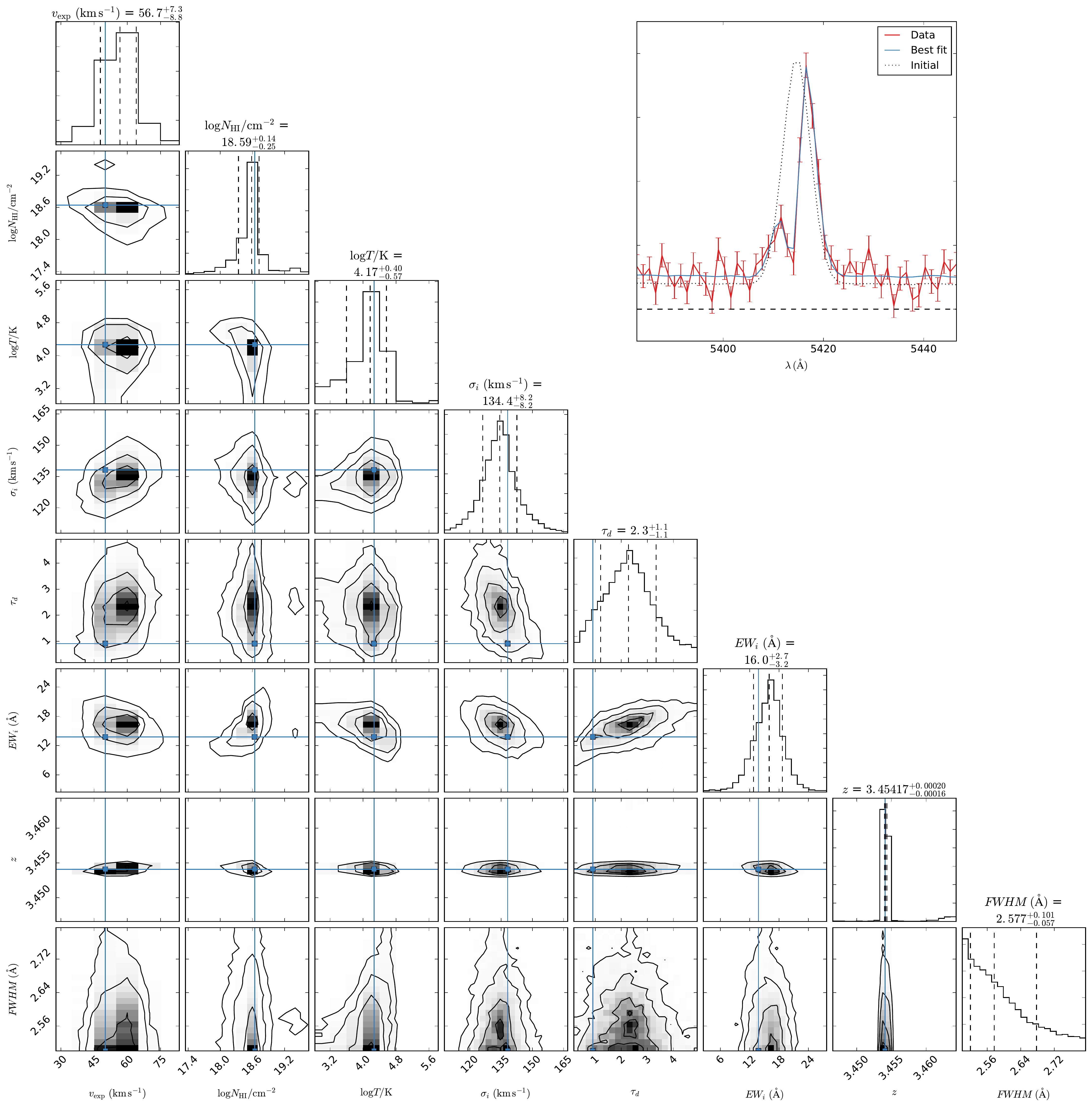}
\caption{Lya modelling results for NE-96. The lines and legend
are the same as in Fig. \ref{fig:lya_112}.\label{fig:lya_117}}
\end{center}
\end{figure*}

\clearpage

\begin{figure*}
\begin{center}
\includegraphics[width=\textwidth]{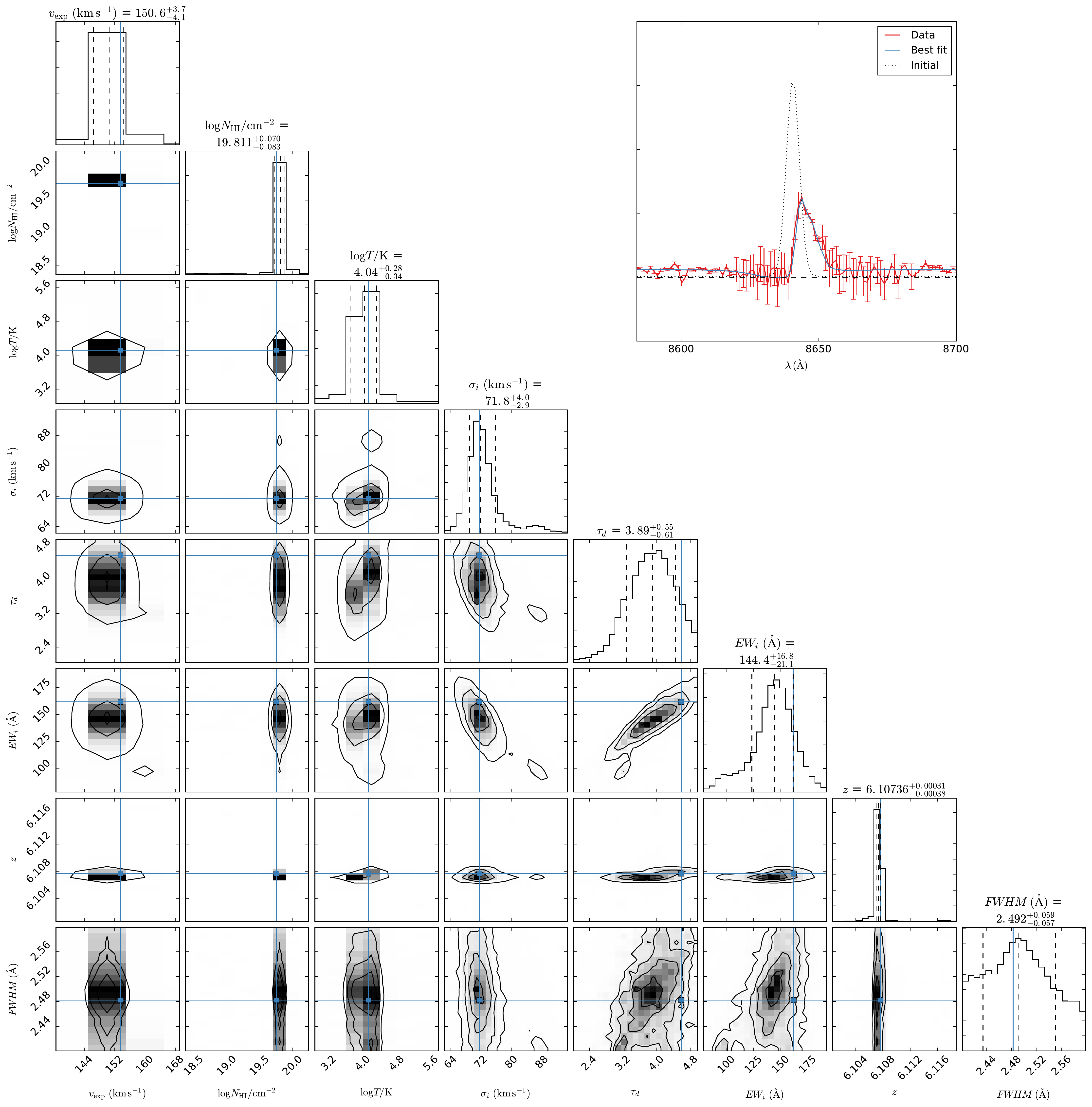}
\caption{Lya modelling results for SW-53a. The lines and legend
are the same as in Fig. \ref{fig:lya_112}.\label{fig:lya_530}}
\end{center}
\end{figure*}

\begin{figure*}
\begin{center}
\includegraphics[width=\textwidth]{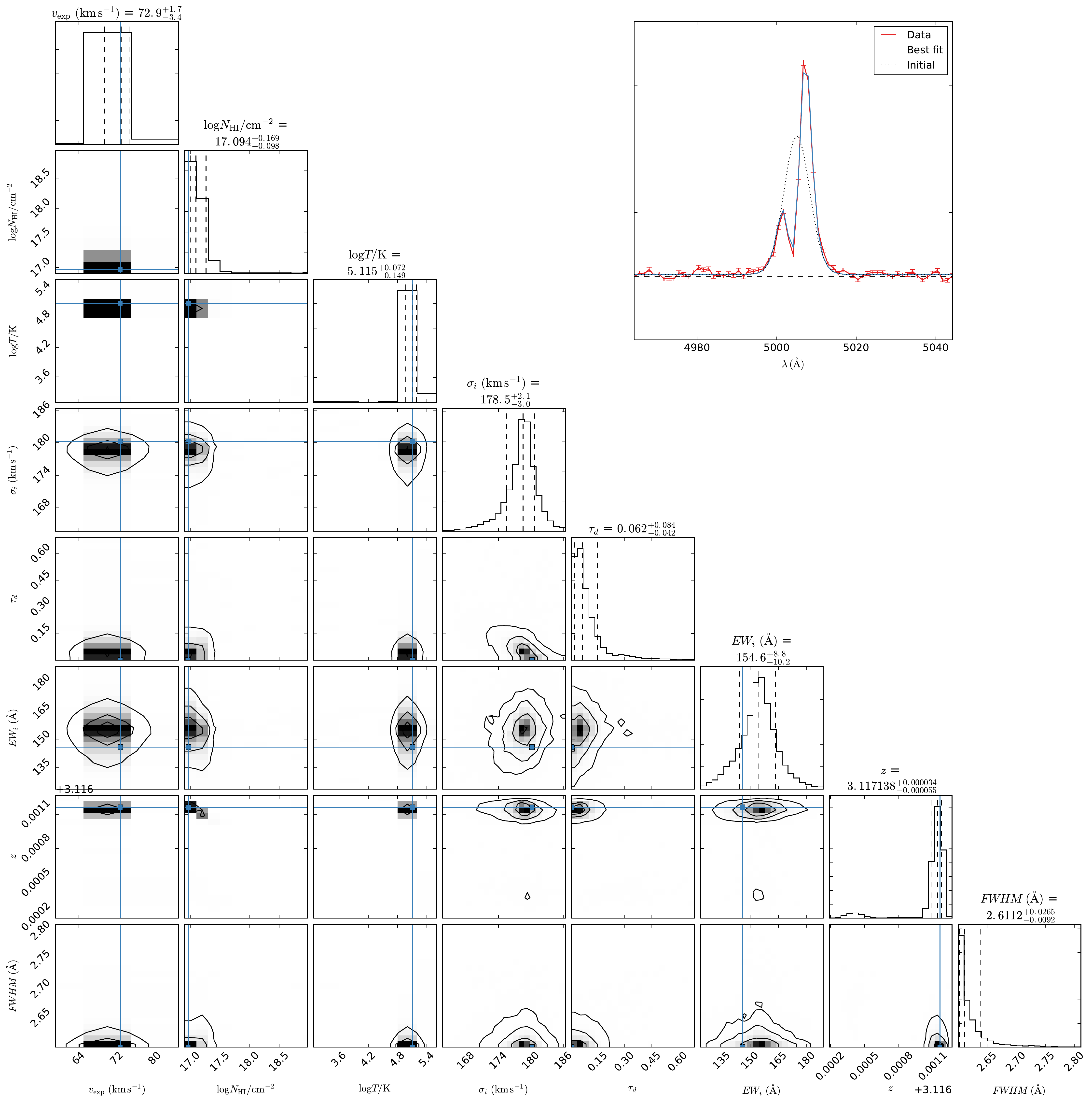}
\caption{Lya modelling results for SW-49a. The lines and legend
are the same as in Fig. \ref{fig:lya_112}.\label{fig:lya_490}}
\end{center}
\end{figure*}

\begin{figure*}
\begin{center}
\includegraphics[width=\textwidth]{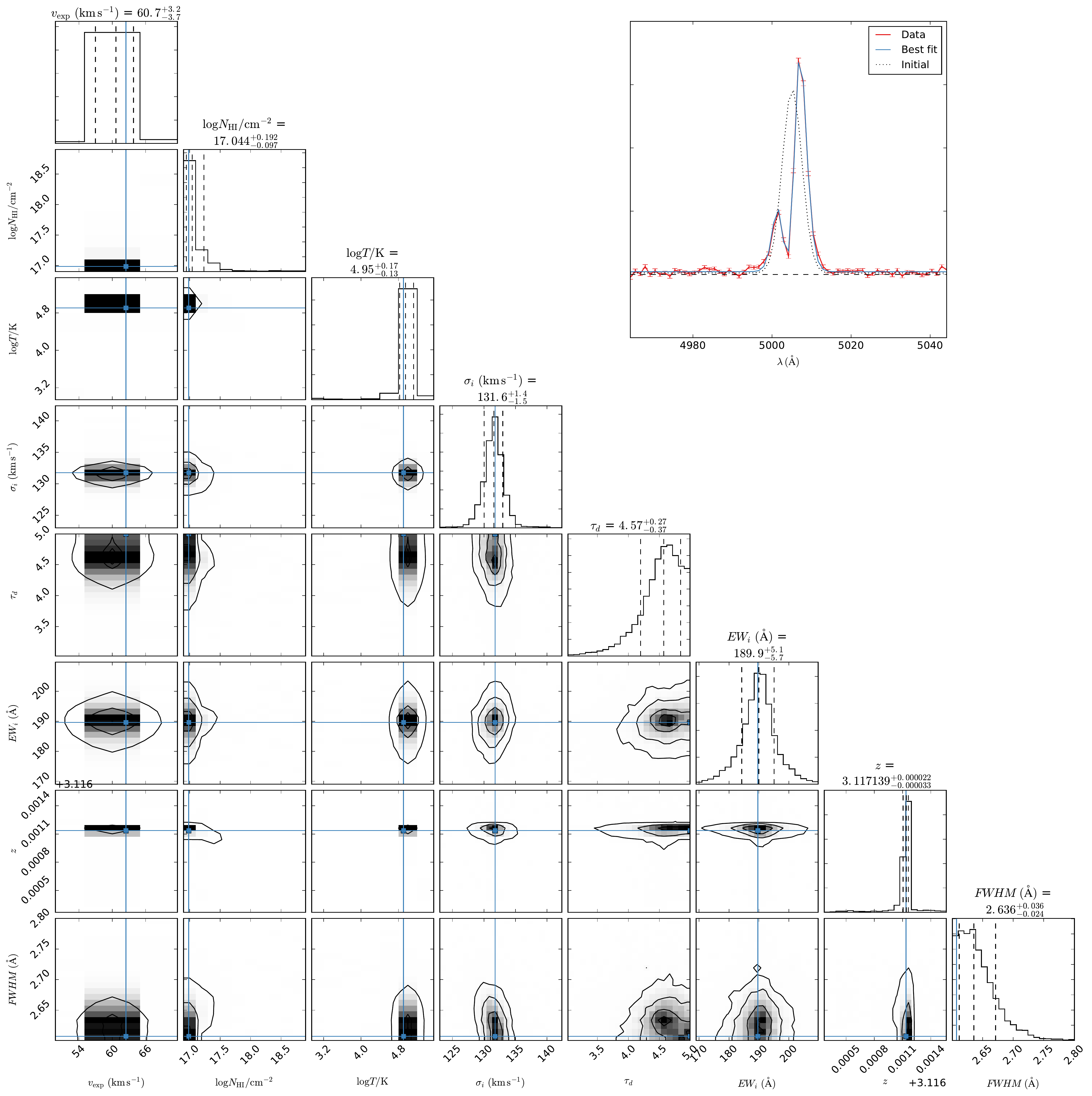}
\caption{Lya modelling results for SW-49b. The lines and legend
are the same as in Fig. \ref{fig:lya_112}.\label{fig:lya_491}}
\end{center}
\end{figure*}

\clearpage

\begin{figure*}
\begin{center}

\includegraphics[width=\textwidth]{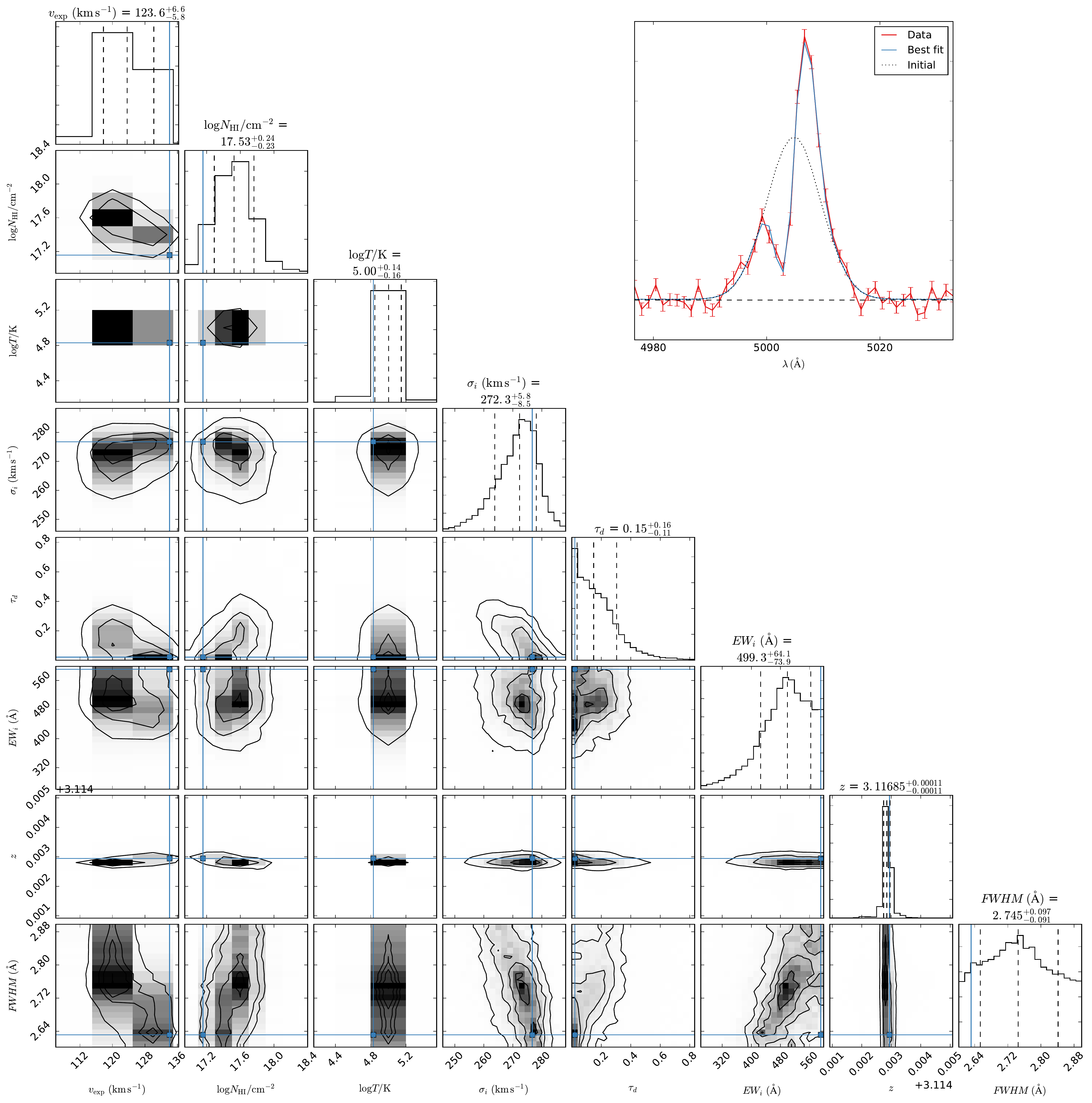}
\caption{Lya modelling results for SW-50.  The lines and legend
are the same as in Fig. \ref{fig:lya_112}.\label{fig:lya_50}}
\end{center}
\end{figure*}

\begin{figure*}
\begin{center}
\includegraphics[width=\textwidth]{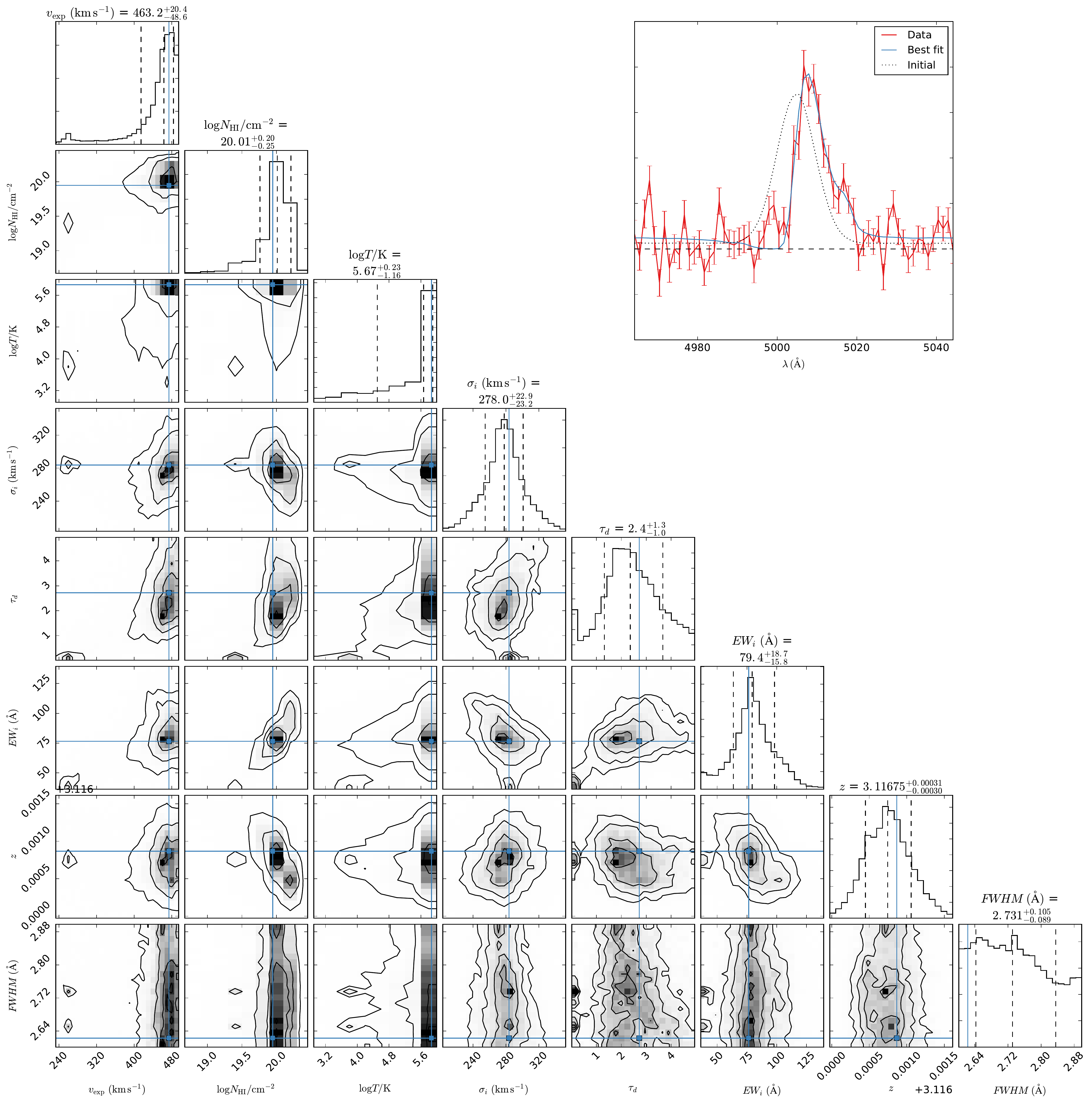}
\caption{Lya modelling results for SW-68a. The lines and legend
are the same as in Fig. \ref{fig:lya_112}.\label{fig:lya_271}}
\end{center}
\end{figure*}

\begin{figure*}
\begin{center}
\includegraphics[width=\textwidth]{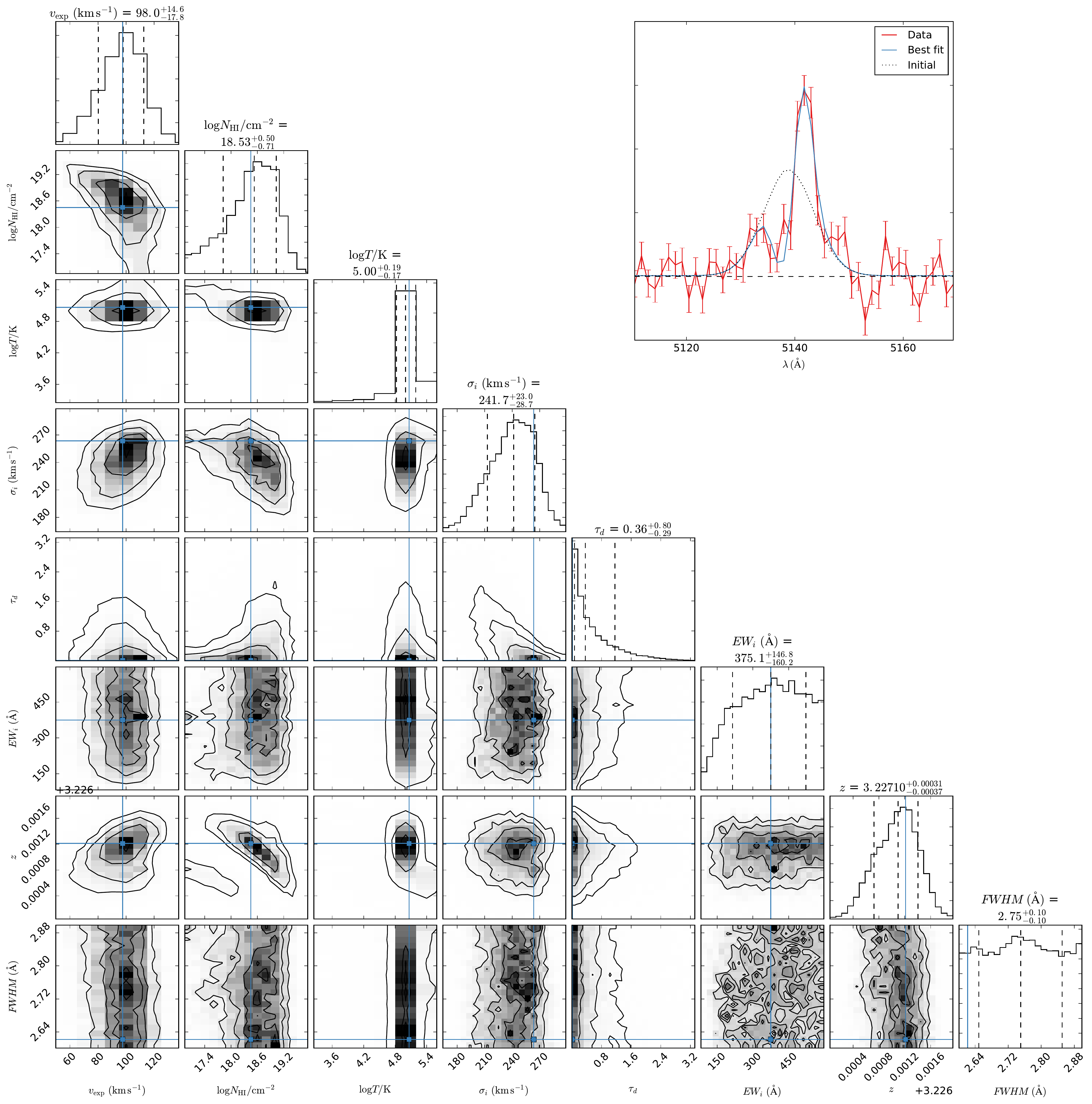}
\caption{Lya modelling results for SW-51. The lines and legend
are the same as in Fig. \ref{fig:lya_112}.\label{fig:lya_51}}
\end{center}
\end{figure*}

\begin{figure*}
\begin{center}
\includegraphics[width=\textwidth]{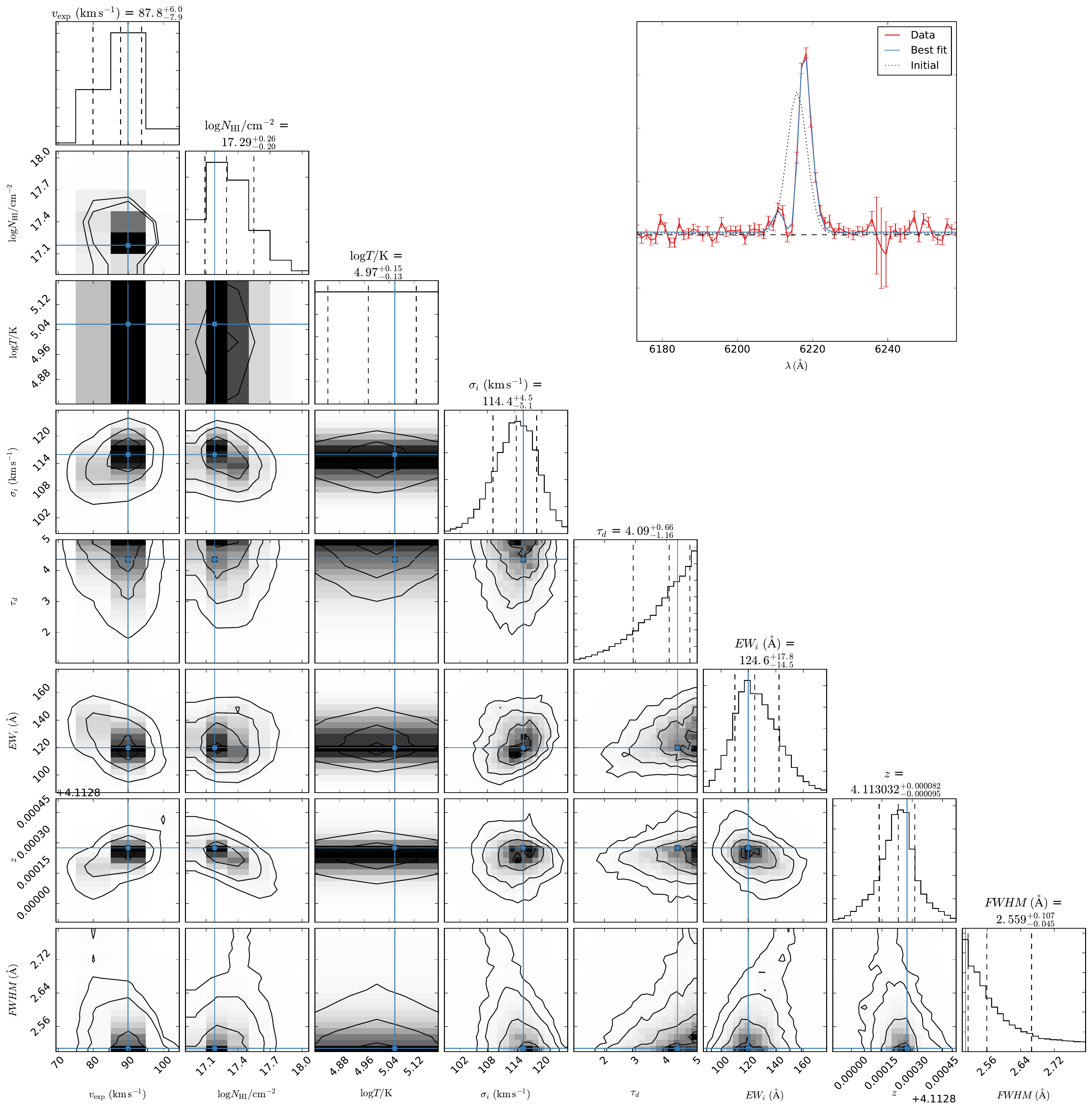}
\caption{Lya modelling results for SW-52a. The lines and legend
are the same as in Fig. \ref{fig:lya_112}.\label{fig:lya_520}}
\end{center}
\end{figure*}

\clearpage

\begin{figure*}
\begin{center}
\includegraphics[width=\textwidth]{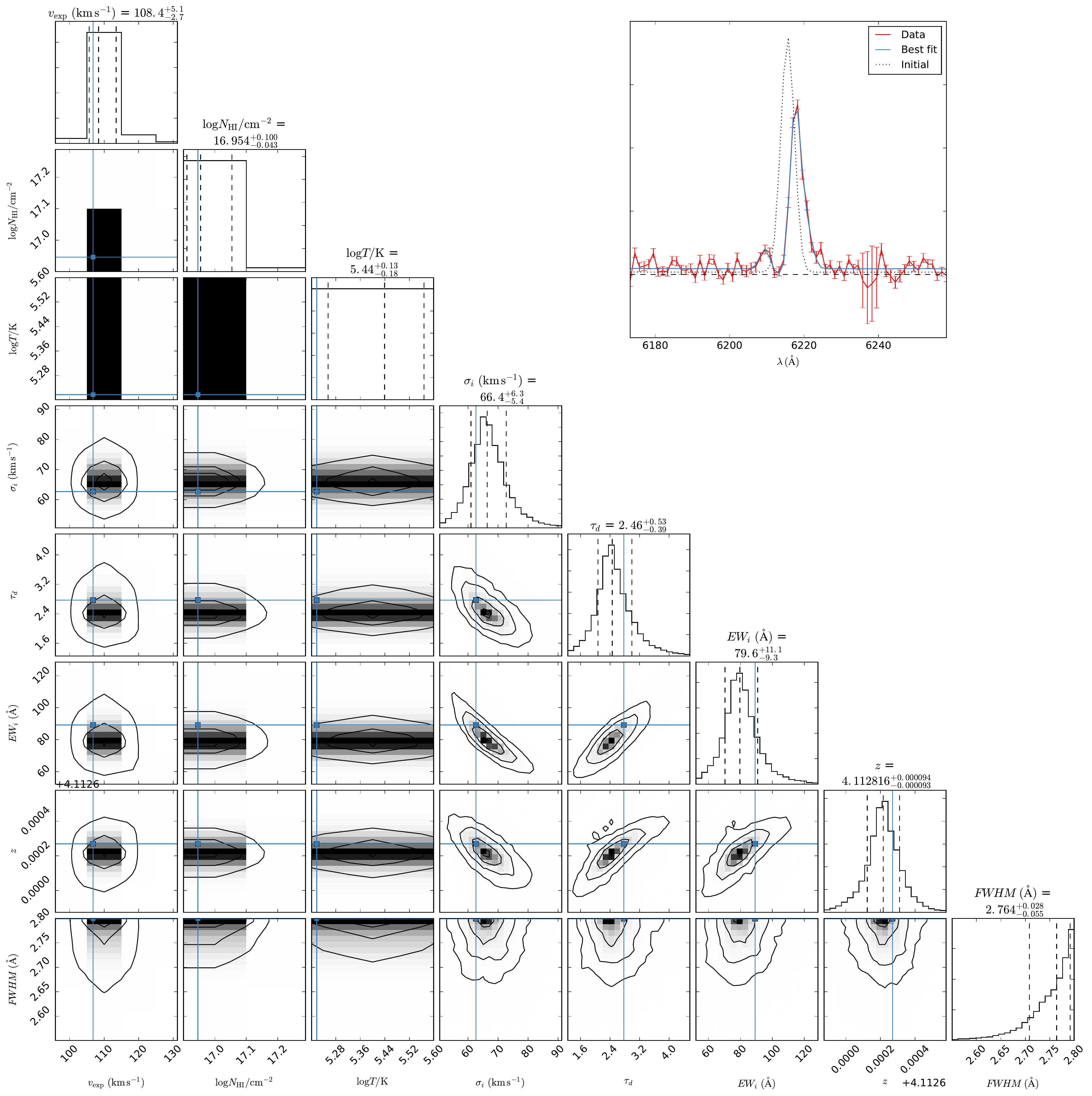}
\caption{Lya modelling results for SW-52b. The lines and legend
are the same as in Fig. \ref{fig:lya_112}.\label{fig:lya_521}}
\end{center}
\end{figure*}

\begin{figure*}
\begin{center}
\includegraphics[width=\textwidth]{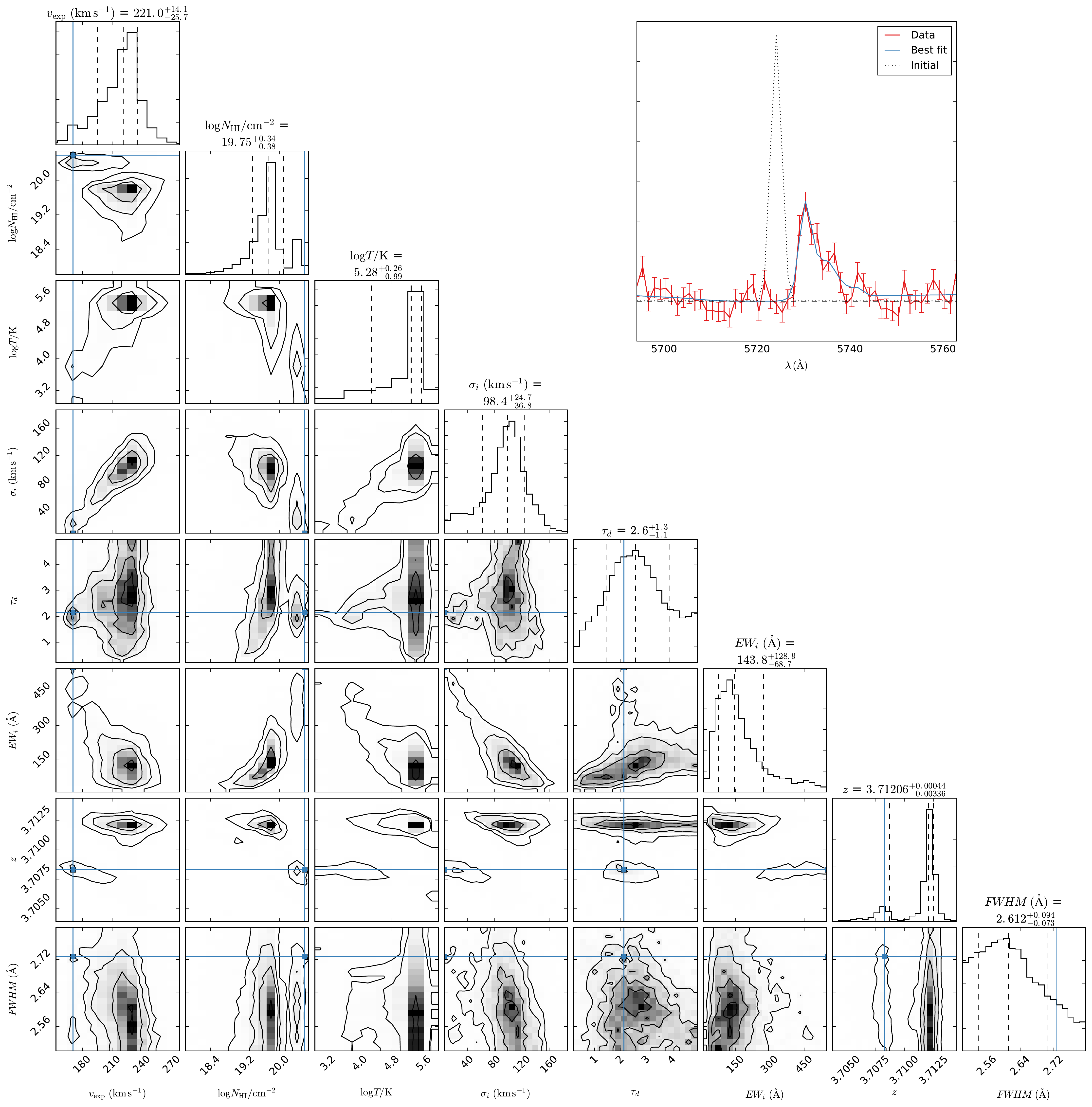}
\caption{Lya modelling results for NE-97.The lines and legend
are the same as in Fig. \ref{fig:lya_112}.\label{fig:lya_804}}
\end{center}
\end{figure*}